\documentclass[twocolumn,showkeys,prd,nofootinbib,floatfix,preprintnumbers]{revtex4-1}

\usepackage{multirow}
\usepackage{array}
\usepackage{hyperref}
\usepackage[utf8]{inputenc}
\usepackage{amsfonts,amsmath,amssymb} 
\usepackage{graphicx,graphics,color}
\usepackage{xcolor} %colore exoticos
\usepackage{gensymb}
\usepackage{longtable}
\usepackage{bbding}

\usepackage{hyperref}

\begin{document}

\title{Black Holes with Abelian and Non-Abelian Charges and Their Impact on Matter Accretion Flows}

\author{Gabriel G\'omez}
\email{gabriel.gomez.d@usach.cl}
\affiliation{Departamento de F\'isica, Universidad de Santiago de Chile,\\Avenida V\'ictor Jara 3493, Estaci\'on Central, 9170124, Santiago, Chile}

\author{\'Angel Rinc\'on}
\email{angel.rincon@ua.es}
\affiliation{Departamento de Física Aplicada, Universidad de Alicante,
Campus de San Vicente del Raspeig, E-03690 Alicante, Spain}

\author{Norman Cruz}
\email{norman.cruz@usach.cl}
\affiliation{Departamento de F\'isica, Universidad de Santiago de Chile,\\Avenida V\'ictor Jara 3493, Estaci\'on Central, 9170124, Santiago, Chile}

\begin{abstract}
In this paper, we study the black hole spacetime structure of a model consisting of the standard Maxwell theory and a $p$-power-Yang-Mills term. This non-linear contribution introduces a non-Abelian charge into the global solution, resulting in a modified structure of the standard Reissner-Nordstr\"{o}m black hole. Specifically, we focus on the model with $p=1/2$, which gives rise to a new type of modified Reissner-Nordstr\"{o}m black hole. For this class of black holes, we compute the event horizon, the innermost stable circular orbit, and the conditions to preserve the weak cosmic censorship conjecture. The latter condition sets a well-established relation between the electric and the Yang-Mills charges.
As a first astrophysical implication, the accretion properties of spherical steady flows around this new modified Reissner-Nordstr\"{o}m black hole are investigated in detail. Concretely, we compute the critical radius that establishes the condition for having stable transonic flow in terms of the local sound speed and the involved charges. Extensive numerical examples of how the Yang-Mills charge affects the accretion process of an isothermal fluid in comparison to the standard 
Reissner-Nordstr\"{o}m and Schwarzschild black holes are displayed. Finally, analytical solutions in the fully relativistic regime, along with numerical computations, of the mass accretion rate for a polytropic fluid in terms of the electric and Yang-Mills charges are obtained. As a main result, the mass accretion rate efficiency is considerably improved, with respect to the standard Reissner-Nordstr\"{o}m and Schwarzschild solutions, for negative values of the Yang-Mills charge.
\end{abstract}

\maketitle
%%%%%%%%%%%%%%%%%%%%%%%%%%%%%%%%%%55
\section{Introduction}
%%%%%%%%%%%%%%%%%%%%%%%%%%%%%%%%%%%

In classical and alternative theories of gravity (i.e. in General Relativity and beyond), the research in the context of Black Hole (BH) physics is quite relevant \cite{Misner:1973prb,Hartle:2003yu}. Given that BHs are simple solutions of the Einstein's field equations, and they incorporate several classical and quantum properties \cite{Grumiller:2022qhx}, they represent an ideal arena to get insights into profound questions still open, as 
i) how we can consistently combine both the classical and the quantum regimen, or 
ii) how to deal with certain singularities present in General Relativity (GR).
Black holes are parameterized at most by three fundamental quantities: 
i) the mass $M$, 
ii) the angular momentum $J$ and finally, 
iii) the charge $Q$, 
statement supported by the "no-hair theorem" \cite{heusler_1996}. 

BHs are the perfect example in which 
classical and quantum effects coexist in a non-trivial way and rule in different regimes. 
At this point, it is essential to mention the 
Hawking radiation \cite{Hawking:1975vcx}. Ignoring the details, Hawking demonstrated that
a black hole, with a surface gravity (labelled by $\kappa$), emits thermal radiation with a temperature given by $T_H =\kappa/(2\pi)$. The simplest case, i.e., a Schwarzschild black hole \cite{Schwarzschild:1916uq}, for a given a black hole mass, $M_0$, the temperature grows up when the black hole emits energy (consequence of its negative specific heat). Thus, albeit Hawking radiation has not been detected yet \cite{Hawking:1975vcx,Hawking:1974rv}, it has a special spot inside (among all the potential signatures related to the BH's observations \cite{EventHorizonTelescope:2019dse}) the different effects present in a black hole.

Numerous black hole solutions have been obtained within and beyond GR since the pioneering Schwarzschild solution, and we have three remarkable examples of four-dimensional black holes.  In addition to the Schwarzschild solution \cite{Schwarzschild:1916uq},
we have: i) the Reissner-Nordstr\"om solution \cite{1916anp...355..106r,1918knab...20.1238n},
ii) the Kerr solution \cite{kerr:1963ud} and, finally,
iii) the Kerr-Newman solution \cite{newman:1965my}. The above-mentioned four solutions are considered as the "black-hole" solutions of general relativity. Such examples have been significantly studied, computing a huge variety of their physical properties. To be more precise, 
the celebrated Reissner-Nordstr\"{o}m solution has been extensively studied in the literature \cite{1916anp...355..106r,1918knab...20.1238n,Wald:1984rg,Carroll:1997ar}, leading, among other things, to interesting similarities with the Kerr solution for rotating BHs: the electric charge plays an analogue role as the spin parameter does in the event horizon radius and in the dynamics of magnetized particles as the magnetar J1745-2900 orbiting around the supermassive BH Sagittarius A$^{\star}$ (Sgr A$^{\star}$) \cite{Juraeva:2021gwb}.

Classical self-gravitating configurations, such as soliton solutions (or boson stars) \cite{Bizon:1994dh,Seidel:1991zh,Herdeiro:2017fhv,Brito:2015pxa,Colpi:1986ye,Martinez:2022wsy}, are of great interest in astrophysics since they can serve as black holes mimickers \cite{Liebling:2012fv}. In general relativity and in the pure Yang-Mills theory, for instance, soliton solutions with localized energy densities do not exist, but in the coupled Eistein-Yang Mills scenario they can arise \cite{Volkov:1998cc}.
It is known, on the other hand, that the static Schwarzschild and Reissner-Nordstr\"om solutions are unique \cite{Israel:1967wq,Israel:1967za}, in the sense that in these solutions staticity implies spherical symmetry whereby regularity of the event horizon and the asymptotic behavior determine the solutions completely in terms of the mass and charge, as appropriate \cite{Ruffini:1971bza}. As a consequence, stationary electrovacuum BHs are axisymmetric and belong to a more general family of BH solutions called the Kerr–Newman BH \cite{Mazur:1982db}. 
Contrary to this claim, the uniqueness and no-hair theorems can be {\bf{violated}} by the presence of a non-Abelian gauge field \cite{Greene:1992fw}, leading to the Einstein Yang-Mills BH solution with non-trivial non-Abelian hair \cite{Volkov:1989fi,Bizon:1990sr,Kuenzle:1990is}) (see also Ref.~\cite{Volkov:1998cc} for other generalized non-Abelian BH solutions). 
Let us mention that the (non-Abelian) Yang–Mills theory is a natural generalization of the Abelian Maxwell theory. As mentioned, the idea of considering Yang–Mills theory with gravity was first studied in Ref. \cite{Bartnik:1988am} where a family of (purely magnetic) particle-like solutions was found.
The inclusion of a non-vanishing cosmological constant $\Lambda$ has been studied in Refs. \cite{Torii:1995wv,Mann:2006jc,Bjoraker:1999yd}.
Other non-trivial solutions involving the Yang–Mills theory can be found in \cite{Brihaye:2006xc,Mazharimousavi:2008ap} in higher spacetime dimensions, and in the context of the generalized SU(2) Proca theory in \cite{Martinez:2022wsy,Gomez:2023wei}.

Subsequently, novel and also non-trivial black hole solutions were found in the Einstein-Yang Mills theory by adding extra terms such as quartic self-interactions \cite{Gomez:2023wei}. An emblematic example emerges when the idea of non-linear electrodynamics is implemented in Einstein–Yang–Mills black hole solutions. 
Albeit simple, the most natural generalization appears when a power-law electromagnetic Lagrangian is included (see \cite{Gurtug:2010dr,Xu:2014uka,Panotopoulos:2019tyg,Hendi:2010bk,Panotopoulos:2020zbj,Gonzalez:2021vwp,Rincon:2017goj,Panotopoulos:2017hns,Rincon:2018sgd,Rincon:2018dsq,Panotopoulos:2018rjx,Rincon:2021hjj,Rincon:2021gwd,Panotopoulos:2022bky,Hendi:2017uau,Panah:2022cay,Hendi:2017lgb,EslamPanah:2021xaf} and references therein). 

The power Yang-Mills solutions are quite promising in the sense that they generalize naturally the well-known Einstein-Yang-Mills solution with the Wu-Yang Ansatz for the gauge field configuration, and allow non-trivial solutions to exist. 
Although Yang-Mills and power Yang-Mills solutions are relevant, they are less popular than the Maxwell and power Maxwell counterpart (see \cite{Meng:2017srs,ElMoumni:2018bkl,HabibMazharimousavi:2008zz,Biswas:2022qyl,Stetsko:2020nxb,Mazharimousavi:2009mb} and references therein).  

The non-linearities of YM invariant have been exploited in different theories, and it has been also coupled either minimally or non-minimally to other theories, resulting in generic BH solutions \cite{Meng:2017srs,ElMoumni:2018bkl,HabibMazharimousavi:2008zz,Stetsko:2020nxb,Gomez:2023wei}.
Nevertheless, one can think of one minimal extension that includes both the electric and magnetic YM charge in the simplest setup. This configuration gives precisely place to what we call the Einstein-Maxwell power Yang-Mills theory. Hence, the incorporation of the power Yang-Mills term to the Maxwell theory naturally generalizes the new spacetime structure of the emerging BH. We can immediately think of this as a modification of the standard Reissner-Nordstr\"{o}m solution: a new kind of modified Reissner-Nordstr\"{o}m BH with new properties. 

A consolidated route to test gravity theories that involve (non-trivial) fields beyond the vacuum case is to study the behavior of matter (and the same fields) and particles test around the emerging BH solutions. A variety of phenomena such as gravitational lensing \cite{Cunha:2018acu}, the movement of the so-called $S$-stars around SgrA$^\star$ in our Galactic center \cite{Ghez:2003qj} and accretion of matter through optically thick accretion disk \cite{EventHorizonTelescope:2019ths,EventHorizonTelescope:2019dse,EventHorizonTelescope:2022wkp}, they all have been the central targets in recent observational projects to understand not only how matter behaves in extreme environments but also to comprehend the spacetime structure itself, which is crucial to study the associated astrophysical processes. This latter aspect, in turn, provides an appealing way to test gravity theories beyond Einstein's general relativity theory. A particular program under this perspective is to study the stability conditions under which accretion of matter onto BHs can take place \cite{Ramirez-Velasquez:2019cgr}, ensuring thus the astrophysical viability of the theory.

Motivated by this, the primary goal of this paper is to investigate the steady accretion flows around this new class of black holes. Specifically, we aim to address the main question of how efficient the mass accretion rate is in this new setting compared to the standard Reissner Nordstr\"{o}m solution. To accomplish this, our study begins by examining the structure of the event horizon and the conditions required to prevent the occurrence of a naked singularity. These aspects are essential to establish the transonic conditions of steady flows and, therefore, the mass accretion rate. 
In this respect, the study of accretion onto black holes for non-neutral solutions in four-dimensional space-time has been studied on several occasions (see e.g. \cite{Babichev:2008jb,Ficek:2015eya,Abbas:2018ygc,Abbas:2019wcg,John:2019was,Perez:2017spz,Feng:2022bst,Ravanal:2023ytp} and references therein). Bondi accretion is an interesting subject in astrophysics serving as a probe of concept of several phenomena \cite{Bondi:1952ni,Richards:2021zbr,Shapiro:1983du}.

This paper is organized as follows: 
after this compact introduction, 
in Section \eqref{sec:2}, we present the main ingredients and basic equations that describe the BH spacetime structure. Subsequently, we present in Sec. \eqref{sec:3} the equations that describe the spherical steady accretion flows in a general theory of gravity. Then, in section \eqref{sec:4}, we particularize the equations for the Einstein-Maxwell $p$-power Yang-Mills theory for the specific case $p=1/2$ and compute the accretion rate for isothermal fluids and, more generally, for polytropic fluids in the fully relativistic regime. Finally, in the last section, a general discussion of our main results and some observational perspectives of this work are presented.
We will use the mostly positive metric signature, $(-,+,+,+)$, and work in geometrical units, i.e., $c=1=G$.

%%%%%%%%%%%%%%%%%%%%%%%%%%%%
\section{Spacetime structure of the Einstein-Maxwell-Power Yang-Mills black} \label{sec:2}

This section describes the main ingredients of the theory that describe a new non-linear black hole solution.  
We work in a 4-dimensional theory which includes three  ingredients:
  i) The Einstein-Hilbert term,
 ii) The Maxwell invariant, and
iii) the Power Yang-Mill invariant.
Accordingly, the total action becomes:
\begin{align}
\begin{split}
     S_0 = \frac{1}{2}\int \sqrt{-g}\ \mathrm{d}^4x 
     \Bigg[ & R -
     \mathcal{F}_{\text{M}}  -
     \mathcal{F}_{\text{YM}}^{p} 
     \Bigg],
     \label{action}
\end{split}
\end{align}
where we have used the conventional definitions, i.e., 
\textcolor{purple}{
%$\kappa \equiv 8\pi G = 1$ is the Einstein's constant, $G$ is the Newton's constant, 
%
}
$R$ is the Ricci scalar, $g$ is the determinant of the metric tensor $g_{\mu \nu}$, $p$ is a real parameter that introduces possible non-linearities in the Yang-Mills theory.
Moreover, we have defined the Maxwell invariant
\begin{align}
    \mathcal{F}_{\text{M}} &= F_{\mu \nu} F^{\mu \nu},
\end{align}
and, subsequently, we have defined the power Yang-Mills term with the help of the following relations
\begin{align}
    \mathcal{F}_{\text{YM}} &=  \mathbf{Tr}  
    \Bigl(
    F_{\lambda \sigma}^{(a)}F^{(a) \lambda \sigma}
    \Bigl),
\\
 \mathbf{Tr} (\cdot) &= \sum_{a=1}^{3}  (\cdot)
\end{align}
As usual, $F_{\mu \nu}$ is the electromagnetic field strength, and $F_{\mu \nu}^{(a)}$ is the gauge strength tensor that are defined in terms of the potentials $A_\nu$ and $A_\nu^{(a)}$, respectively, as follows:
\begin{align}
F_{\mu \nu} & \equiv \partial_\mu A_\nu - \partial_\nu A_\mu\,,
\\ 
F_{\mu \nu}^{(a)} & \equiv \partial_\mu A_\nu^{(a)} - \partial_\nu A_\mu^{(a)} + \frac{1}{2\sigma} C_{(b)(c)}^{(a)} 
A_{\mu}^{(b)} A_{\nu}^{(c)}\,,
\end{align}
where the Greek indices run from 0 to $3$ and  $a$ is the internal gauge index running from $1$ to $3$. 
Also, notice that 
$C_{(b)(c)}^{(a)}$ symbolize the structure constants of 3 parameter Lie group $G$, $\sigma$ is an arbitrary coupling constant, $A^{(a)}_{\mu}$ are the $SO(3)$ gauge group Yang–Mills potentials, and finally $A_{\mu}$ is the conventional Maxwell potential.
To be more precise, for the YM field, we use the well-known magnetic Wu-Yang ansatz 
\cite{Mazharimousavi:2008ap,Yasskin:1975ag}
\begin{align}
    \mathbf{A}^{(a)} = \frac{q_{\text{YM}}}{r^2}(x_i d x_j - x_j dx_i),
\end{align}
with
\begin{align}
& 2 \le j+1	\le i 	\le 3 ,  
\\
& 1 \le a \le 3,
\end{align}
and
\begin{align}
    r^2 &= \sum_{i=1}^{3} x_i^2.
\end{align}
The Maxwell potential 1-form is given by
\begin{align}
        \mathbf{A} &= \frac{Q}{r}dt,
\end{align}
where $Q$ is the electric charge and $q_{\text{YM}}$ is the YM charge.
Moreover, the Maxwell field and the non-Abelian gauge field are decoupled to each other but, of course, they are coupled linearly through gravity. 
Notice that the general solution (for an N-dimensional case) was first studied in \cite{Zhang:2014eap} in the presence of a cosmological constant. In that work, the main thermodynamics, including a review of the first law of black hole thermodynamics in the extended phase space, was also computed. 
To maintain the discussion self-contained, we present the corresponding modified Einstein's field equations, i.e., 
\begin{align}
    G_{\mu \nu} + \Lambda g_{\mu \nu} &= T_{\mu \nu},
\end{align}
where we have defined two contributions to the energy-momentum tensor, i.e., 
\begin{align}
    T_{\mu \nu} \equiv T_{\mu \nu}^{\text{M}} + T_{\mu \nu}^{\text{YM}},
\end{align}
with the above tensors defined as
\begin{align}
    T_{\mu \nu}^{\text{M}} &= 2 F_{\mu}^{\lambda} F_{\nu \lambda} - \frac{1}{2} F_{\lambda \sigma} F^{\lambda \sigma} g_{\mu \nu}, 
    \\
    T_{\mu \nu}^{\text{YM}} &= -\frac{1}{2} g_{\alpha \mu}
    \Bigg[
    \delta^{\alpha}_{\nu} \mathcal{F}_{\text{YM}}^p - 4p \mathbf{Tr} 
    \Bigl(
    F_{\nu \lambda}^{(a)}F^{(a) \alpha \lambda}
    \Bigl)
    \mathcal{F}_{\text{YM}}^{p-1}.
    \Bigg]
\end{align}
%
%where we take advantage of the expressions:
%
By Variation with respect to the gauge potentials $\bf{A}$ and $\bf{A}^{(a)}$ provides the Maxwell and Yang Mills equations, respectively
\begin{align}
  \mathrm{d}\Bigl( {} ^{\star} \mathbf{F}  \Bigl) &= 0,
\\
    \mathbf{d}\Bigl( {} ^{\star} \mathbf{F}^{(a)} \mathcal{F}_{\text{YM}}^{p-1}  \Bigl)   +  \frac{1}{\sigma} C^{(a)}_{(b)(c)} \mathcal{F}_{\text{YM}}^{p-1} \mathbf{A}^{(b)} \wedge^{\star} \mathbf{F}^{(c)} &= 0,
\end{align}
where $\star$ means duality. For the Wu-Yang ansatz, the trace of the Yang-Mills takes the form:
\begin{align}
    \mathcal{F}_{\text{YM}} =  \frac{q^2_{\text{YM}}}{r^4},
\end{align}
which is positive, allowing us thus to consider all rational numbers for the $p$-values. It is evident that for $p = 1$, the formalism reduces to the standard Einstein-Yang Mills theory.
In the same direction, we have remarkable examples of which Yang-Mills term is included, for instance, \cite{HabibMazharimousavi:2008zz,Mazharimousavi:2009mb}.
Now, considering a spherically symmetric spacetime in Schwarzschild coordinates, we can write the line element as
\begin{equation}
ds^{2} = - f(r) dt^{2} + f(r)^{-1} dr^{2} + r^{2}  (d\theta^2 + \sin^{2}\theta d\phi^{2}),
\label{lineelement}
\end{equation}
with $r$ being the radial coordinate. 
Ignoring the unnecessary details, the differential equation for the lapse function, coming from the Einstein field equation, is given by
\begin{align}
\frac{\mathrm{d}f(r)}{\mathrm{d}r} + \frac{1}{r}f(r) 
&=
\frac{1}{r} - \frac{Q^2}{r^3} - \frac{2^{p-1}q^{2p}_{\text{YM}}}{r^{4p-1}},
\end{align}
and, identifying the total derivative on the left-hand side we simply write the well-known form
\begin{align}
    \frac{\mathrm{d}}{\mathrm{d}r} \Bigl( r f(r) \Bigl) 
    &=
    1 - \frac{Q^2}{r^2} - \frac{2^{p-1}q^{2p}_{\text{YM}}}{r^{4p-2}}.
\end{align}
Thus, the metric function admits the general solution (setting the constant of integration as $-2M$):
\begin{equation}
f(r)=1 - \frac{2M}{r} + \frac{Q^{2}}{r^{2}} + \frac{Q_{\rm YM}}{r^{4p-2}}.
\label{metricfunction}
\end{equation}
Notice that the power Yang-Mills solution\footnote{Keeping only the power Yang-Mills solution with $p=1$ leads also to the Reissner-Nordstr\"{o}m solution.}  assembles linearly the standard Reissner-Nordstr\"{o}m solution, and the Yang-Mills charge $q_{\rm YM}$ is related to its normalized version as \cite{Mazharimousavi:2009mb}
\begin{align}
  Q_{\rm YM} &\equiv \frac{2^{p - 1}} {4 p - 3}   q_{\rm YM}^{2p},
\end{align}
for $p\neq3/4$.  For $p=3/4$ a radial logarithmic dependency  appears from the solution which overclouds analytical treatments, so we left with the case $p\neq 3/4$ for simplicity. We also have to restrict this study to some values of $p$ that satisfy some of the energy conditions of general relativity (in the pure Yang-Mills case) and provide nearly asymptotic flat solutions for a wide range of Yang-Mill charge values\footnote{For instance, taking $Q=0.6M$, and $p=1/3,1/4$, $Q_{YM}\sim \mathcal{O}(\pm 10^{-3})$ whereas $p=1/2$ allows $Q_{YM}\sim \mathcal{O}(\pm 1)$. The nearly asymptotic flat solution with a residual Yang-Mills charge is achieved either from above or below, depending on the Yang-Mills charge sign since the associated term dominates at large radial coordinate.}. One may ask whether such a Yang-Mills charge is related to Noether currents or any symmetry in the theory. The answer is not. The magnetic charge is topological and comes from the Bianchi identity\footnote{We thank J. F. Rodriguez for clarifying this point.} \cite{Abbott:1982jh}.

After exploring some possible $p$-exponents, we deal concretely with a simple, by still intriguing, case $p=1/2$ because of: 
i) it is consistent with the well-known energy conditions of general relativity and the causality condition \cite{Mazharimousavi:2009mb}
and 
ii) it modifies the Reissner-Nordstr\"{o}m spacetime structure in a non-trivial but still manageable way, unlike other cases explored that lead to obscure tremendously the solution.
Moreover, this case provides an illuminating analytical solutions for the inner (Cauchy) $r_{-}$ and external (event horizon) $r_{+}$ radii:
\begin{equation}
r\pm = \frac{M\pm \sqrt{M^{2}-Q^{2}(1+Q_{\rm YM})}}{1+Q_{\rm YM}}.
\label{horizons}
\end{equation}
We call this solution henceforth modified Reissner-Nordstr\"{o}m (MRN) solution with $Q_{\rm YM}\neq-1$. 
It is interesting to see that the modification to the standard Reissner-Nordstr\"{o}m is, in fact, effective on account of the denominator quantity; otherwise, a simple charge redefinition would be carried out like for the case $p=1$. Notice that for the exponent $p=1/2$, $Q_{\rm YM}$ is a dimensionless parameter.
The precise form for the event horizon radius is of great importance because, among other things, it sets a well-defined relation between both charges for preventing the naked singularity. It yields\footnote{Observational inferences of the EHT horizon-scale image of SgrA$^{\star}$ set at $1\sigma$ and $2\sigma$, respectively, $Q/M\lesssim0.8$ and $Q/M\lesssim0.95$. So extremal RN BH and the naked singularity regime $1 < Q/M\lesssim\sqrt{9/8}$ are ruled out \cite{Vagnozzi:2022moj}. Since such constraints are not applicable to our MRN BH, we allow extremal solutions in our description, but not naked singularities, for merely astrophysical purposes.}
\begin{equation}
    Q_{\rm YM}>-1\; \land\; 0<\frac{Q}{M}<\sqrt{\frac{1}{1+Q_{\rm YM}}}.\label{conditionsWCC}
\end{equation}
Naturally, the standard constraint for the RN BH is contained in the previous expression  and fully recovered  in the limit $Q_{\text{YM}} \rightarrow 0$, leading to $Q/M < 1$ as it should be.
For the allowed range of values of both charges, the corresponding horizons $r_{+}$ are completely determined. Some particular cases within the ranges 
 $-1<Q_{\rm YM}<1$ and derived constraint $Q\in(0,Q_{\rm Max})$ from Eq.~(\ref{conditionsWCC})
are shown in table \ref{tab:charges} to illustrate the emergence of the new structure of the MRN  black hole in terms of both charges as compared with Schwarzschild and RN cases. Another appealing solution embedded here is a kind of modified  Schwarzschild BH, or a bit more precisely, a trivially charged Yang-Mills solution, which is attained for $Q=0$. For this solution clearly $r_{-}=0$ and the event horizon becomes
\begin{equation}
   r_{+}^{\rm YM}=\frac{2 M}{1+Q_{\rm YM}},\label{eqn:MSch} 
\end{equation}
with the restriction $Q_{\rm YM}\neq-1$. On the contrary, for $Q=Q_{\rm Max}$ (upper value in the second column of table \ref{tab:charges} for a given $Q_{\rm YM}$), $r_{+}$ and $r_{-}$ each coincides with, which leads to a family of extremal BH solutions. This particular case corresponds to the lower possible value of $r_{+}$. It is interesting to notice that when $Q_{\rm YM}$ goes from the lowest possible value to larger ones, $Q$ must be reduced consistently to preserve the cosmic censorship conjecture, leading thereby to different extremal cases as compared to the standard RN BH solution:
%$r_{+}^{\rm RN}=1$ 
$r_{+}^{\rm RN}/M=1$
with $Q/M=1$. For instance, for $Q_{\rm YM}=1$, the corresponding maximum electric charge is now $Q/M=0.70710$ with a horizon $r_{+}=r_{+}^{\rm RN}/2$. For very large values $Q_{\rm YM}\gg1$, $r_{+}=r_{-}\to0$. 
In the opposite limit, $Q_{\rm YM}/M\to -1$ which is the smallest asymptotic value, $r_{+}\to \infty$. This is, of course, unphysical as a result of the divergence appearing in Eq.~(\ref{horizons}). 
Taking (the non-conservative) lower value $Q_{\rm YM}/M=-0.9$, results in the extremal case $r_{+}=10 r_{+}^{\rm RN}$. In general, for $Q_{\rm YM}<0$, the event horizon can be formed far beyond the standard Schwarzschild case with a maximum value corresponding to the Einstein-YM power case ($Q=0$). In the opposite range $Q_{\rm YM}>0$, the horizon can be below the Cauchy horizon of the standard RN case when $Q\neq0$ and matches the extreme RN case for $Q_{\rm YM}=1$ and $Q=0$. This analysis is summarized in Table.~\ref{tab:charges} where other combinations of charges give place to extremal solutions as well. 

Notice that the fact of having two charges leads to two-fold degeneracy of the event horizon. This is not illustrated  in Table.~\ref{tab:charges}. Let us then expose some concrete examples for completeness in our analysis. The case $Q_{\rm YM}=-0.25$ with $Q/M=1$ leads to $r_{+}/M=2$, as in the Schwarzschild case, but with a smaller Cauchy horizon $r_{-}/M=0.6666$. Likewise, the case $Q_{\rm YM}=-0.36$ with $Q/M=1.2$ also leads to $r_{+}/M=2$, but this time with a larger Cauchy horizon $r_{-}/M=1.125$. So there are multiple choices of charge values that provide similar results. In general, for a given $Q$ value, the event horizon Eq.~(\ref{horizons}) is a monotonically decreasing function of $Q_{\rm YM}$ with small variations in the range $Q_{\rm YM}>0$.

Finally, it is important to note that some studies on the BH shadow size have constrained the electric charge, $Q$, of a RN black hole to satisfy roughly $Q/M < 1$, effectively ruling out the extremal case \cite{EventHorizonTelescope:2021dqv,EventHorizonTelescope:2022xqj,Zakharov:2014lqa}. In Table \eqref{tab:charges}, we consider large values of $Q$ since they still prevent the formation of a naked singularity, especially when negative Yang-Mills charge values are taken into account. This serves as an illustrative example of how the presence of the Yang-Mills charge can modify the properties of the standard RN black hole. Therefore, our numerical results do not conflict with the aforementioned constraints derived for the standard RN black hole, as they are not directly applicable to our scenario. 
However, it is worth mentioning that we have recently conducted a similar study in which we obtained constraints on the electric charge in terms of the Yang-Mills charge, and vice versa, by using the observed shadow size \cite{Rincon:2023hvd}.

\begin{table}
\caption{Yang-Mills charge values $-1<Q_{\rm YM}<1$ and derived constraint $Q\in(0,Q_{\rm Max})$ by demanding the fulfillment of the the weak cosmic censorship conjecture Eq.~(\ref{conditionsWCC}). The existence of $Q_{\rm YM}$ changes the possible values of $r_{+}$. In particular, for a given $Q_{\rm YM}$, we take the maximum electric charge $Q_{\rm Max}$ that leads to the extremal case which is always equal to one-half the purely power YM case ($Q=0$).}
\begin{tabular}{| c | c| c | }
\hline
$Q_{YM}$ & $Q/M$ & $r_{+}/M$ \\ 
\hline
-0.9 & (0, 3.16228) & (20, 10)\\ 
-0.8 & (0, 2.23607) & (10, 5) \\
-0.7 & (0, 1.82574) & (6.66667, 3.33333)\\
-0.6 & (0, 1.58114) & (5, 2.5)\\ 
-0.5 & (0, 1.41421) & (4, 2)\\
-0.4 & (0, 1.29099) & (3.33333, 1.66667)\\
-0.3 & (0, 1.19523) & (2.85714, 1.42857) \\
-0.2 & (0, 1.11803) & (2.5, 1.25) \\
-0.1 & (0, 1.05409) & (2.22222, 1.11111) \\
0 & (0, 1) & (2 , 1) \\
0.1 & (0, 0.95346) & (1.81818, 0.90909) \\
0.2 & (0, 0.91287) & (1.66667, 0.83333) \\
0.3 & (0, 0.87705) &  (1.53846, 0.76923) \\
0.4 & (0, 0.84515) & (1.42857, 0.71428) \\
0.5 & (0, 0.81649) & (1.33333, 0.66666) \\
0.6 & (0, 0.79056) & (1.25, 0.625) \\
0.7 & (0, 0.76696) & (1.17647, 0.58823) \\
0.8 & (0, 0.74535) & (1.11111, 0.55555) \\
0.9 & (0, 0.72547) & (1.05263, 0.52631) \\
1 & (0, 0.70710) & (1, 0.5) \\
\hline
\end{tabular}\label{tab:charges}
\end{table}

On the other hand, another useful parameter in astrophysical accretion is the radius of the innermost stable circular orbit (ISCO). 
Albeit well-known, we will briefly summarize the main ingredients behind the computation of ISCO for spherically symmetric four-dimensional backgrounds. 
We follow Ref.~ \cite{Boonserm:2019nqq} closely for the main derivations.
By using the line element \eqref{lineelement} and following Ref.~\cite{Garcia:2013zud}, the equations of motion for test particles are obtained by using the geodesic equation
\begin{equation}
\frac{d^2x^\mu}{ds^2} + \Gamma^\mu_{\rho \sigma} \frac{dx^\rho}{ds} \frac{dx^\sigma}{ds} = 0,
\end{equation}
where $s$ is the corresponding proper time. The Christoffel symbols, $\Gamma^\mu_{\rho \sigma}$, are related with the metric and its derivatives by \cite{Landau:1975pou}
\begin{equation}
\Gamma^\mu_{\rho \sigma} = \frac{1}{2} g^{\mu \lambda} \left( \frac{\partial g_{\lambda \rho}}{\partial x^\sigma} + \frac{\partial g_{\lambda \sigma}}{\partial x^\rho} - \frac{\partial g_{\rho \sigma}}{\partial x^\lambda} \right).
\end{equation}
To rewrite the equation in a more suitable form, we take advantage of the existence of two conserved quantities (two first integrals of motion), precisely as in the Keplerian problem in classical mechanics. In such a sense, we notice that, for $\mu=1=t$ and $\mu=4=\phi$, the geodesic equations take the simple form
\begin{eqnarray}
0 & = & \frac{d}{ds} \left( f(r) \frac{dt}{ds} \right), \\
0 & = & \frac{d}{ds} \left( r^2 \frac{d\phi}{ds} \right). 
\end{eqnarray}
Introducing the corresponding conserved quantities 
\begin{equation}
E \equiv f(r) \frac{dt}{ds}, \; \; \; \; \; \; L \equiv r^2 \frac{d\phi}{ds},
\end{equation}
we can, effectively, parameterize the problem easily. Thus, 
the last two quantities, $E$ and $L$, are usually related to the energy and angular momentum, respectively. 
Assuming a motion on the $(x-y)$ plane (namely, studying motions on the equatorial plane: $\theta = \pi/2$), the geodesic equation for the $\theta$ index is also satisfied automatically and, therefore, such an equation does not provide  information. The only non-trivial equation is obtained when $\mu=2=r$ (see \cite{Garcia:2013zud} for more details)
\begin{equation}
\left( \frac{dr}{ds} \right)^2 = \left[ E^2 - f(r) \left( \epsilon + \frac{L^2}{r^2} \right) \right],
\end{equation}
which may also be obtained from \cite{Garcia:2013zud}
\begin{equation}
g_{\mu \nu} \frac{dx^\mu}{ds} \frac{dx^\nu}{ds} = \epsilon.
\end{equation}
Notice at this level that we have more than one possibility, depending on the value of the $\epsilon$-parameter. Thus, when $\epsilon = 1$ we deal with massive test particles, and when $\epsilon = 0$ with light rays.
In what follows, we will consider the case with $\epsilon = 1$ (massive test particle with mass $m$).
Then the geodesic equation can be written accordingly to
\begin{equation}
\left( \frac{dr}{ds} \right)^2 = \left[ E^2 - f(r) \left( 1 + \frac{L^2}{r^2} \right) \right],
\end{equation}
and identifying the effective potential
\begin{align}
V(r) &= f(r) \left( 1 + \frac{L^2}{r^2} \right),
\end{align}
we can obtain ISCO radius. 
In concrete, from the condition $V'(r)=0$, $L^2$ can be derived and replaced subsequently in $V''(r)=0$ to ultimately obtain the (real) root,
\begin{widetext}
\begin{equation}
    r_{\rm ISCO}=\frac{4 M^4 - 3 M^2 Q^2 (1 + Q_{\rm YM}) + 
    M^{4/3} \mu^{1/3} (2 M^{4/3} + \mu^{1/3})}{M^{5/3} (1 + Q_{\rm YM}) \mu^{1/3}},
\end{equation}
\end{widetext}
with $\mu=\xi^{1/2} + 8 M^4 - 9 M^2 Q^2 (1 + Q_{\rm YM}) + 2 Q^4 (1 + Q_{\rm YM})^2$ and $\xi= Q^4 (1 + Q_{\rm YM})^2 (5 M^4 - 9 M^2 Q^2 (1 + Q_{\rm YM}) + 
         4 Q^4 (1 + Q_{\rm YM})^2)$.
After careful examination of the impact of both charges on the ISCO structure, some values of the ISCO radius are reported in Table \ref{tab:isco} to illustrate the richness of this class of BH. 
The pair of values ($Q,Q_{\rm YM}$) also fulfill the constraint Eq.~(\ref{conditionsWCC}). The first and third rows correspond, respectively, to the Schwarzschild and extremal Reissner-Nordstr\"{o}m cases. For a fixed electric charge, say, $Q=0.9$, negative $Q_{\rm YM}$ leads to a larger ISCO radius, even larger than the Schwarzschild case (see e.g. sixth row). On the contrary, keeping $Q_{\rm YM}$ fixed and positive, large values of $Q$ can reduce the ISCO radius even below the extremal Reissner-Nordstr\"{o}m case (see seventh row). Hence, the ISCO radius can be smaller than the extremal Reissner-Nordstr\"{o}m case and larger than the Schwarzschild solution. 
\begin{figure}
\centering
\includegraphics[width=1\hsize,clip]{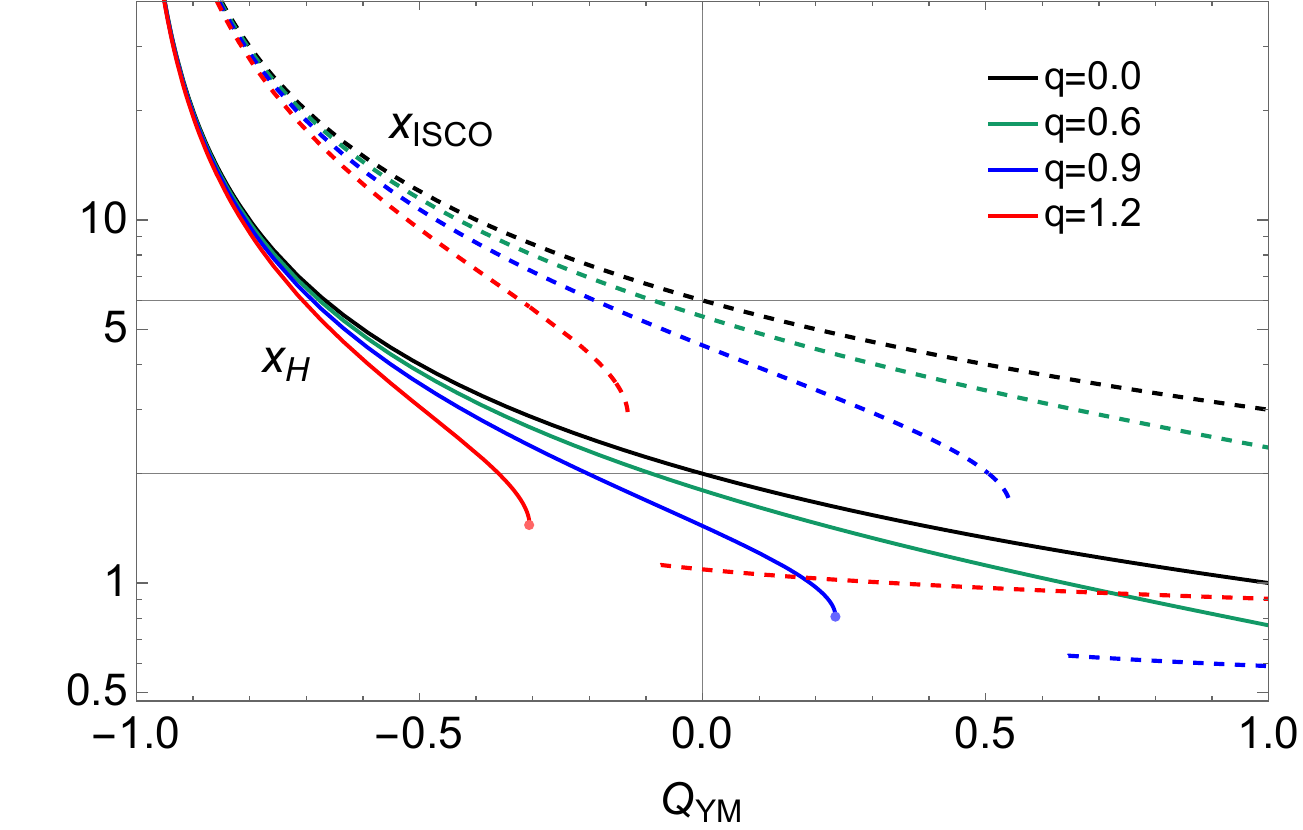}
\caption{The position of the event horizon (solid curves) and the innermost stable circular orbits (dashed curves) is plotted against the Yang-Mills charge $Q_{\rm YM}$ for several fixed values of the (normalized) electric charge $q\equiv Q/M$, as indicated in the legend. The event horizon for the solutions denoted by the solid blue and red curves exists only in a certain region of the parameter space before the naked singularity becomes apparent. This region is delineated by the extremal cases (points on the curves), for which $Q_{\rm YM}$ takes the maximum possible value for a given $q$. Conversely, there are solutions in which the innermost stable circular orbits exhibit discontinuities. This occurs in a very small region of the parameter space (see red and blue dashed curves), where the solution itself cannot be guaranteed. Grey horizontal lines indicate, as appropriate, the event horizon and the innermost stable circular orbit for the Schwarzschild solution.} \label{fig:horizon}
\end{figure}

Interestingly, it is also possible to replicate the ISCO radius of the Schwarzschild and extremal Reissner-Nordstr\"{o}m cases for the selected values of charges:  
$Q_{\rm YM}=1.732050$ and $Q/M= -0.58333$; and $Q_{\rm YM}=1.41421$ and $Q/M= -0.3750$,
respectively, in agreement with Table \eqref{tab:isco}. For the purely power Yang-Mill case ($Q\to 0$), the ISCO reads
\begin{equation}
    r_{\rm ISCO}^{\rm YM}=\frac{6M}{1+Q_{\rm YM}},
\end{equation}
which is, again, a modified version of the Schwarzschild solution. Notice that the ISCO radius is affected in the same fashion as the event horizon (see Eq.~(\ref{eqn:MSch})) by the Yang-Mills charge, that is, scaled by the factor $(1+Q_{\rm YM}$). 

All the aforementioned features for the event horizon and ISCO can also be observed in Fig.~\ref{fig:horizon}. In particular, it is depicted how these quantities behave as functions of the Yang-Mill charge for fixed values of the electric charge and how the extremal cases delimit the existence of the horizon (see points on the blue and red solid curves), thereby giving rise to the naked singularity beyond such critical values.

Having investigated the general properties of the non-trivial structure of the event horizon and ISCO radii in terms of both charges, and derived an useful relation between them that guarantees the cosmic sensor conjecture,  we proceed now to investigate the behavior of spherical steady accretion flows onto this class of black hole. It will be studied after a brief description of the hydrodynamics equations.
\begin{table}
\caption{Radius of the ISCO in terms of both electric and Yang-Mill charge for some reference values according to the relation Eq.~(\ref{conditionsWCC}) that preserves the cosmic censorship conjecture. For comparison the Schwarzschild and Reissner-Nordstr\"{o}m cases have been included in the first and third rows, respectively.}
\begin{tabular}{| c | c| c | }
\hline
$Q_{YM}$ & $Q/M$ & $r_{\rm ISCO}/M$ \\ 
\hline
0 & 0 & 6\\ 
1.732050 & -0.583333 & 6\\ 
0 & 1 & 4 \\
1.414213 & -0.3750 & 4 \\
0.234567 & 0 & 4.860\\
-0.2 & 0 & 7.50\\ 
0.234567 & 0.9 & 3.24\\
-0.2 & 0.9 & 6.09274\\
-0.2 & 1 & 5.67198\\
\hline
\end{tabular}\label{tab:isco}
\end{table}
%

%%%%%%%%%%%%%%%%%%%%%%%%%%%%
\section{Spherical steady accretion flows in a general theory of gravity}
\label{sec:3}
%%%%%%%%%%%%%%%%%%%%%%%%%%%%%%%

Accretion processes of an ideal and polytropic fluids onto black holes in an arbitrary spacetime have been extensively investigated as astrophysical probes to reveal any deviation from general relativity. Although we are working in the framework of Einstein's gravity we derive general accretion equations of steady flows in spherical symmetry spacetime\footnote{We consider the replacement $f(r)^{-1}=g(r)$ in the line element Eq.~(\ref{lineelement}) to be more general in the description. The standard accretion equations are simply recovered by returning to the original metric function.}  but in a general background metric following the general description of Ref.~\cite{Bauer:2021atk}.  Here, the gravitational backreaction of the
accreting fluid on the metric function is neglected. We consider a perfect fluid with total density $\rho$, mass density $\rho_{0}$, internal energy density $\epsilon$, such that $\rho=\rho_{0}+\epsilon$. For isentropic fluids, the pressure can be defined as $P=\omega\rho^{\gamma}$ where $\omega$ is a constant and $\gamma$ is the adiabatic index. The stress energy-momentum tensor is given by
\begin{equation}
   T^{\mu\nu}= (\rho+P)u^{\mu}u^{\nu}+P g_{\mu\nu},\label{streetensor} 
\end{equation}
where $u^{\mu}=(u^{t},u^{r},0,0)$ is the four velocity of the fluid with radial accretion (and wind) flow. From the normalization condition one gets the relation $u^{t}=\sqrt{\frac{g(u^{2}+f)}{f}}$, where we have defined for simplicity $u \equiv u^{r}$. From the baryon conservation and energy-momentum conservation
\begin{align}
    \nabla_{\mu}(\rho_{0}u^{\mu})&=0,\\
    \nabla_{\mu}T^{\mu\nu}&=0,
\end{align}
one obtain two master equations, respectively
\begin{align}
    &\frac{\rho^\prime_{0}}{\rho_{0}}+\frac{f^\prime}{2f}+\frac{g^\prime}{2g}+\frac{u^\prime}{u}+\frac{2}{r}=0,\label{densityeqn}
    \\
    &
    u u^\prime + \Bigg( \frac{f^\prime}{2 fg} +  \frac{c_{s}^{2}}{g}\frac{\rho^\prime_{0}}{\rho_{0}} \Bigg) (1 + g u^{2})+\frac{g^\prime}{2g} u^{2} = 0
    ,\label{velocityeqn}
\end{align}
where prime denotes radial derivative. Here the sound speed is defined as $c_{s}^{2}\equiv \frac{dP}{d\rho}$ at constant entropy. We have also used the first law of thermodynamics in the form $\frac{d\rho}{d\rho_{0}}=\frac{\rho+P}{\rho_{0}}$ to obtain the useful relation $P^\prime=\frac{(\rho+P)}{\rho_{0}}c_{s}^{2}\;\rho^\prime_{0}$. Integration of Eqs.~(\ref{densityeqn}) and (\ref{velocityeqn}) 
gives, respectively, the mass accretion rate
\begin{equation}
    \dot{M}=4\pi r^{2} u \rho_{0},
\end{equation}
and the relativistic version of the Bernoulli equation 
\begin{equation}
    f(1 + g u^{2})\left(\frac{\rho+P}{\rho_{0}}\right) = C,
\end{equation}
where $C$ is an integration constant that must be defined, as well as the equation of state, in order to solve for the infall radial velocity.

%%%%%%%%%%%%%%%%%%%%%%%%%%%%%%%%%%%%%%%%%%%%%%%%%%%%%%%%%%%%%%%%%%%%%%%%%%%%
\section{Accretion process in Einstein-Maxwell-Power Yang-Mills theory}
\label{sec:4}
%%%%%%%%%%%%%%%%%%%%%%%%%%%%%%%%%%%%%%%%%%%%%%%%%%%%%%%%%%%%%%%%%%%%%%%%%%%%

We start by solving the system of differential equations Eqs.~(\ref{densityeqn})-(\ref{velocityeqn}) that governs the dynamics of the fluid flow for the Einstein-Maxwell-Power Yang-Mills theory and metric function given by Eq.~(\ref{metricfunction}) with a power $p=1/2$. There should exist a critical radius that guarantees the monotonic increase of the velocity as decreasing $r$ and avoid singularities in the flow. At such a point, the flow speed equals the sound speed even in the Newtonian treatment of the problem. This condition imposes regularity in both equations at some critical point $r_{c}$, resulting in
\begin{align}
    u_{c}^{2} &= - \frac{Q^{2}-M r_{c}}{2r_{c}^{2}}, \label{eqn:criticalvel}
    \\
    c_{s,c}^{2} &= \frac{-Q^{2}+ Mr_{c}}{Q^{2}+r_{c}(-3 M + 2 (1 + Q_{\rm YM}) r_{c})},\label{eqn:criticalsound}
\end{align}
for the critical velocity of the fluid and the sound speed, respectively. In spite of having explicitly the Yang-Mills charge that changes the structure of the fluid flow, it is possible, after some algebra, to write the critical velocity as $u_{c}^{2}=-\frac{c_{s,c}^{2}f}{-1+c_{sc}^{2}}$ as in the Schwarzschild case. With this, the critical radius can be determined:
\begin{widetext}
\begin{equation}
    r_{c} = \frac{M + 3 c_{s,c}^{2} M \pm \sqrt{(M + 3 c_{s,c}^{2} \;M)^{2} - 
  8 c_{s,c}^{2} (1 + c_{s,c}^{2}) Q^{2} (1 + Q_{\rm YM})}} {4 c_{s,c}^{2} (1 + Q_{\rm YM})}.\label{eqn:criticalradius}
\end{equation}
\end{widetext}
The critical radial velocity matches the Reissner-Nordstr\"{o}m velocity at leading order whereas the critical sound speed receives contributions from the Yang-Mills charge which makes the accretion flow differently. The Reissner-Nordstr\"{o}m solutions is recovered in the limit $Q_{\rm YM}\to0$ and the Schwarzschild case taking, in addition, $Q\to0$. 

The existence of the critical radius is ensured provided that the conditions
\begin{align}
\begin{split}
  & Q_{\rm YM} < -1\; \lor\; 
  \\
  -1 < & Q_{\rm YM} < -1 + \frac{(1 + 3 c_{s,c}^{2})^2}{8 (Q/M)^{2} c_{s,c}^{2} (1 +c_{s,c}^{2}) },\label{eqn:condcritradius}
\end{split}
\end{align}
are satisfied. Such constraints are compatible with $Q>0$ and the causality condition $c_{s,c}^{2}<1$. 
Nevertheless, to prevent the naked singularity, the region $Q_{\rm YM}<-1$ must be discarded in agreement with the condition set by Eq.~(\ref{conditionsWCC}).
The critical radius for the pure power Yang-Mill case ($Q=0$) is
\begin{equation}
 r_{c}^{\rm YM}=  \frac{1 + 3 c_{s}^{2} + \sqrt{(1 + 3  c_{s}^{2})^{2}}}{4  c_{s}^{2} (1 + Q_{\rm YM})},
\end{equation}
whose existence relies only on the nature of the matter fluid, that is, $c_{s,c}^{2} > -1/3$, which is plainly guaranteed and consistent with the causality condition. 

\begin{figure}
\centering
\includegraphics[width=1\hsize,clip]{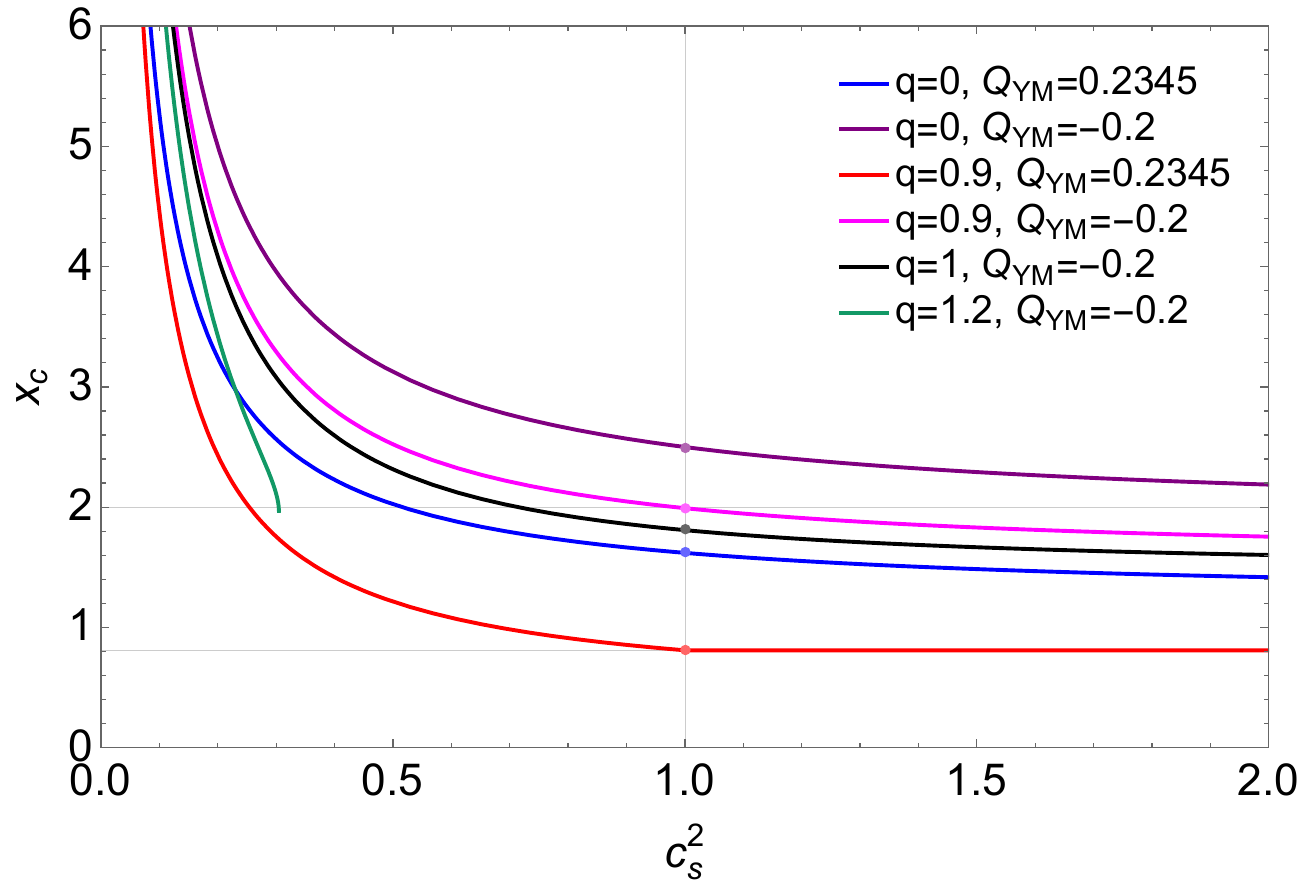}
\caption{Position of the critical point in terms of the critical sound speed for different values of the electric and YM charges as described by the legend. Dots over curves point out the match of the critical radius with the event horizon which happens for $c_{s,c}^{2}=1$. Green curve corresponds to a naked singularity whose associated critical radius exists only for $c_{s,c}^{2}\lesssim0.3$
and the condition  Eq.~(\ref{eqn:condcritradius}) can be preserved.} \label{fig:criticalradius}
\end{figure}

It is instructive to see how the variables involved affect the position of the critical radius in both cases. In doing so, it is convenient to introduce the (electric) charge-to-mass ratio $q=Q/M$ and the dimensionless critical radius $x_{c}=r_{c}/M$. Since we have many variables at play, our strategy is to take certain values of the electric and YM charges and allow $c_{s,c}^{2}$
varies within some reasonable range. Although, values within $c_{s,c}^{2}<1$ are the ones of physical interest, we relax this condition knowingly to get a more complete picture of the accretion problem. As we shall see, for $c_{s,c}^{2}>1$ accretion is apparently possible but breaks down the causal propagation of the sound speed. In particular, we take values of charges so that the Cauchy and event horizons exist simultaneously, with the exception of the case $q=1.2$ and $Q_{\rm YM}=-0.2$ (naked singularity) as can be seen in Fig.~\ref{fig:criticalradius}, where the electric charge overcomes its maximum allowed value as described in Tab.~\ref{tab:charges}. The latter also violates the condition Eq.~(\ref{eqn:condcritradius}) for  $c_{s,c}^{2}\gtrsim0.3$ whereas the other cases behaves regularly for any $c_{s,c}^{2}$. This is the reason why the green curve is cut at a certain sound speed value as can be seen. 

For subsonic sound speeds the critical radius can be accommodated far away from their respective event horizons and in the opposite range, it can be located between the event horizon and the Cauchy horizon unless it corresponds to the extremal case: $q=0.9$ and $Q_{\rm YM}=-0.2$ (red curve), where the position of the critical radius is unaltered for $c_{s,c}^{2}\geq1$, namely, once it reaches the event horizon. 

For $c_{s}^2=1$, the position of all critical points coincides with their respective position of the event horizon, as in the Schwarzschild and RN cases. This is marked with dots over all curves. Finally, in the limit of $Q_{\rm YM}\to0$, our results agree fully with accretion of perfect fluids in the RN metric \cite{Babichev:2008jb}.

%

%%%%%%%%%%%%%%%%%%%%%%%%%%%%%%%%%%%%%%%%%%%%%%%%%%%%
\subsection{Accretion flow for isothermal fluids}
%%%%%%%%%%%%%%%%%%%%%%%%%%%%%%%%%%%%%%%%%%%%%%%%%%%%

In order to simplify our analysis and get some physical insights before proceeding in a more general way, we investigate first accretion of isothermal test fluids. The equation of state is then of the form $P= \omega\rho$, from which the sound speed is simply derived: $c_{s,c}^{2}=\omega$. This choice, in addition to reducing the analysis and the numerical computations considerably, gives us a general idea of the behavior of the mass density and radial velocity through the critical points for a given constant $\omega$ and charges values. So, the implemented numerical strategy will cover some discrete values of the full parameter space that capture the general behavior of the accreting matter flow.

We remind that the YM charge leads to an enhancement of the electric charge up to $\sim3.2$ that still preserves the cosmic weak censorship. It makes, as a consequence, that the critical radius can be located far away from the BH in comparison with the standard Schwarzschild and Reissner Nordstr\"{o}m cases. So, for suitable comparison and representation, we take $q\leq 1$ in the subsequent analysis.

\begin{figure*}
\centering
\includegraphics[width=0.47\hsize,clip]{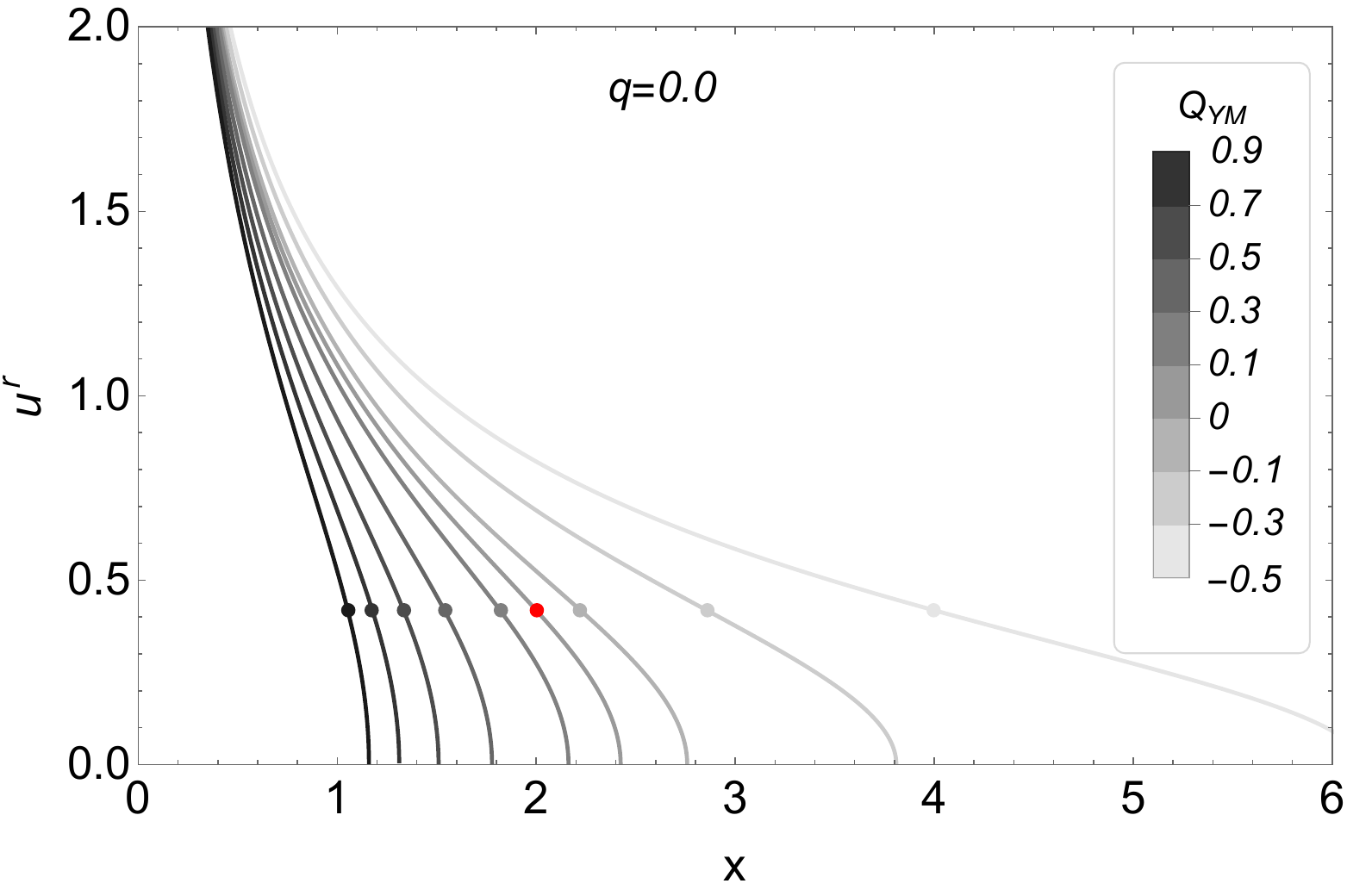}
\includegraphics[width=0.47\hsize,clip]{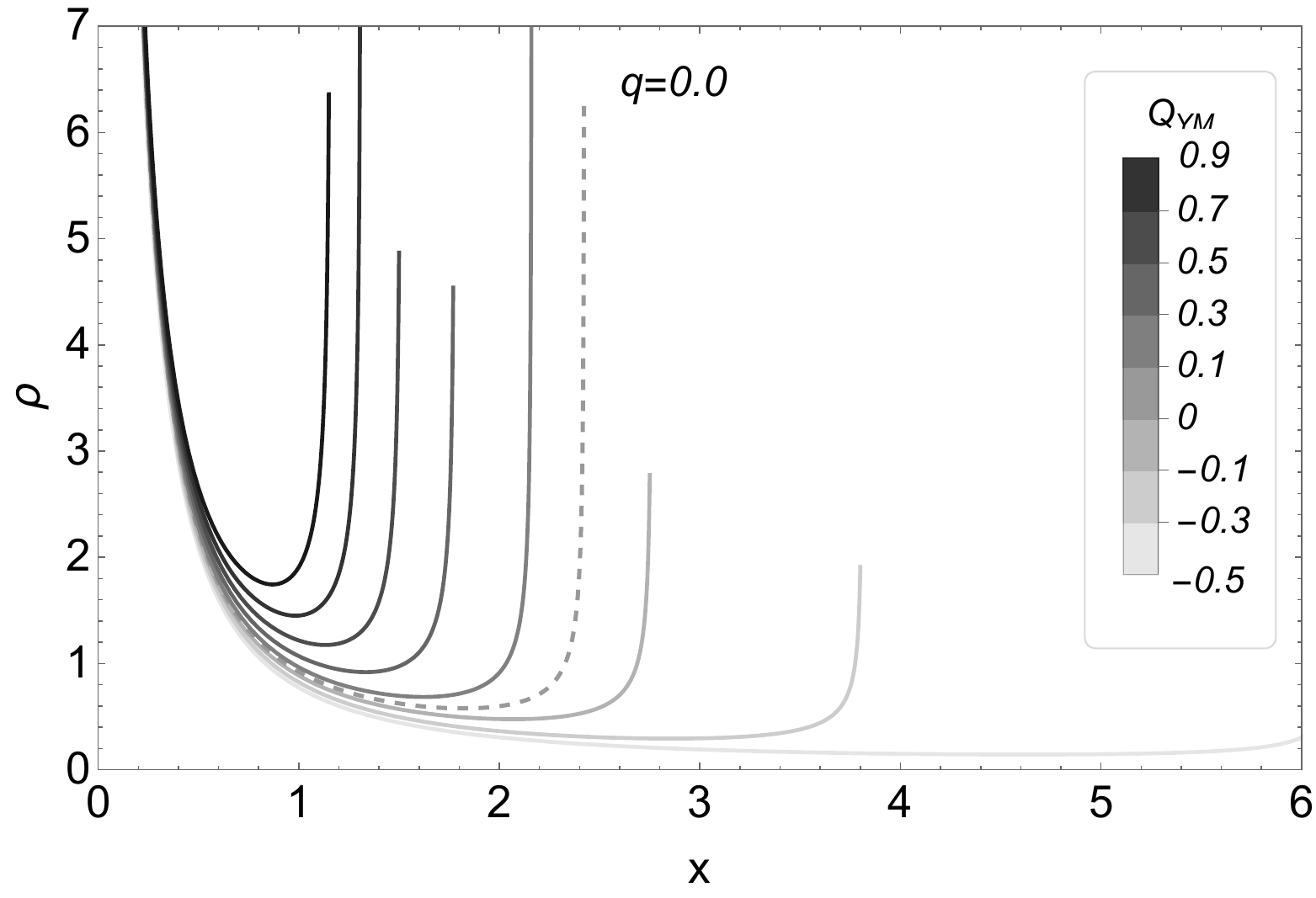}
\includegraphics[width=0.47\hsize,clip]{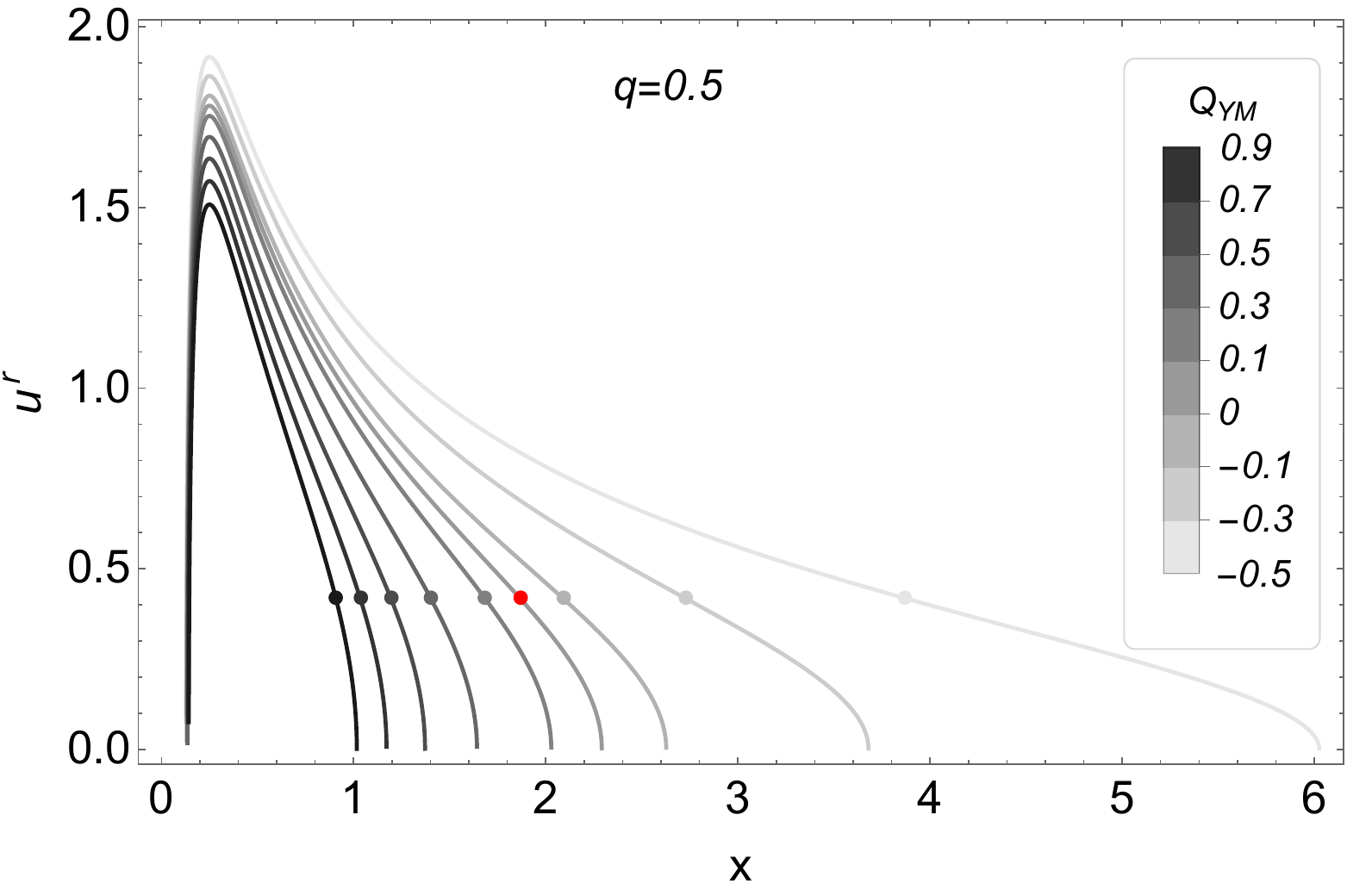}
\includegraphics[width=0.47\hsize,clip]{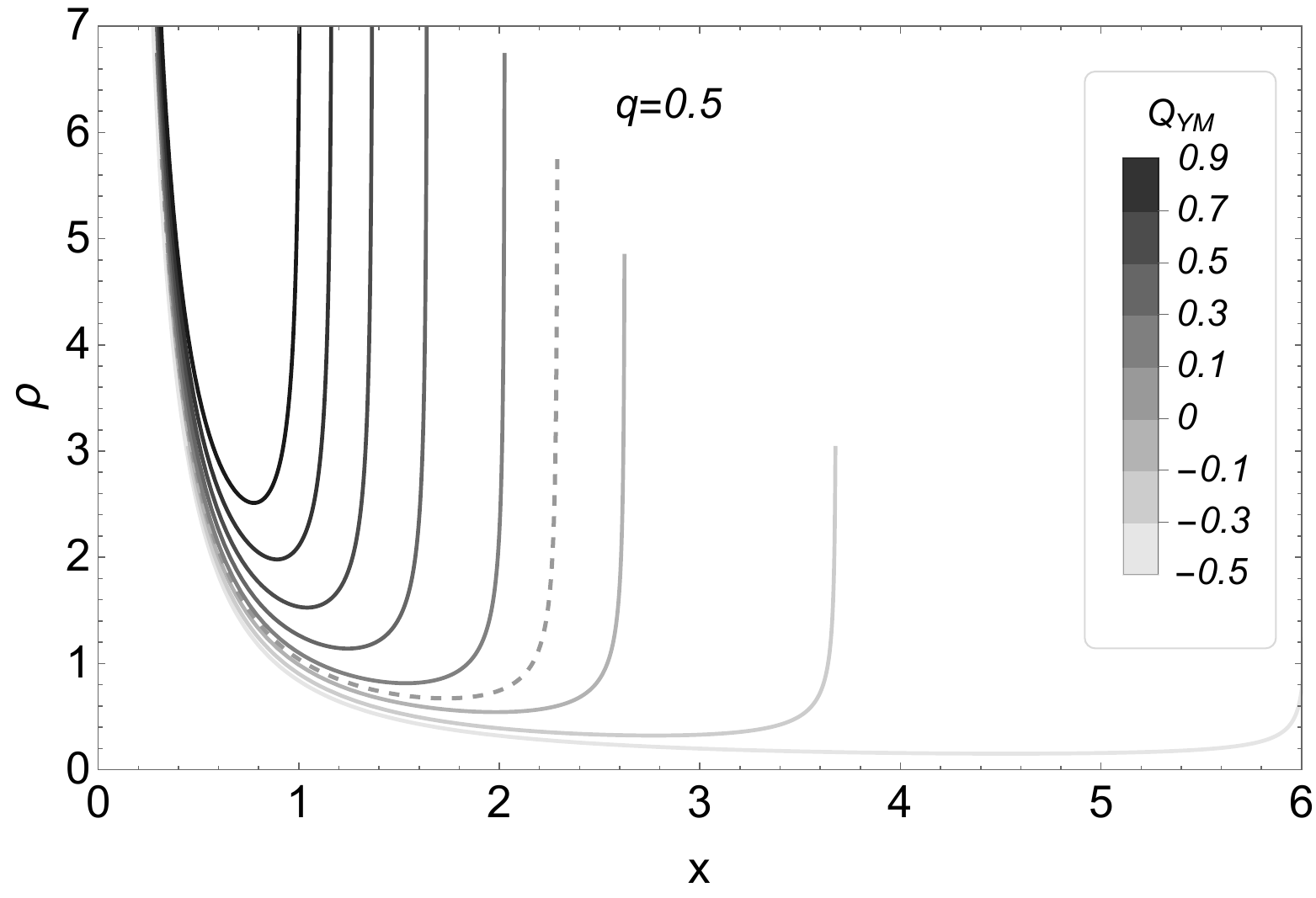}
\includegraphics[width=0.47\hsize,clip]{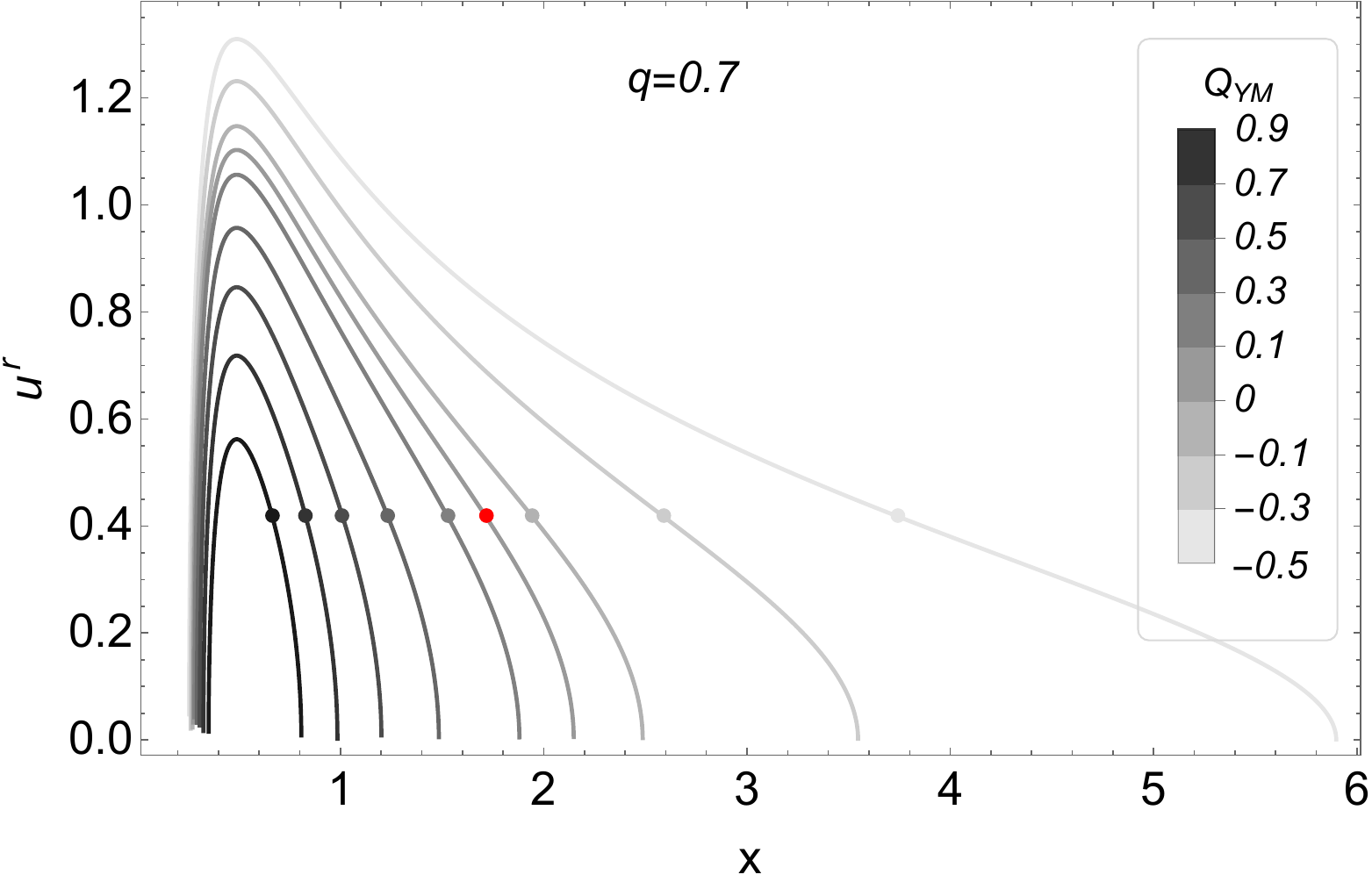}
\includegraphics[width=0.47\hsize,clip]{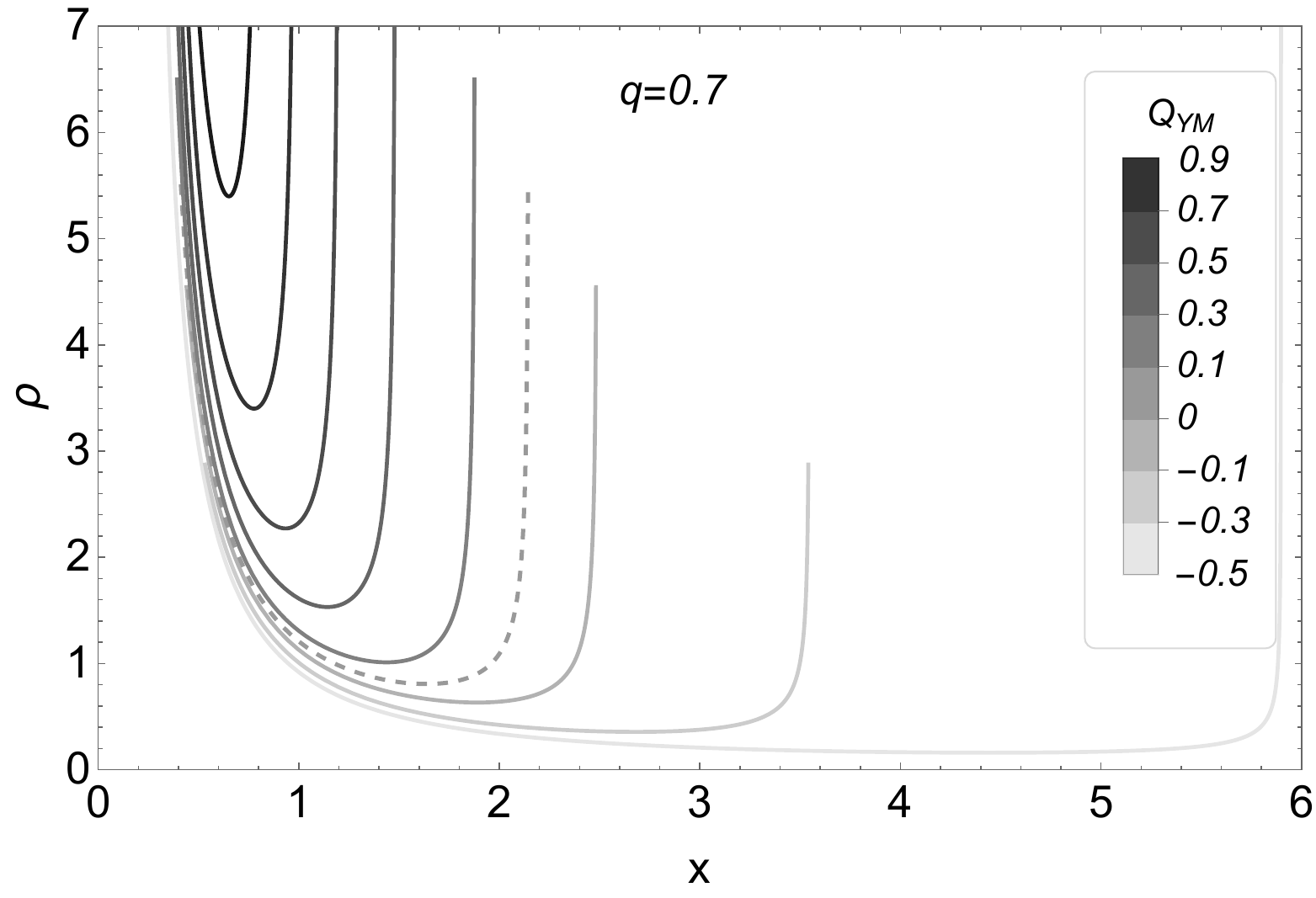}
\includegraphics[width=0.47\hsize,clip]{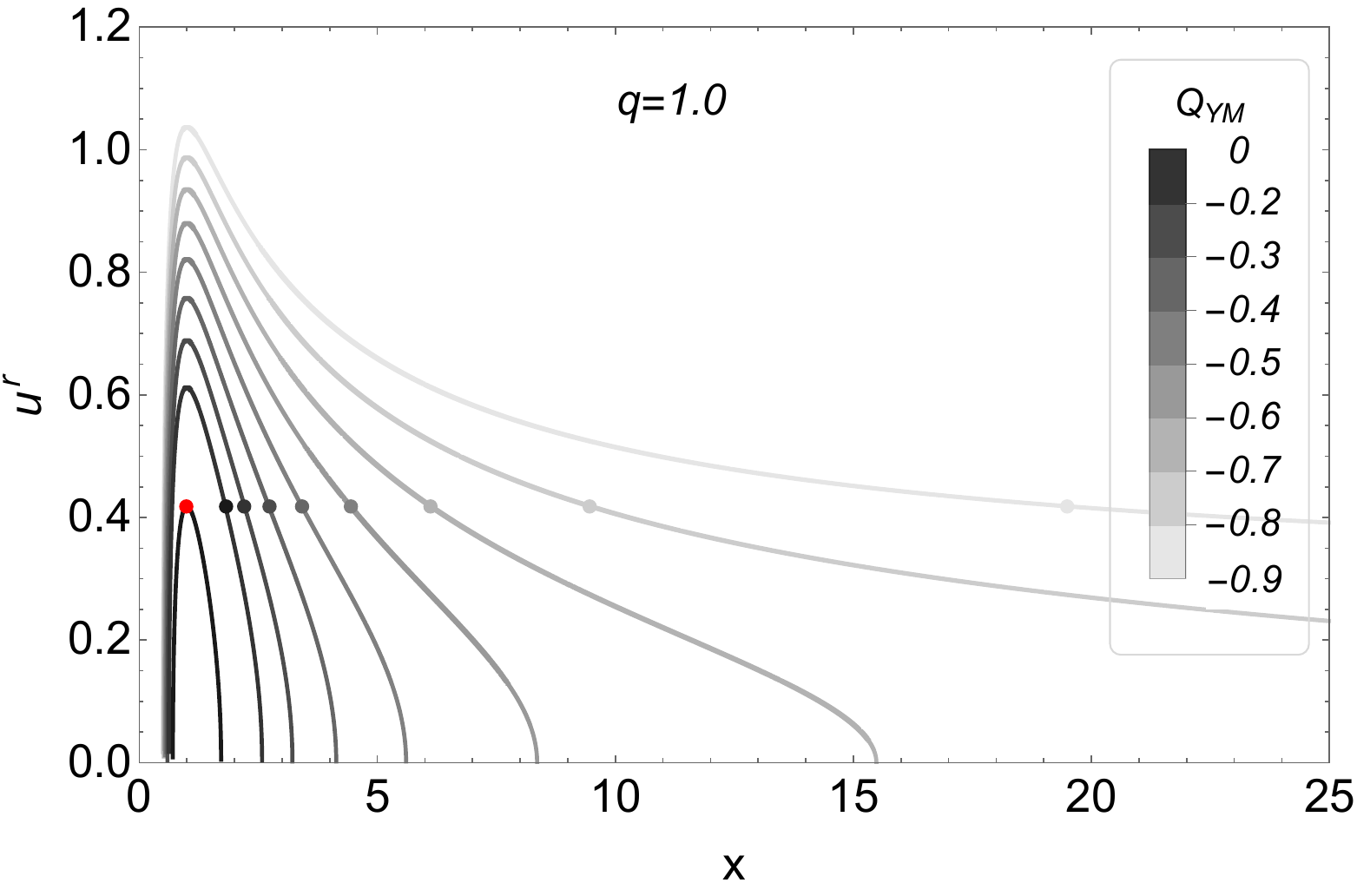}
\includegraphics[width=0.47\hsize,clip]{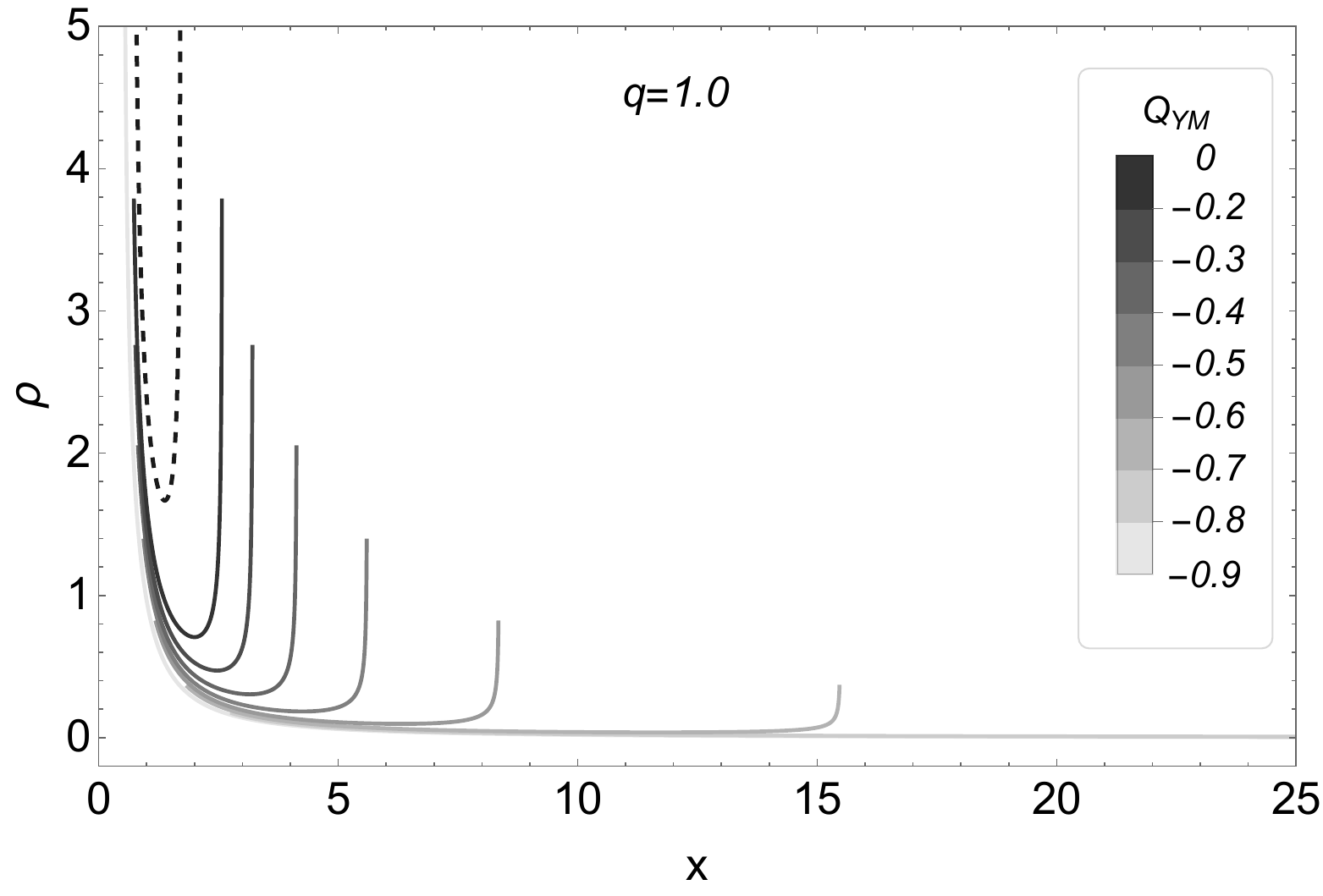}
\caption{\textbf{Stiff case} $\omega=1$. \textit{Left panel}: infall radial velocity for specific values of the electric charge $q$ along with the effect of changing simultaneously the Yang-Mill charge $Q_{\rm YM}$, as described by the bar legend. Red points mark the critical point for the standard RN BH case $Q_{\rm YM}=0$. \textit{Right panel}: mass density distribution due to the BH gravitational potential for the same values of the electric charges as in the radial velocity. Notice that in bottom panels negative values of $Q_{\rm YM}$ have been considered only to allow $q=1$ to exist according to the table 1.} \label{fig:stiff}
\end{figure*}

Let us start by considering a stiff fluid $\omega=1$. For certain values of the electric charge $q$, a range of values for the YM charge that satisfies Eq.~(\ref{eqn:condcritradius}) is derived. The results are shown in Fig.~\ref{fig:stiff} with bar legends depicting the YM charge. Left panels correspond to the radial velocities from top $q=0$ to bottom $q=1$ while right panels display the mass densities for the same values of charges. To read this plot correctly, notice that as long as the color tone intensifies, which is dictated by the enhancement of 
$Q_{\rm YM}$ up to some positive upper value that is constrained by the existence of the solution itself according to Eq.~(\ref{eqn:criticalvel}), the trajectory of the infall radial velocity approaches to the BH. Fair comparison can be done between the cases $q=0$, $q=0.5$ and $q=0.7$ where we have chosen the same range of values for the YM charge unlike the case $q=1$ where negative values of $Q_{\rm YM}$ have been taken instead. Red points, as well as their associated curves, correspond to the case with vanishing YM charge that coincides of course with the Schwarzschild case for $q=0$. Turning on the Yang-Mills charge,  but still keeping $q=0$, leads to the purely power Yang-Mills case. Notice that from $q=0$ till $q=0.7$, several accretion flows are allowed to transit around the standard RN case (red points) provided that $Q_{\rm YM}\neq0$. Hence, there do exist as many possible transonic solutions as values for the YM charge. The main effect of $Q_{\rm YM}$ on the critical point is to bring the transonic flow closer to the BH as it increases. As to the case $q=1$, the critical point (red color) coincides with the event horizon in agreement with our previous discussion about $c_{s,c}^{2}=1$ found below Eq.~(\ref{eqn:condcritradius}). Notice that the abscissa now covers large values of the radial coordinate compared to the other cases. So, a comparison with the above cases must be taken with caution.

\begin{figure*}
\centering
\includegraphics[width=0.47\hsize,clip]{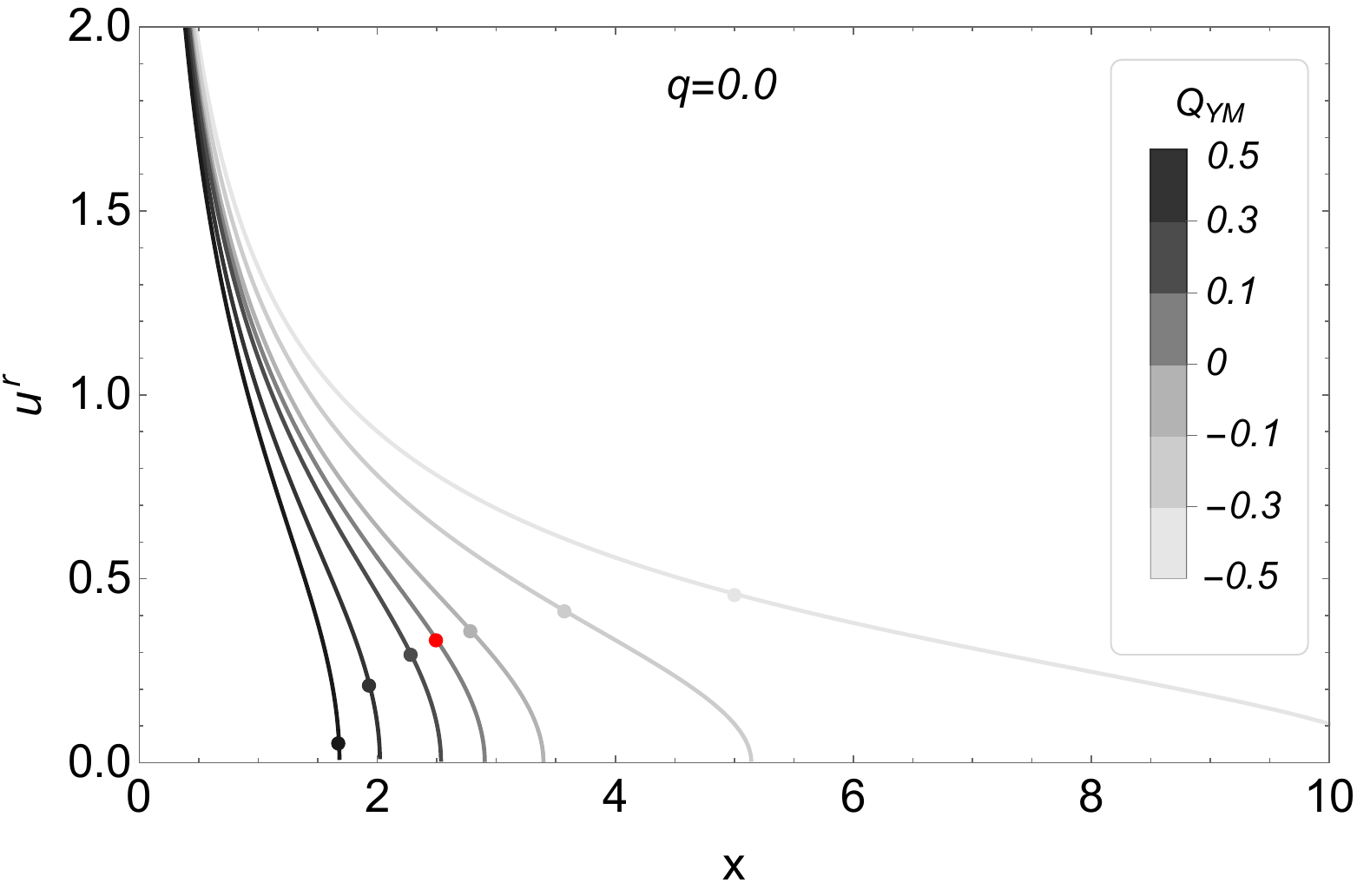}
\includegraphics[width=0.47\hsize,clip]{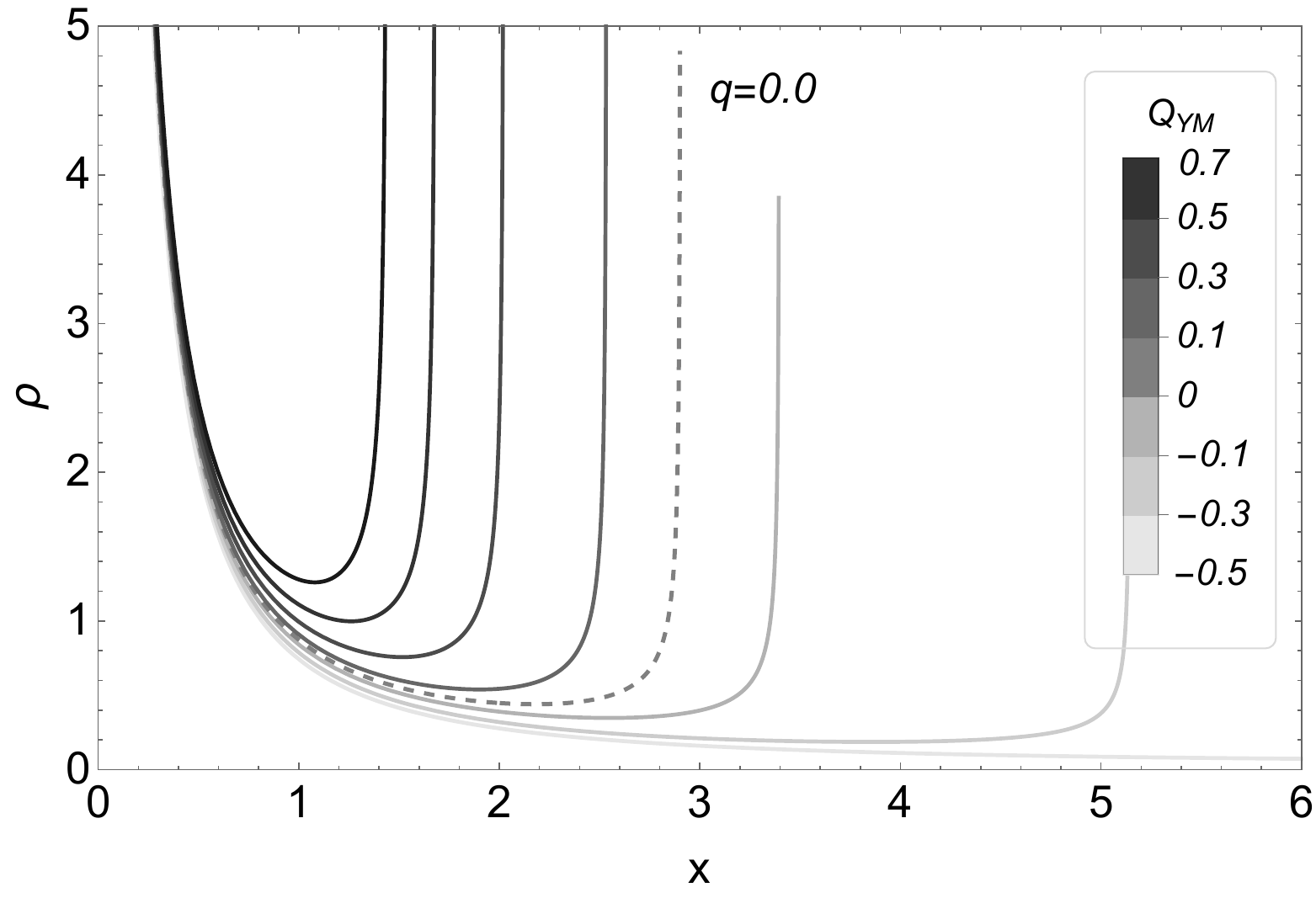}
\includegraphics[width=0.47\hsize,clip]{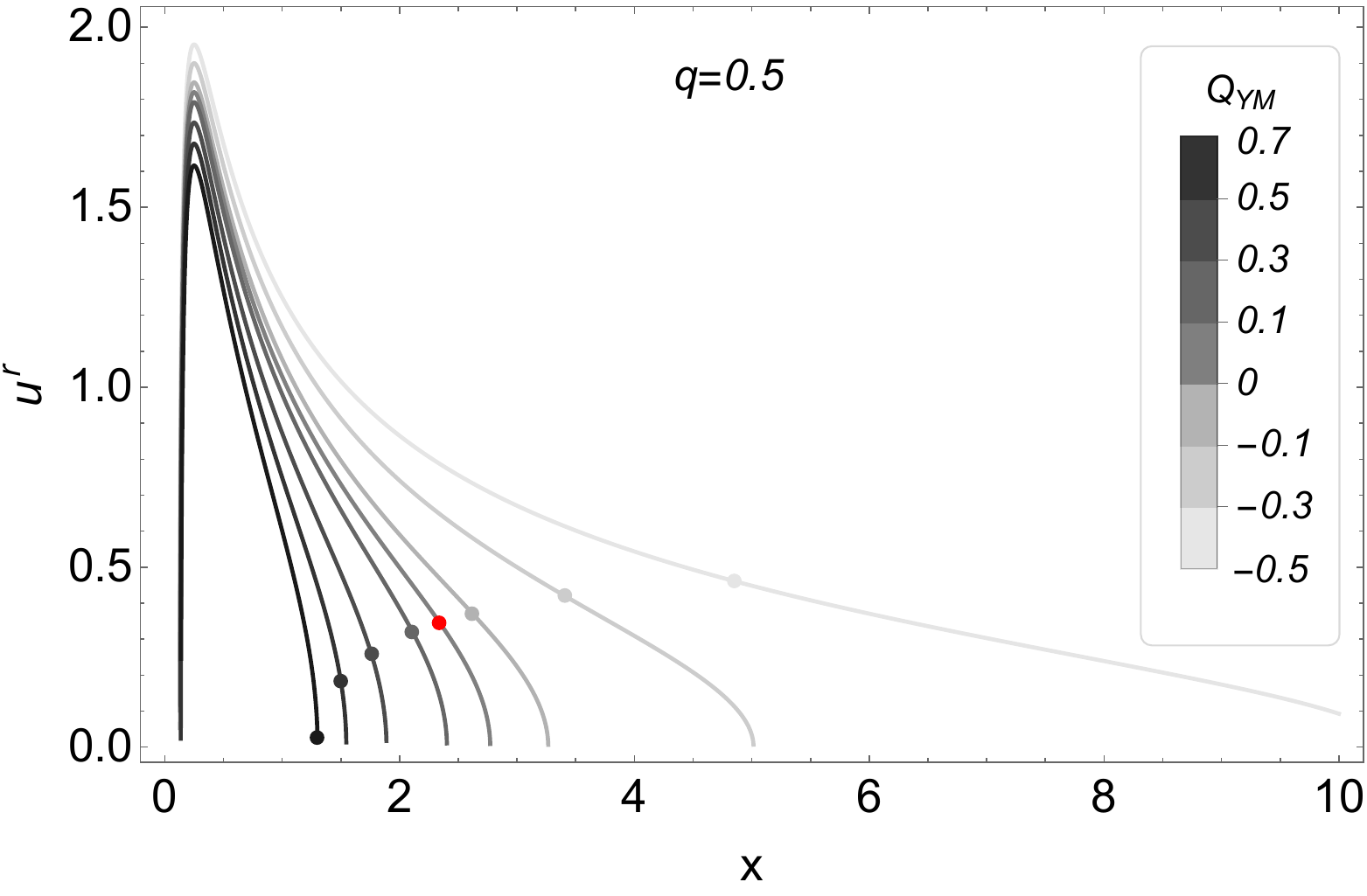}
\includegraphics[width=0.47\hsize,clip]{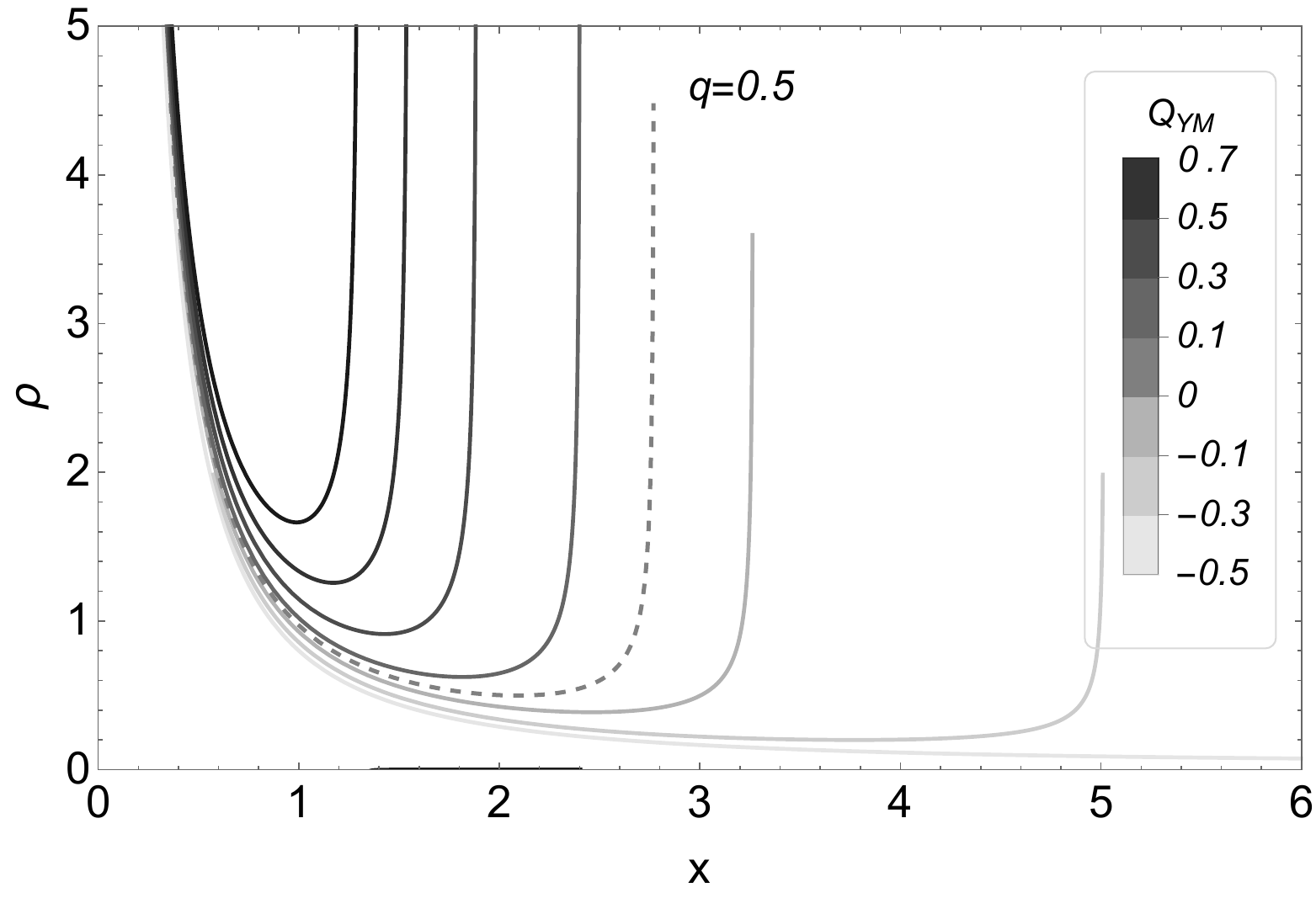}
\includegraphics[width=0.47\hsize,clip]{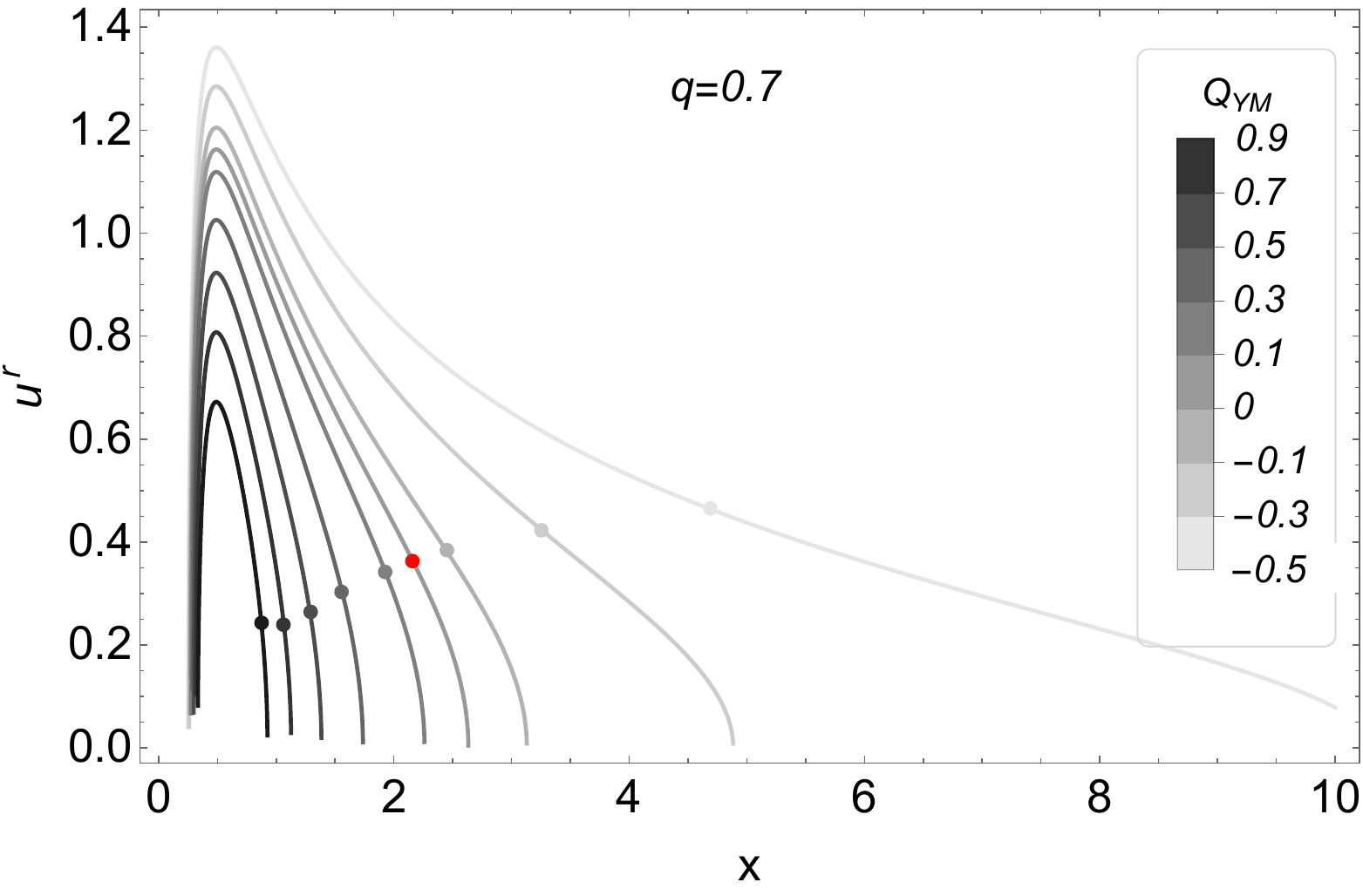}
\includegraphics[width=0.47\hsize,clip]{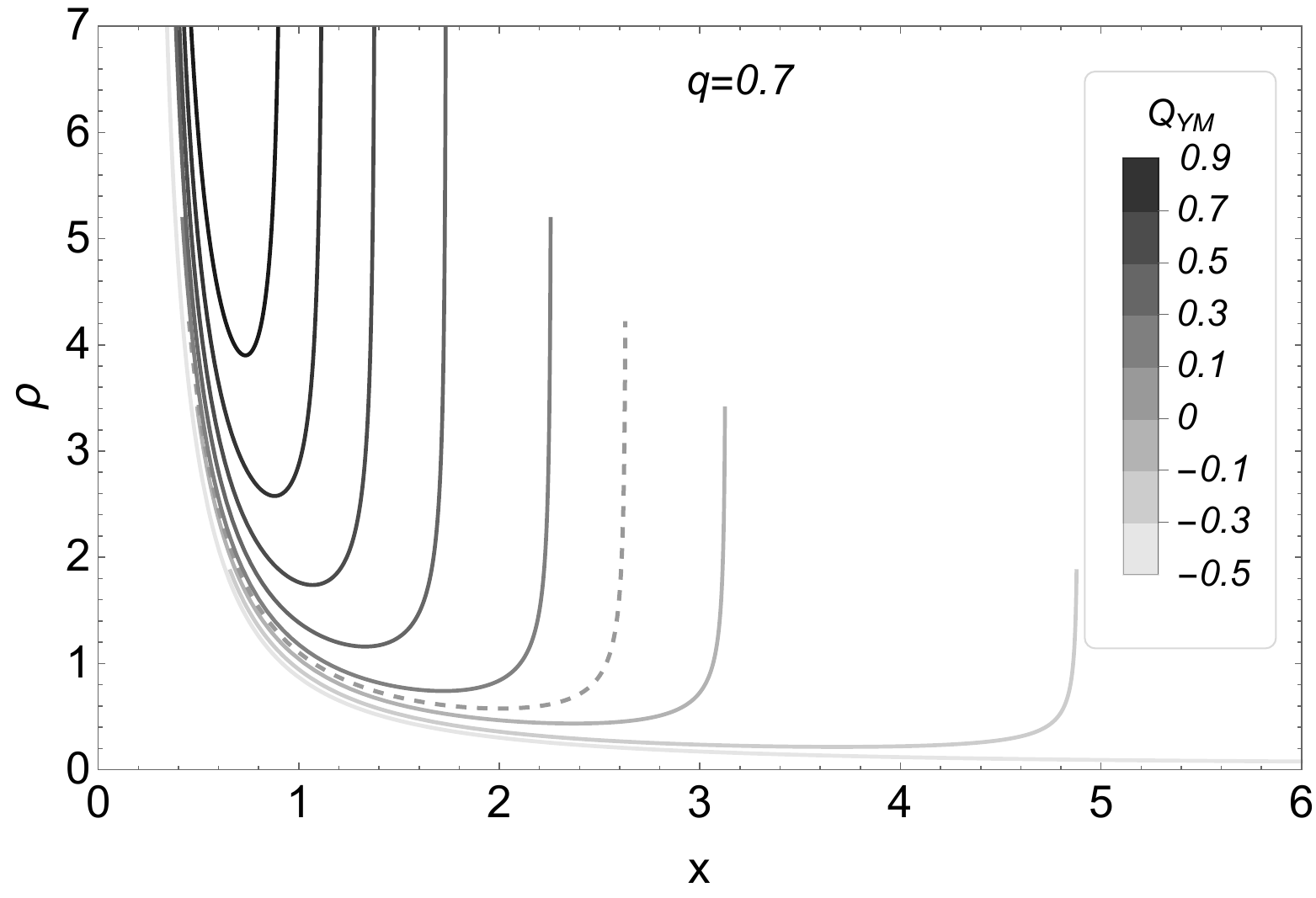}
\includegraphics[width=0.47\hsize,clip]{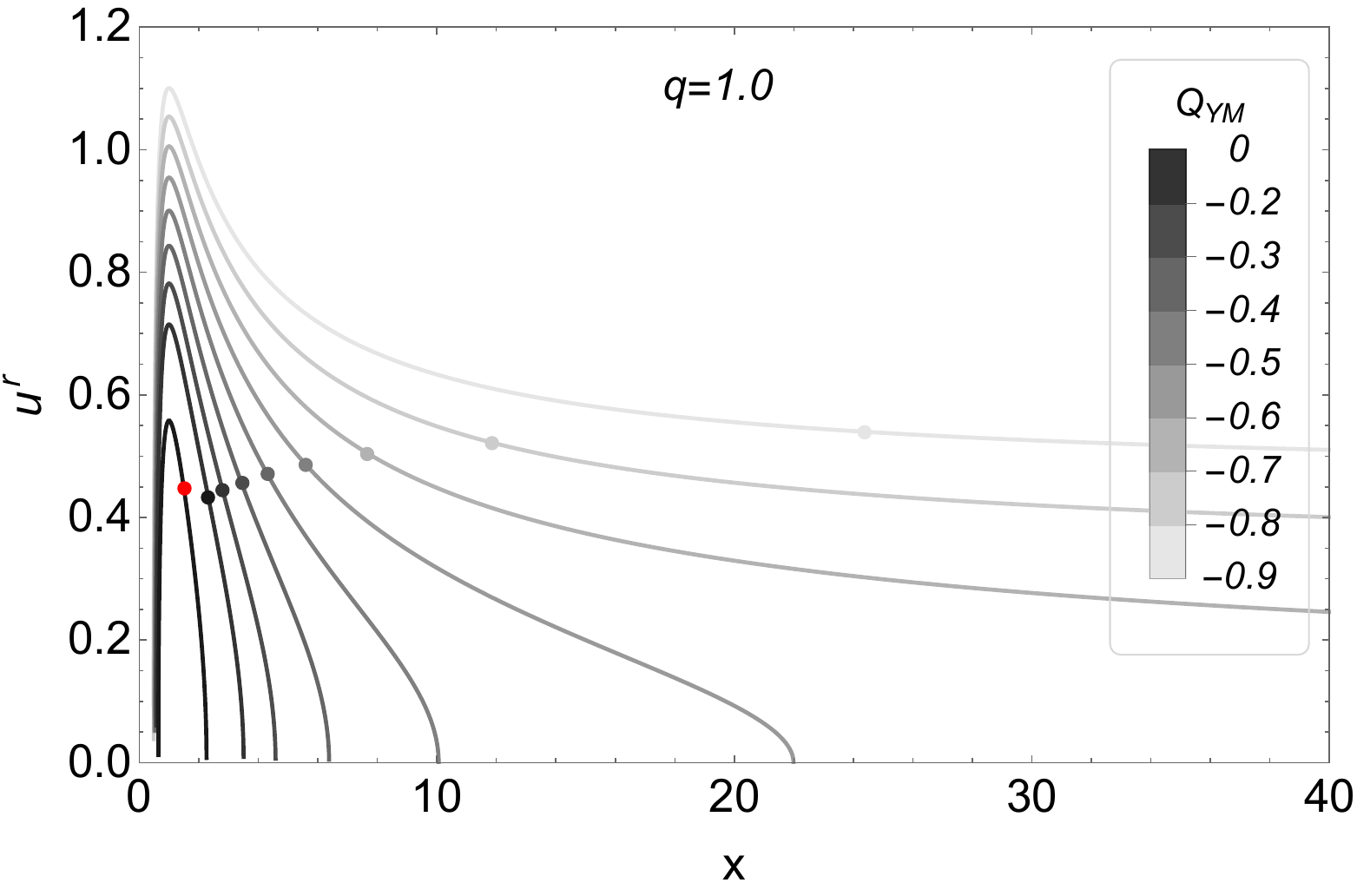}
\includegraphics[width=0.47\hsize,clip]{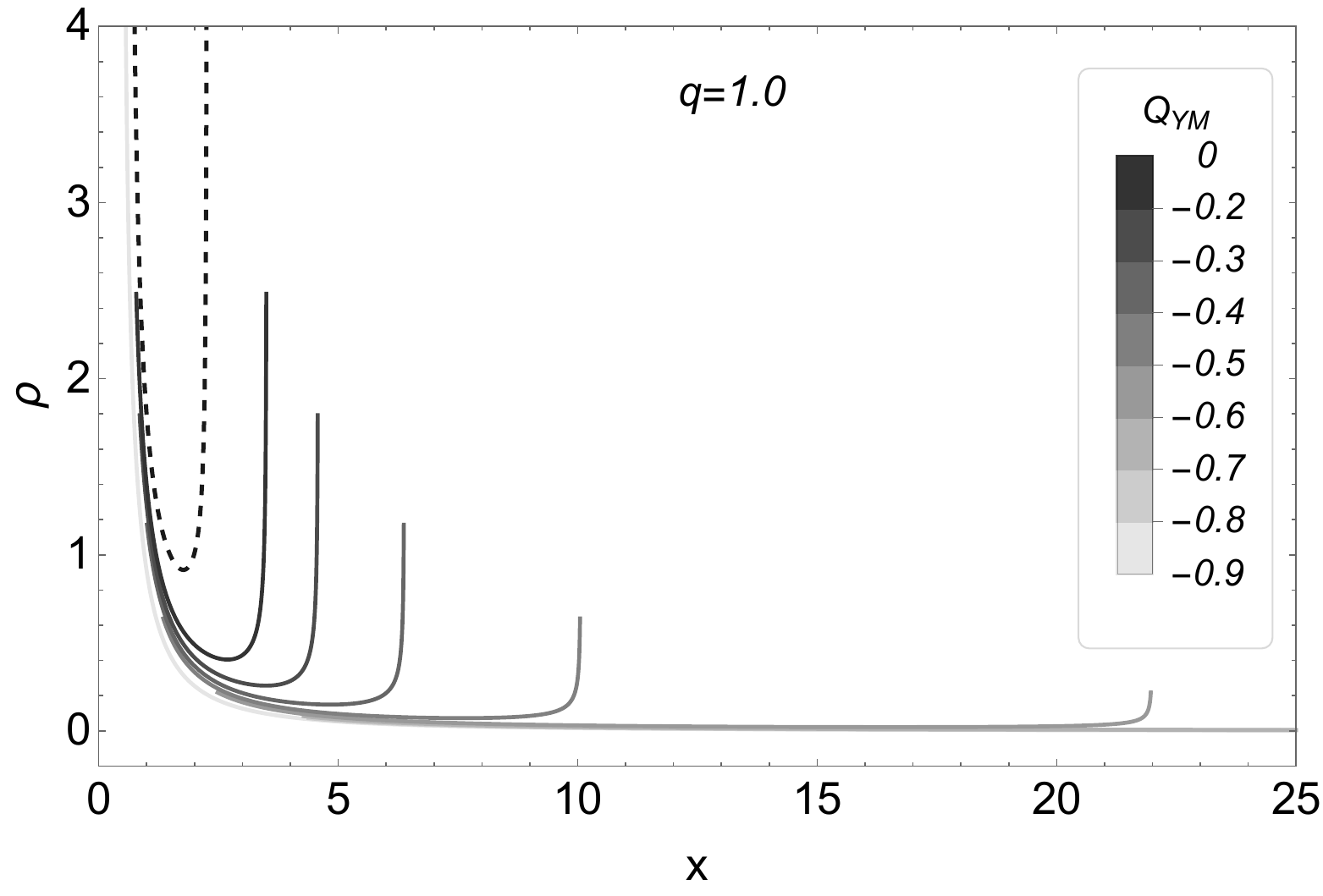}
\caption{\textbf{Ultra-relativistic case} $\omega=1/2$. Left panels show the infall radial velocity while right panels describe the mass density distribution due to the BH gravitational potential for a specific value of the electric charge $q$ along with the effect of changing simultaneously the Yang-Mill charge $Q_{\rm YM}$ as described by the bar legend. Notice that in bottom panels negative values of $Q_{\rm YM}$ have been considered only to allow $q=1$ according to the table 1.} \label{fig:ultrarel}
\end{figure*}

\begin{figure*}
\centering
\includegraphics[width=0.47\hsize,clip]{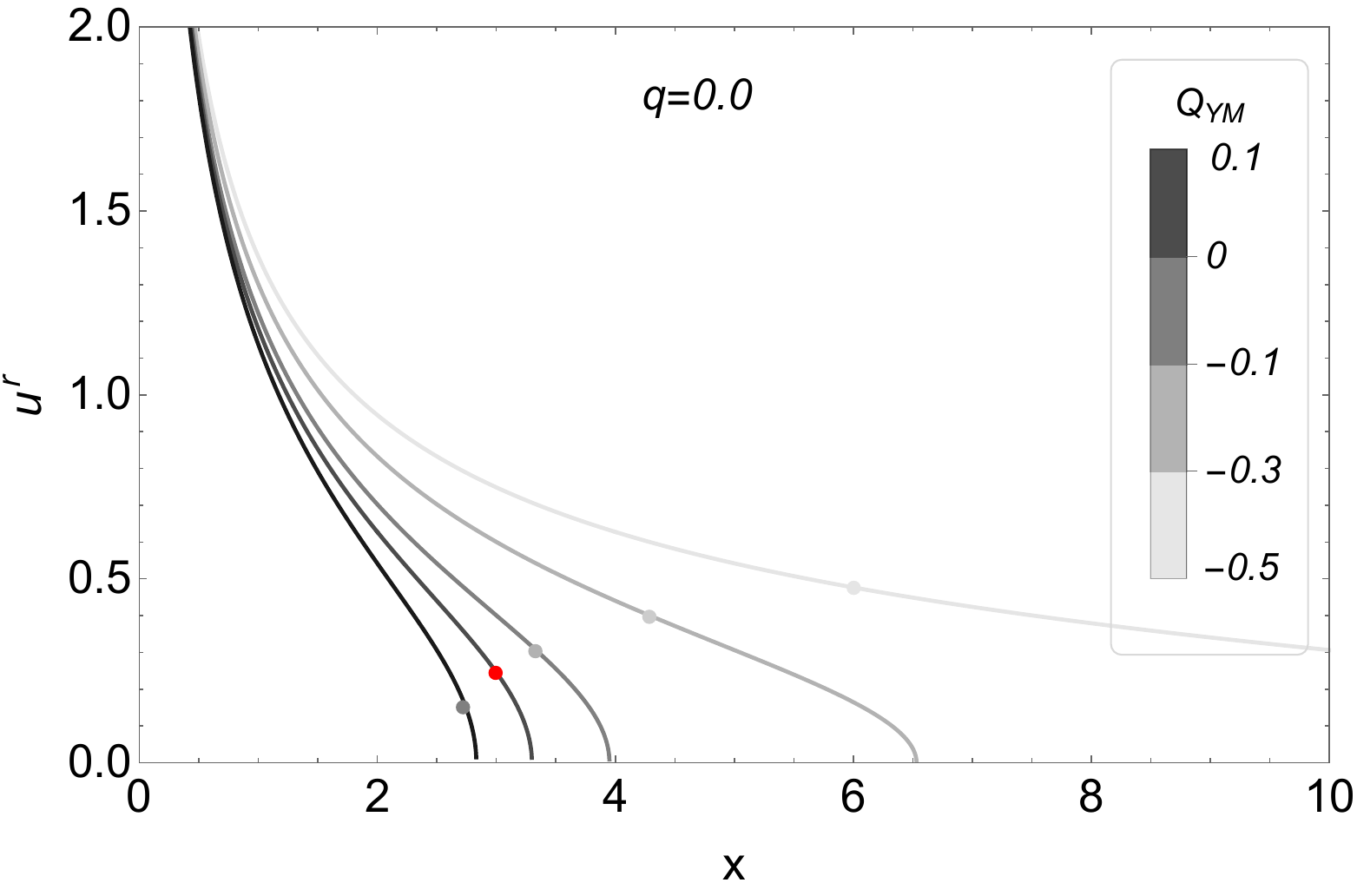}
\includegraphics[width=0.47\hsize,clip]{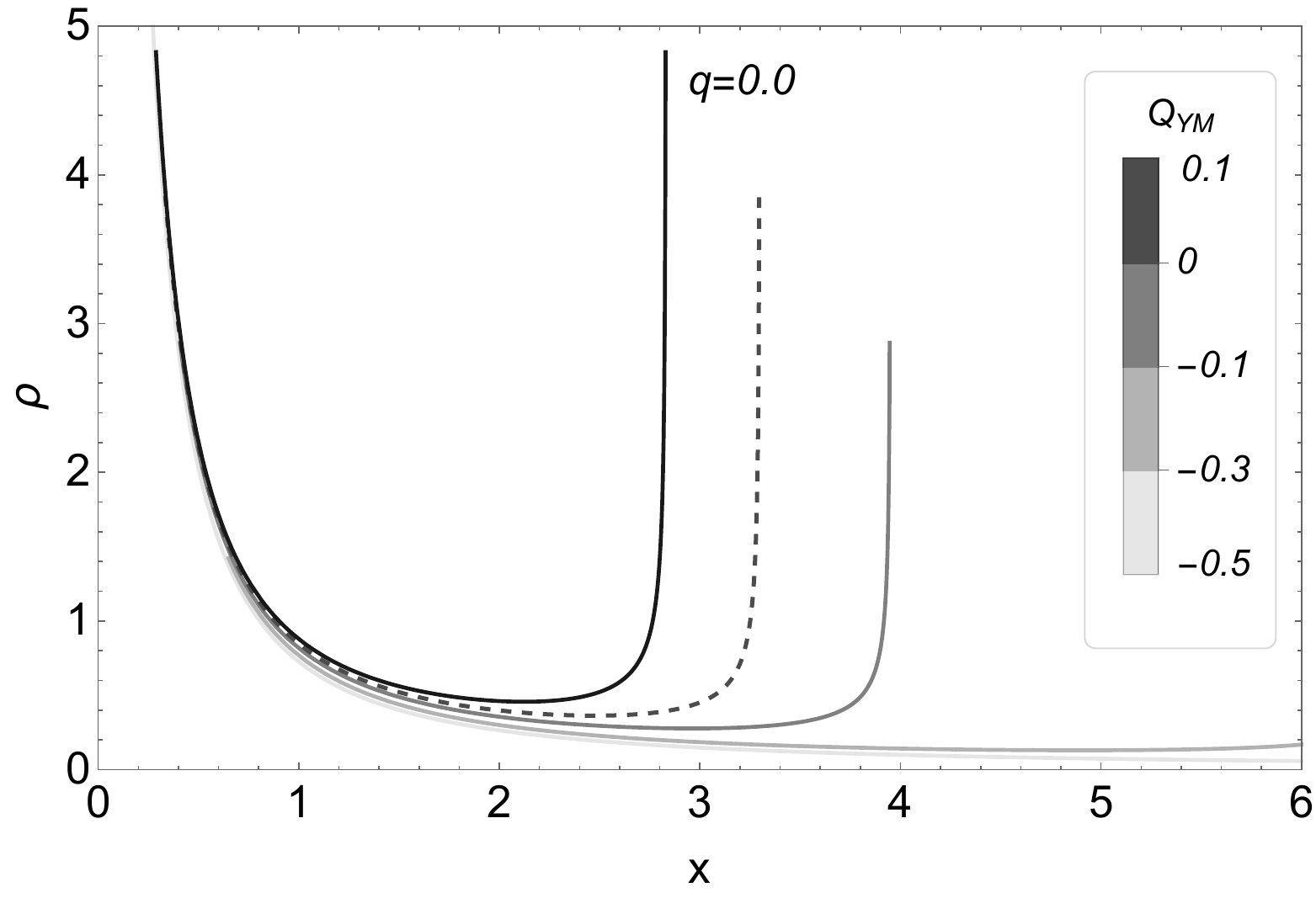}
\includegraphics[width=0.47\hsize,clip]{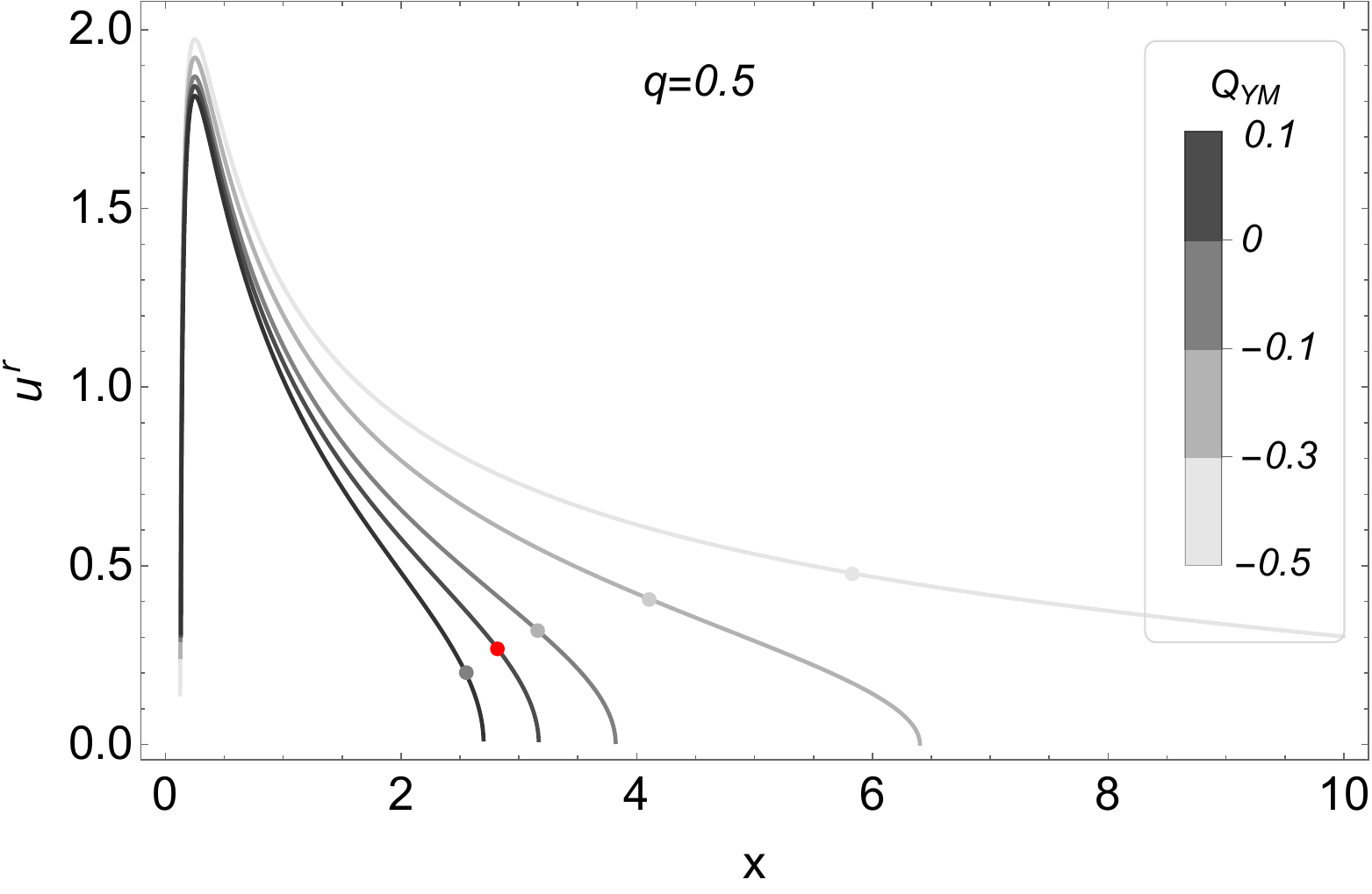}
\includegraphics[width=0.47\hsize,clip]{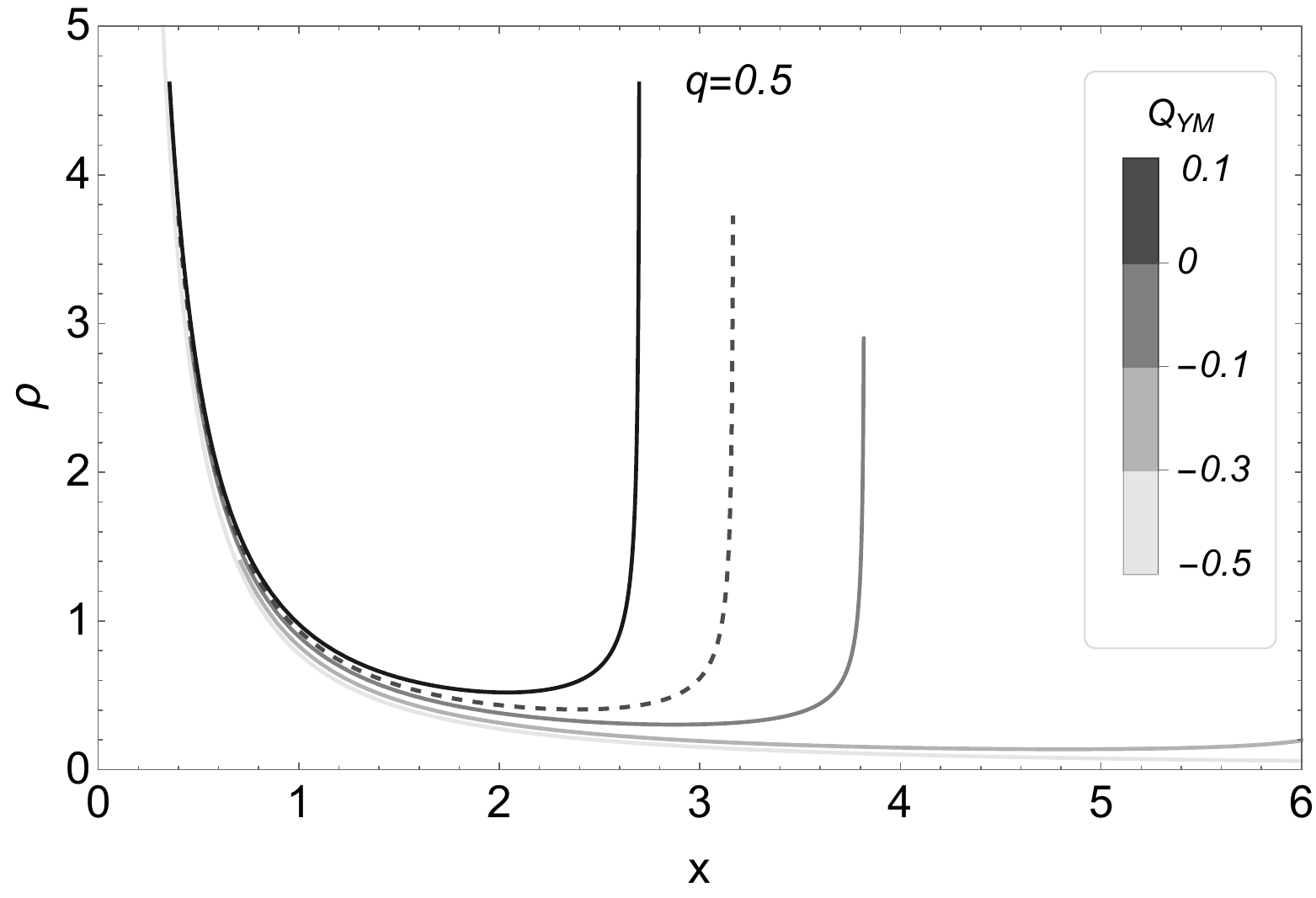}
\includegraphics[width=0.47\hsize,clip]{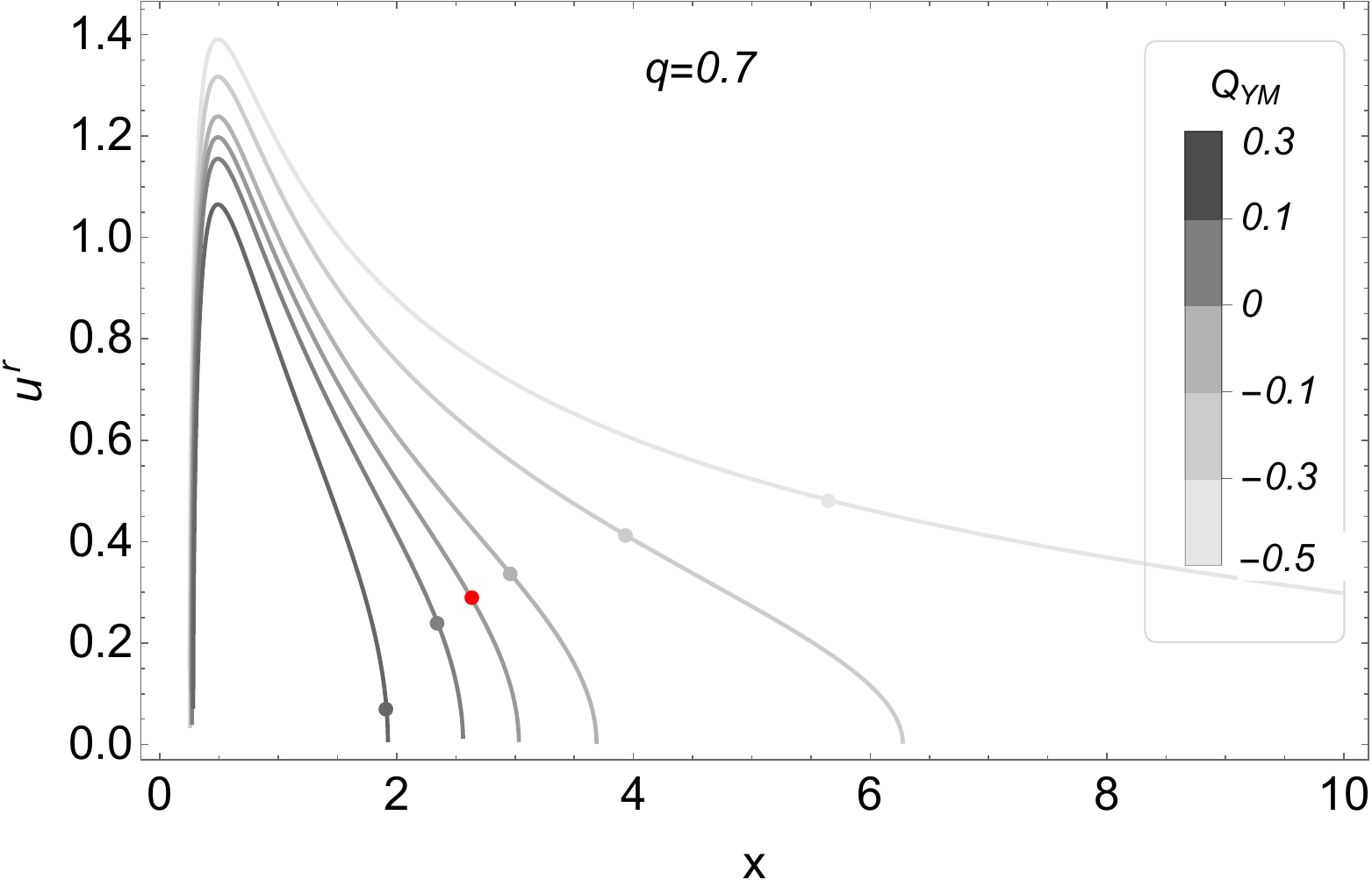}
\includegraphics[width=0.47\hsize,clip]{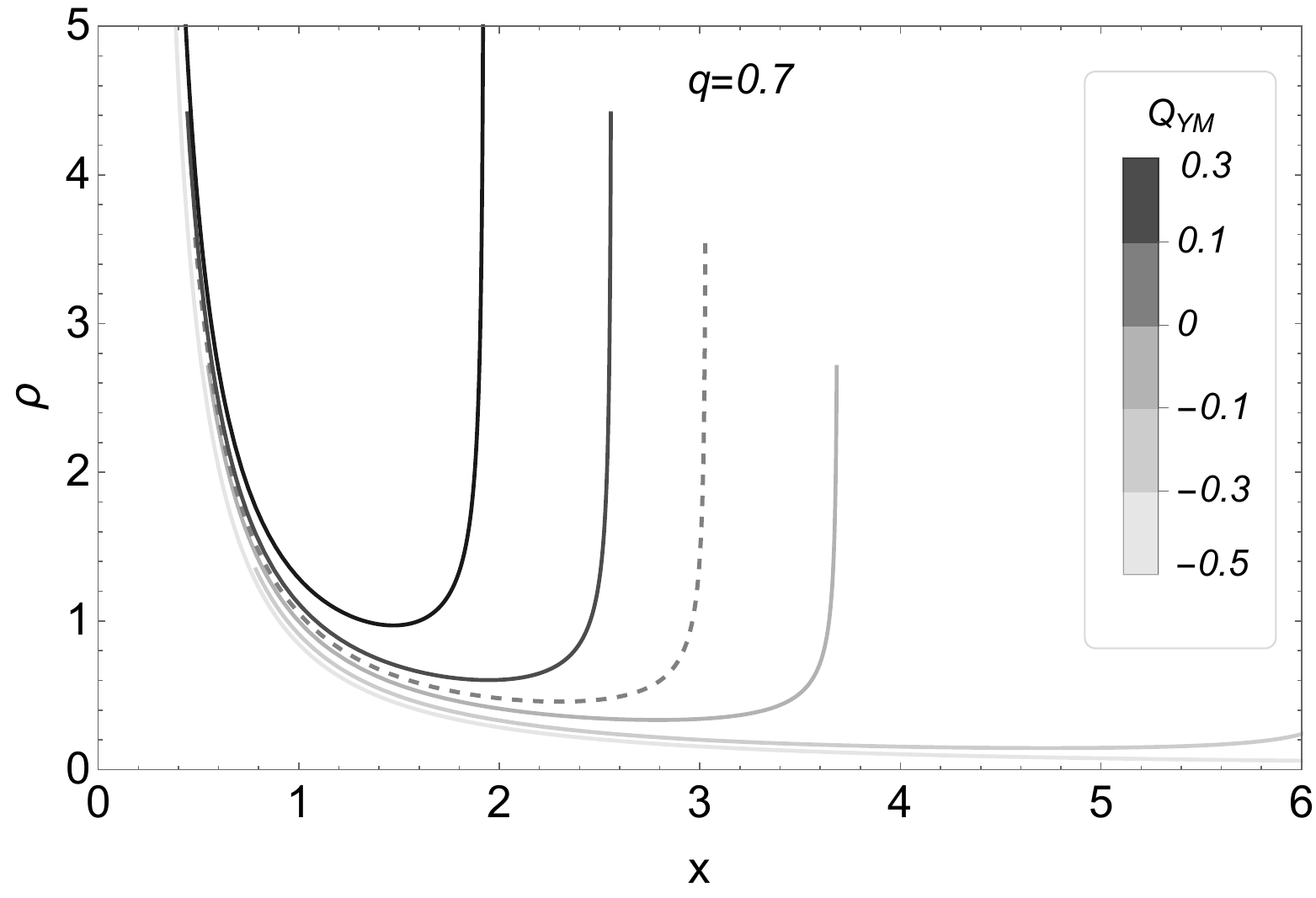}
\includegraphics[width=0.47\hsize,clip]{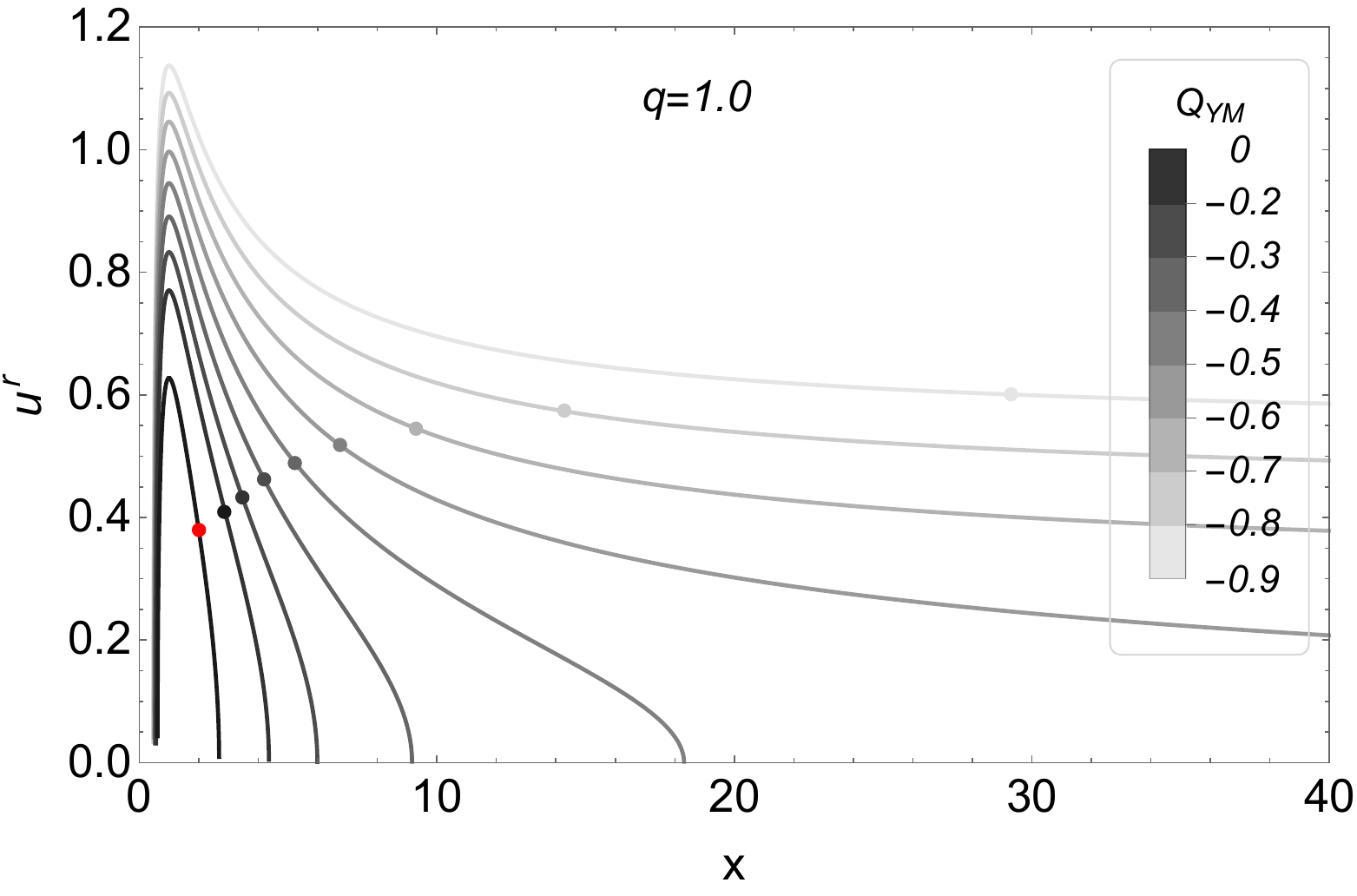}
\includegraphics[width=0.47\hsize,clip]{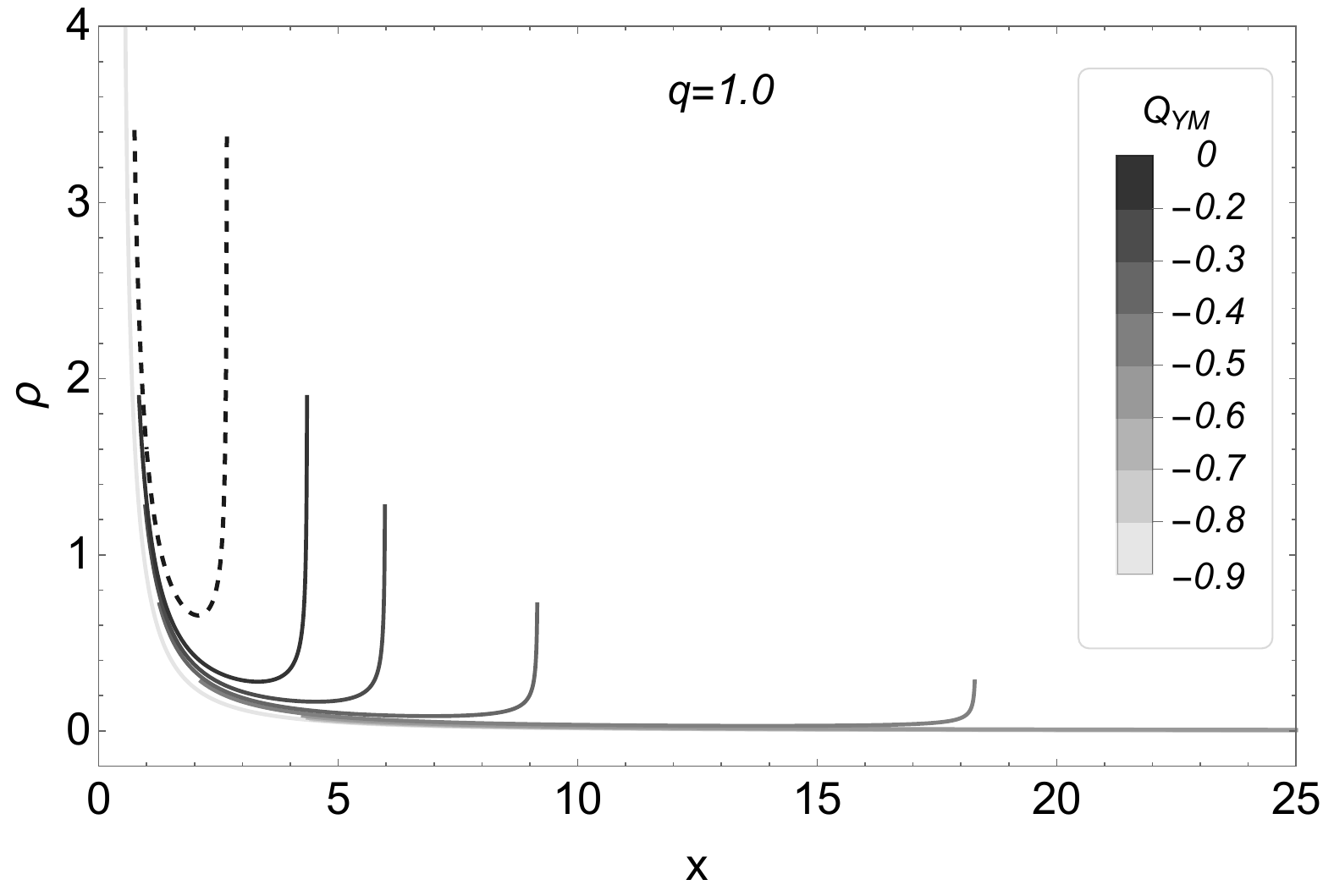}
\caption{\textbf{Relativistic case} $\omega=1/3$. Left panels show the infall radial velocity while right panels describe the mass density distribution due to the BH gravitational potential for a specific value of the electric charge $q$ along with the effect of changing simultaneously the Yang-Mill charge $Q_{\rm YM}$, as described by the bar legend. The delimited range $0<Q_{\rm YM}<-0.5$ in the three first arrows is because those values do provide transonic flows. Notice that in bottom panels negative values of $Q_{\rm YM}$ have been considered only to allow $q=1$ according to the table 1.} \label{fig:rel}
\end{figure*}

\begin{figure*}
\centering
\includegraphics[width=0.47\hsize,clip]{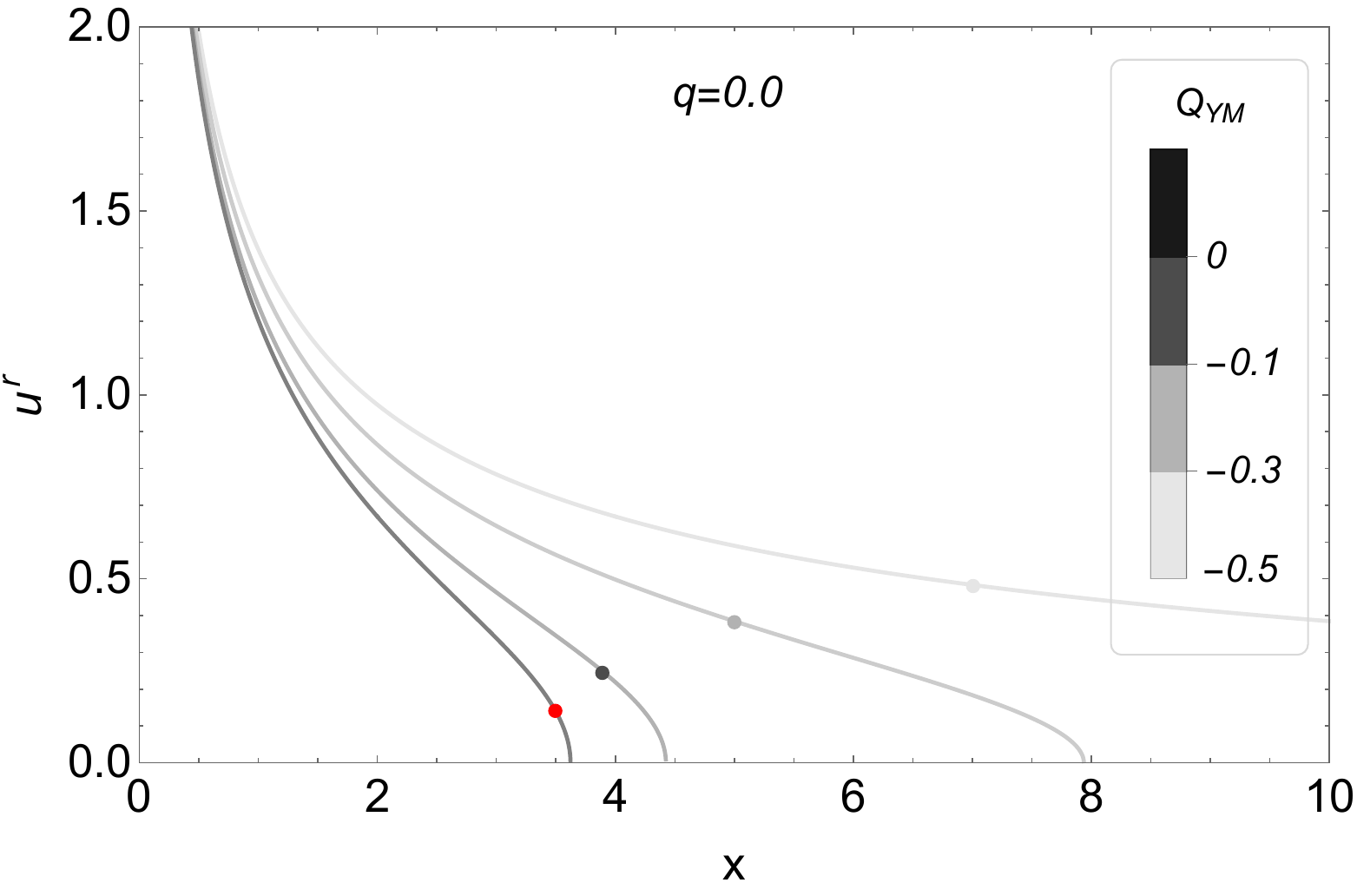}
\includegraphics[width=0.47\hsize,clip]{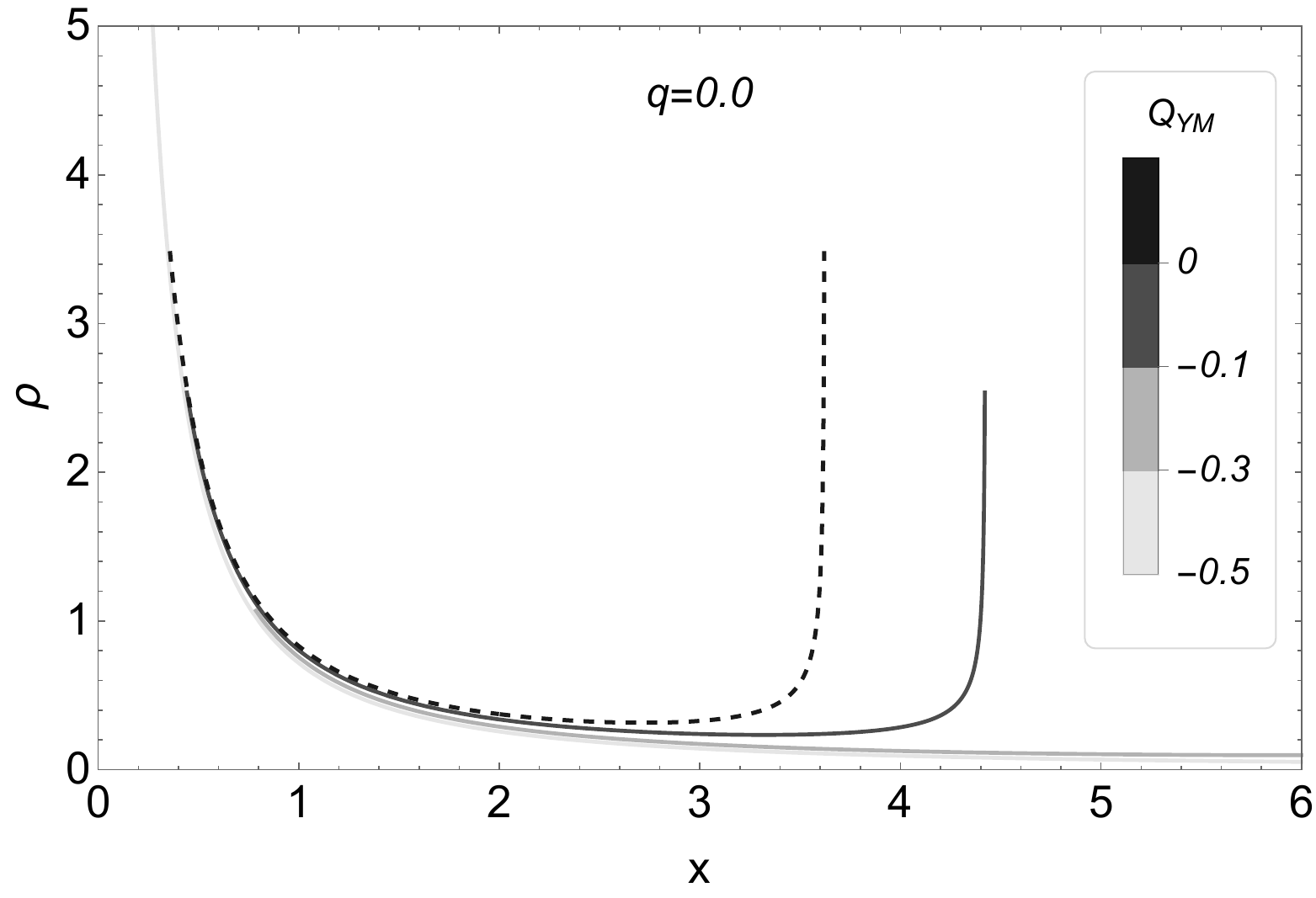}
\includegraphics[width=0.47\hsize,clip]{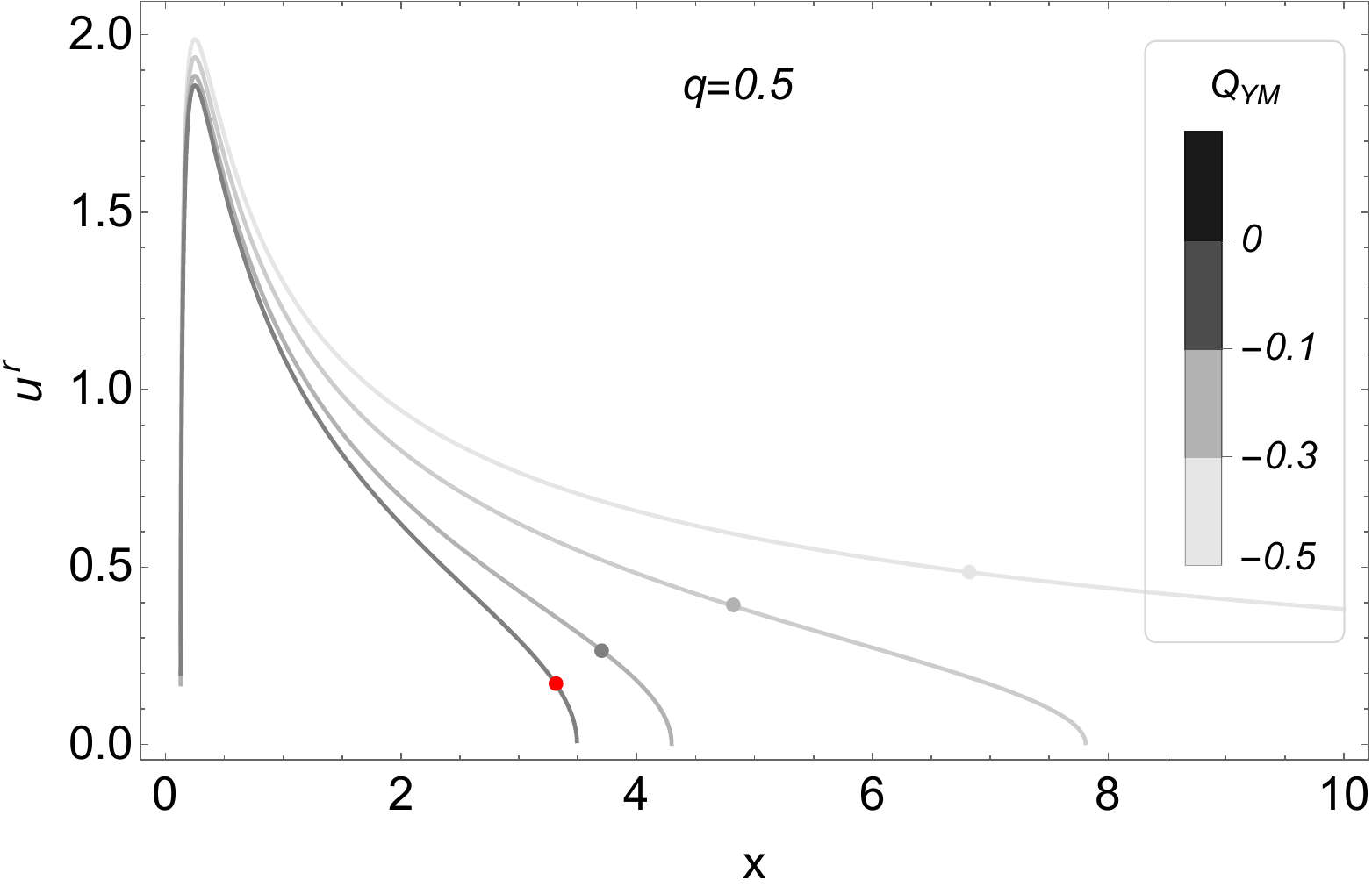}
\includegraphics[width=0.47\hsize,clip]{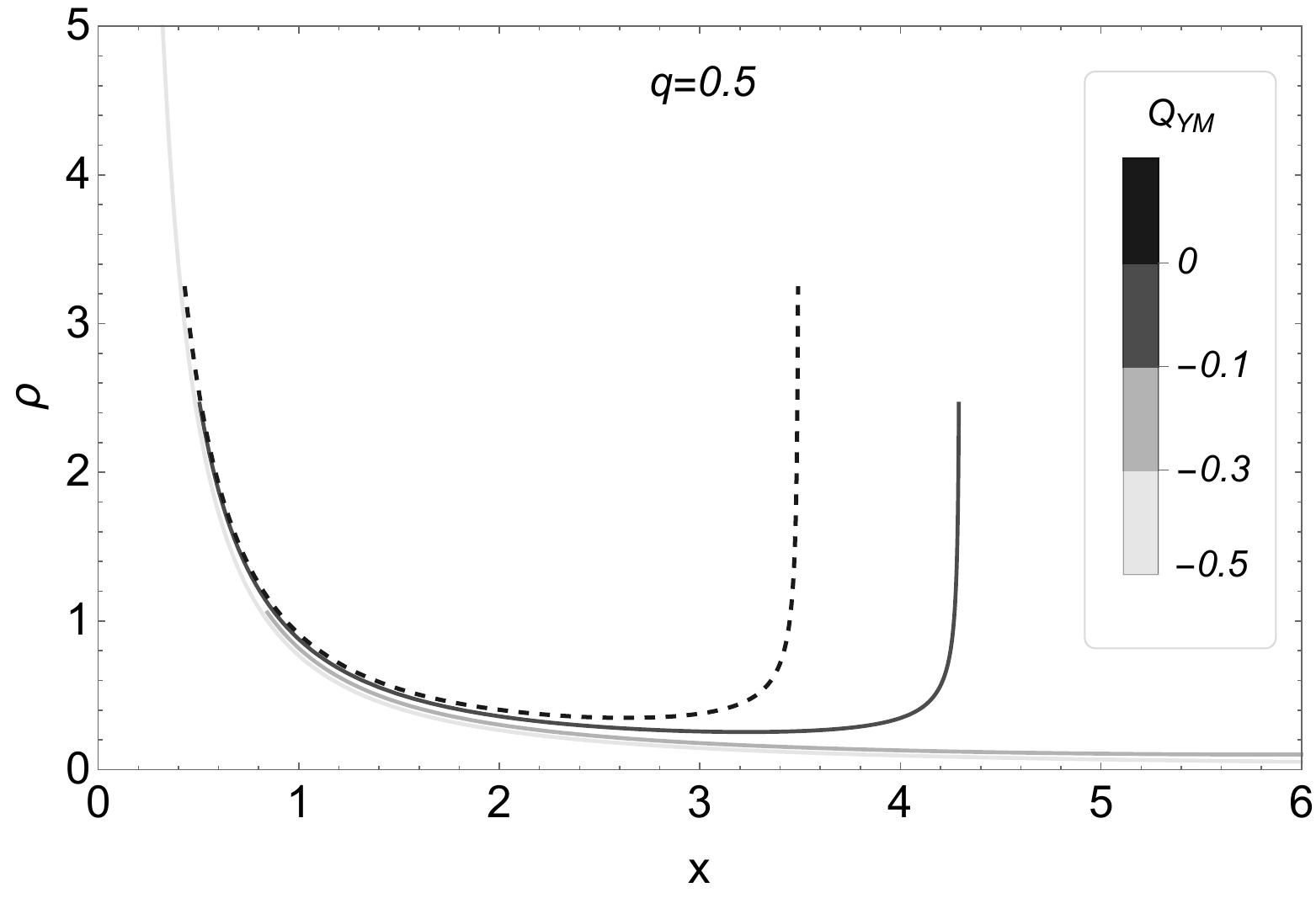}
\includegraphics[width=0.47\hsize,clip]{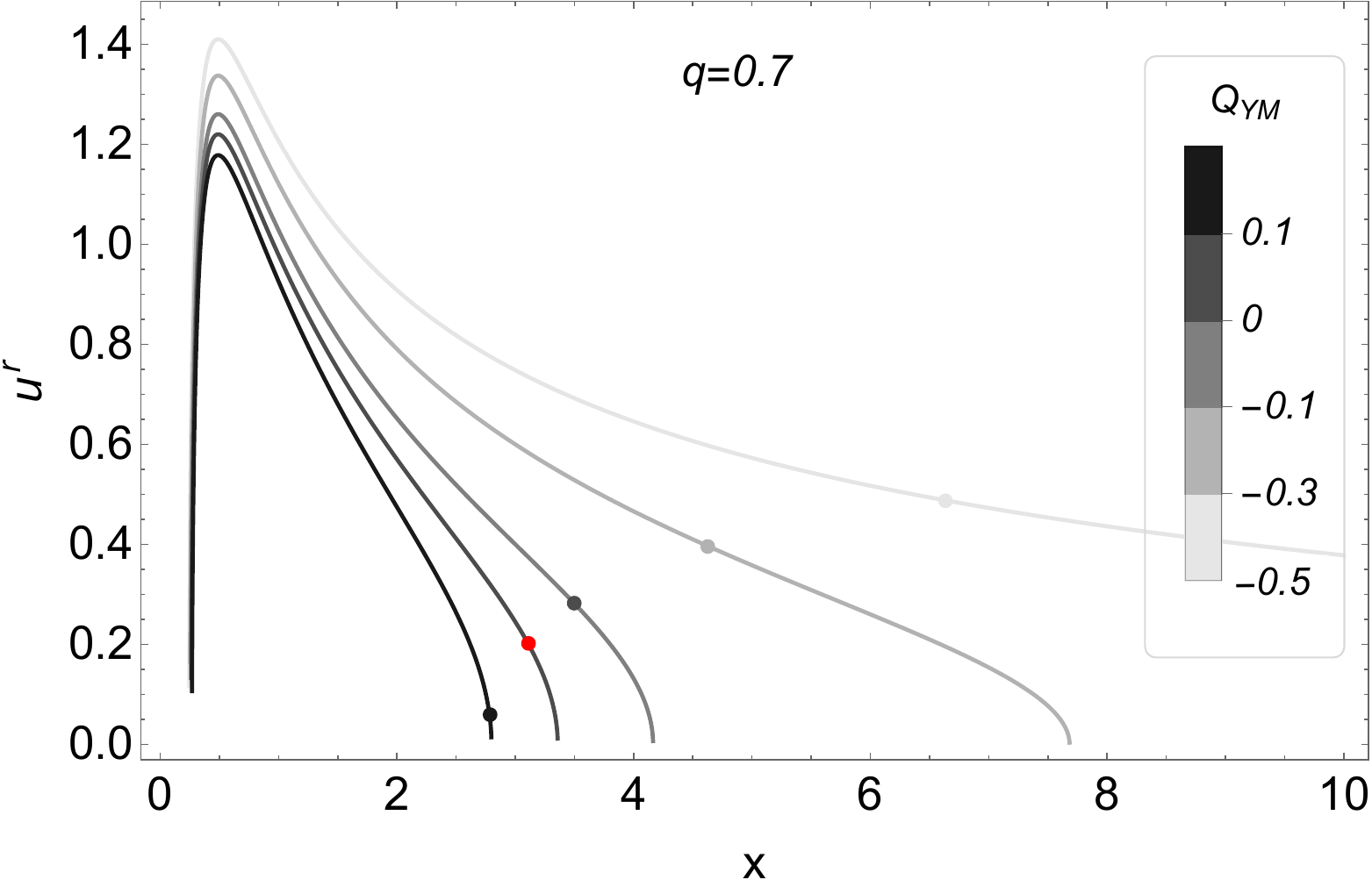}
\includegraphics[width=0.47\hsize,clip]{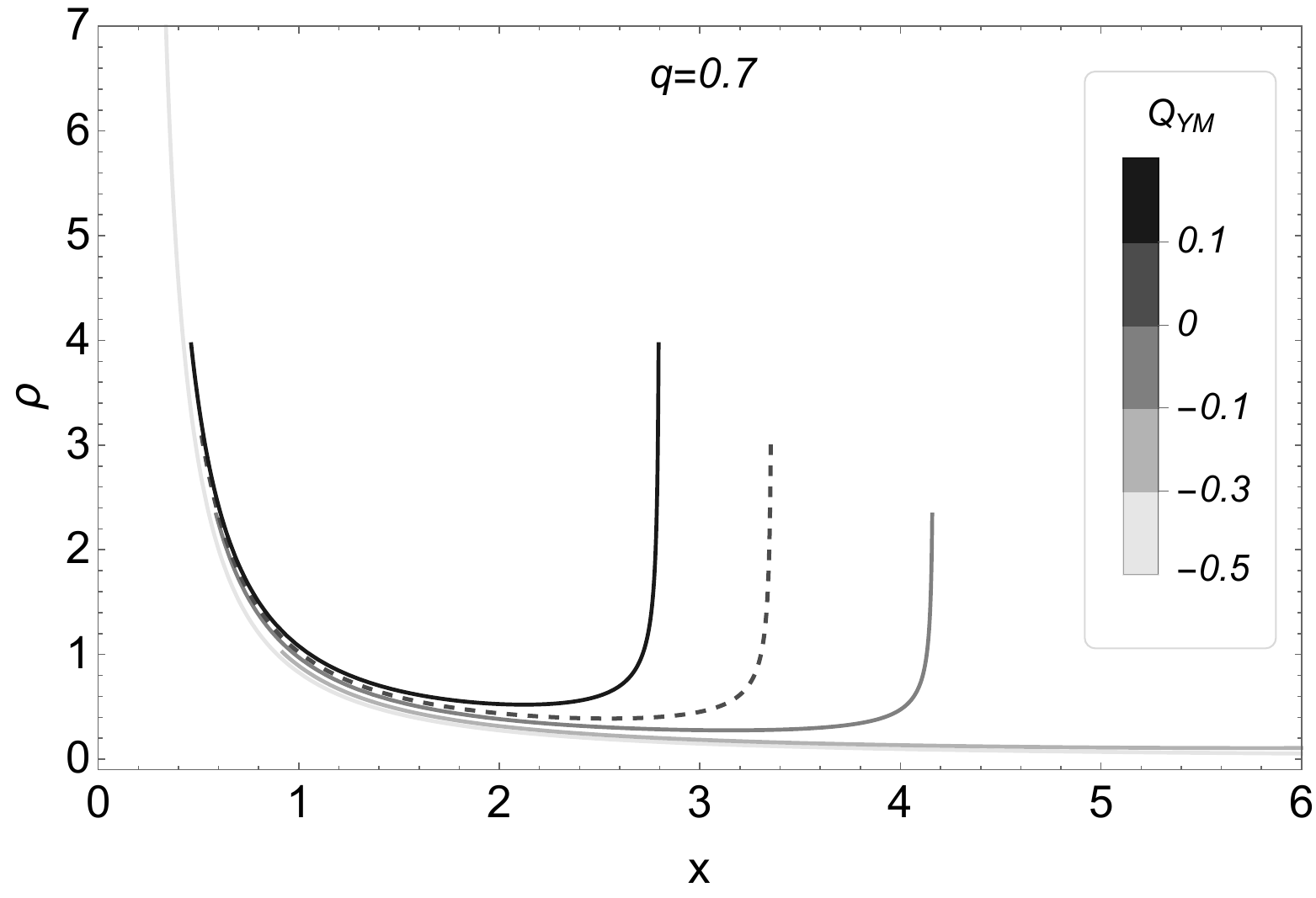}
\includegraphics[width=0.47\hsize,clip]{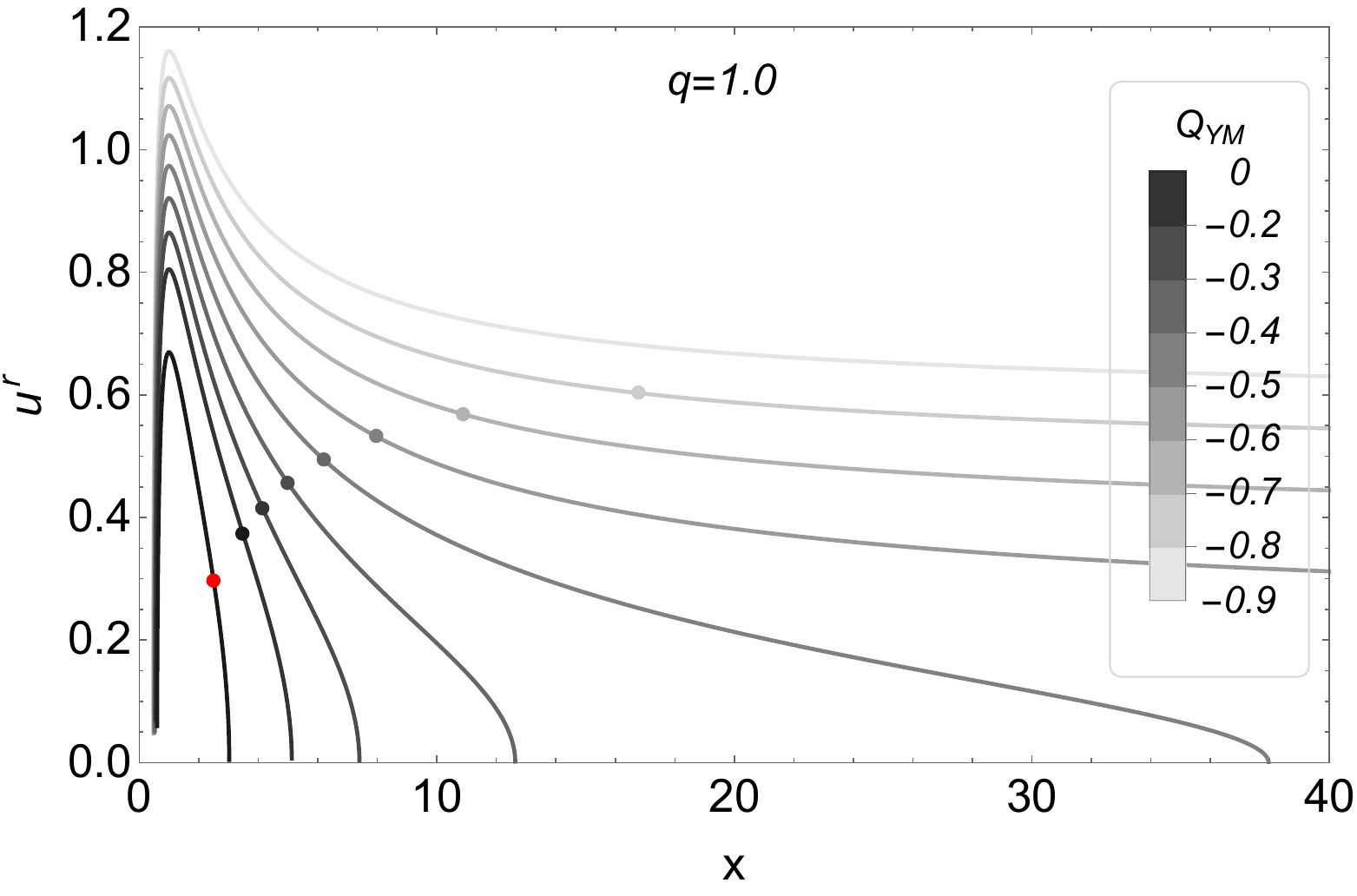}
\includegraphics[width=0.47\hsize,clip]{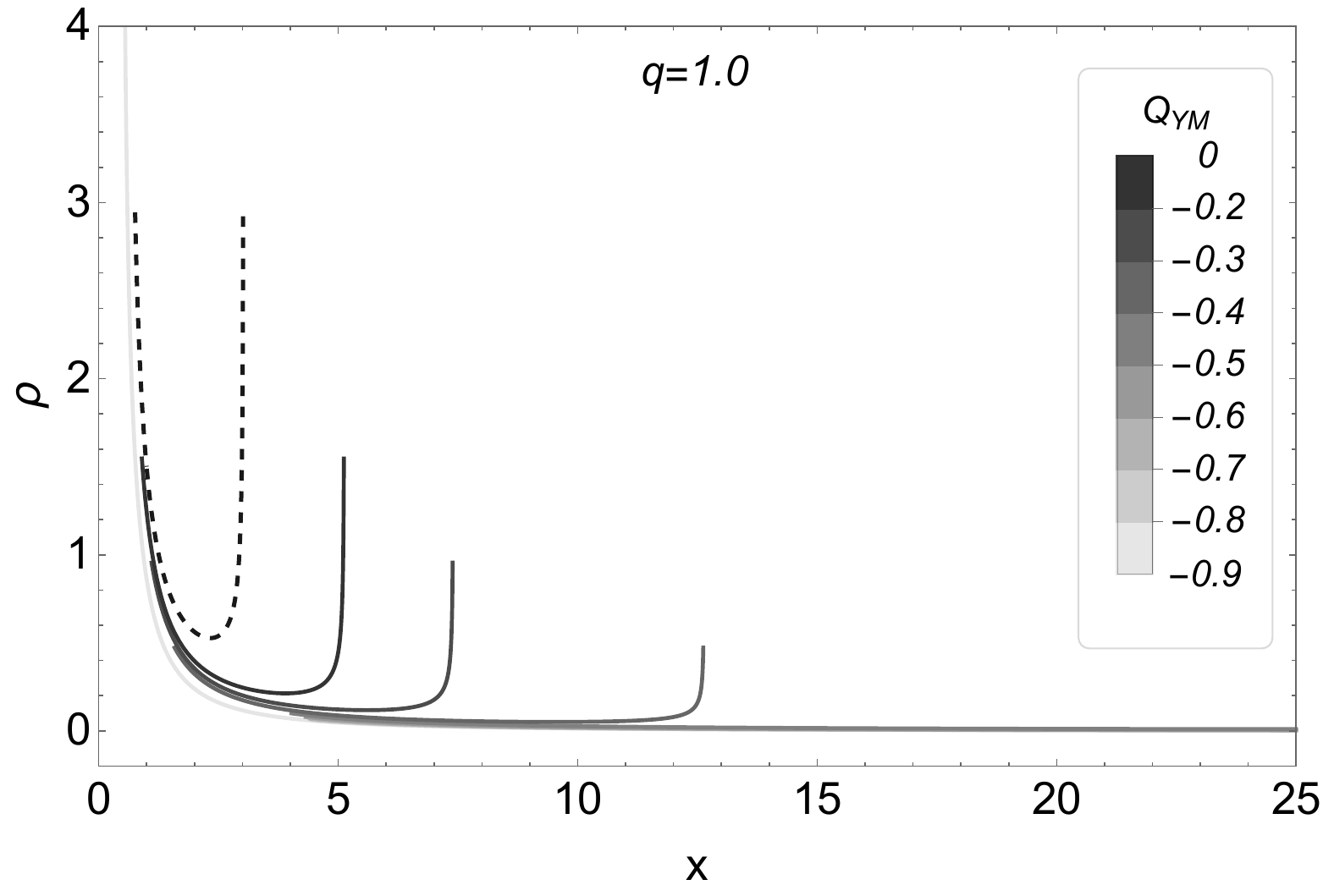}
\caption{\textbf{Case} $\omega=1/4$. Left panels show the infall radial velocity while right panels describe the mass density distribution due to the BH gravitational potential for a specific value of the electric charge $q$ along with the effect of changing simultaneously the Yang-Mill charge $Q_{\rm YM}$, as described by the bar legend. We have  delimited the range $0<Q_{\rm YM}<-0.5$ for the same reason as the relativistic case.
Notice that in bottom panels negative values of $Q_{\rm YM}$ have been considered only to allow $q=1$ according to the table 1.} \label{fig:2}
\end{figure*}

This first comprehensive inspection tells us then that the Yang-Mills charge remarkably enriches the physical condition behind the transonic flow of accreting matter, thus allowing new critical or sound points to existing.
We do not explain in detail the other cases because they can be described, considering the aforementioned properties for the stiff case, in a general way as follows. As $\omega$ gets smaller, passing from the ultra-relativistic case $\omega=1/2$ (Fig. \ref{fig:ultrarel}), the relativistic case (Fig. \ref{fig:rel}) to the case $\omega = 1/4$ (Fig. \ref{fig:2}),
and considering the same values of charges, all critical points move away from the BH. This is the reason why the sequence of curves spreads out so that they must be relocated in the new position of the critical points. This aspect becomes more noticeable as the electric charge $q$ increases. 
Another noteworthy feature is that, for $\omega=1/3$ and $\omega=1/4$, as well as small electric charges $q<1$, the transonic condition is no longer guaranteed. As a result, few sequence of curves are allowed to transit through the critical points, in particular those characterized by $Q_{\rm YM}>0$.

\begin{figure*}
\centering
\includegraphics[width=0.47\hsize,clip]{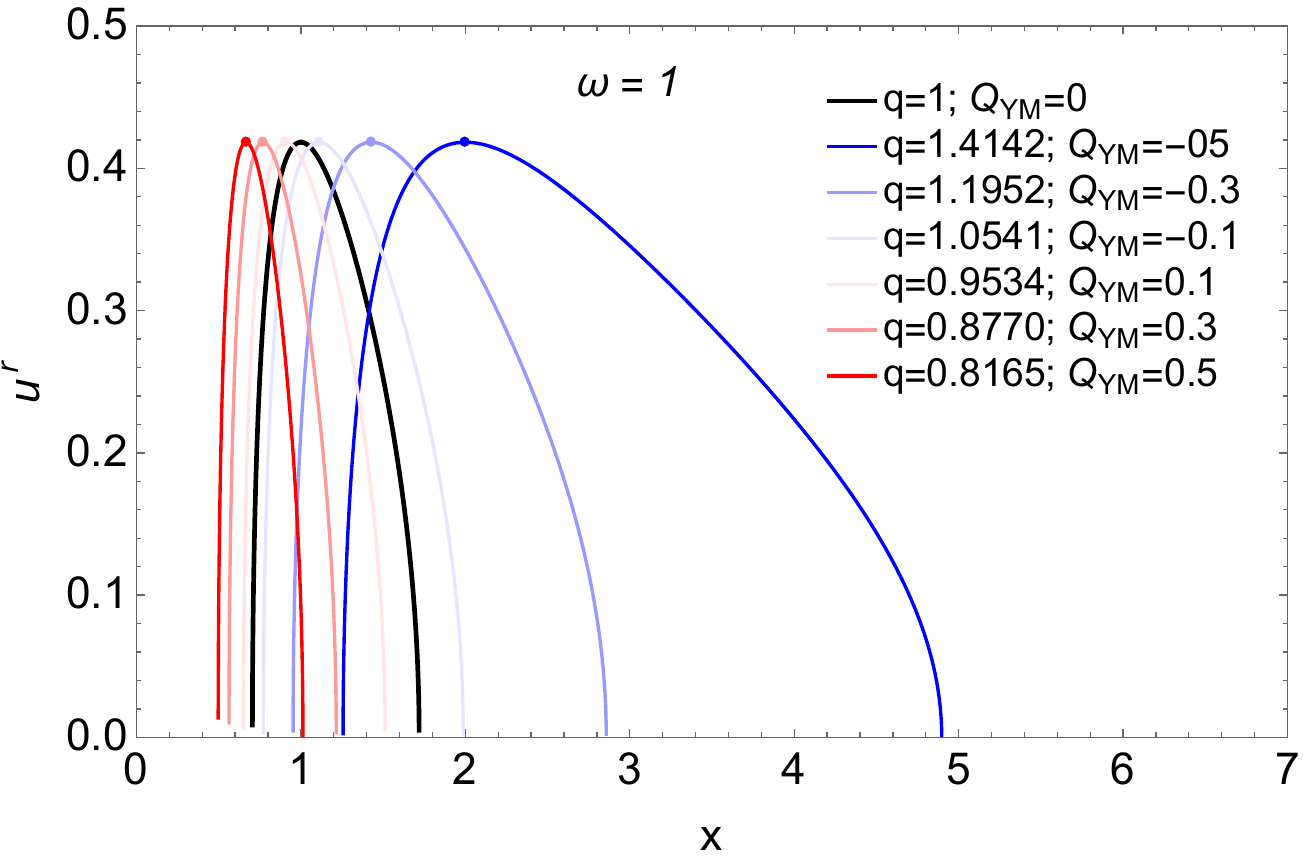}
\includegraphics[width=0.47\hsize,clip]{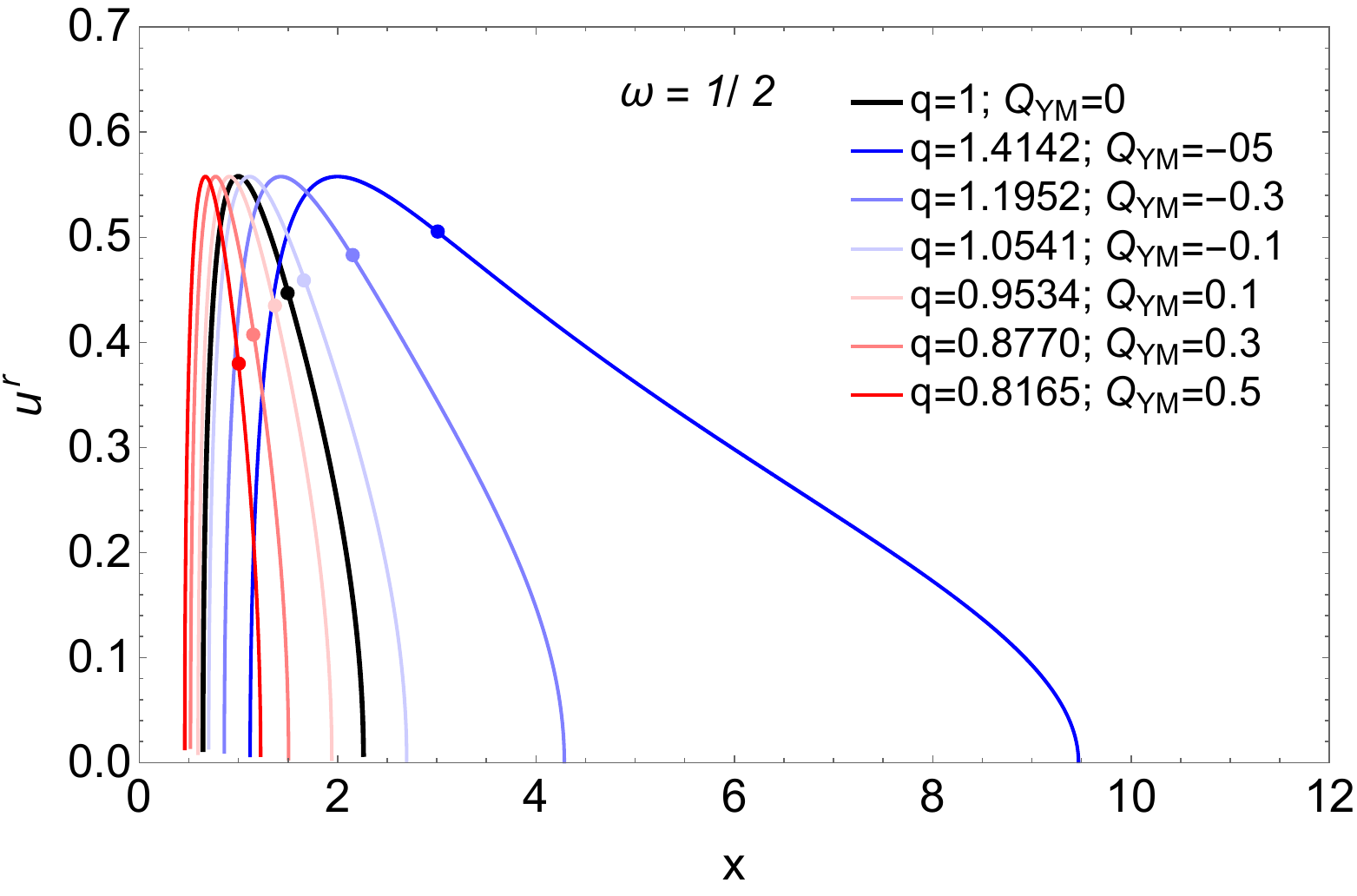}
\includegraphics[width=0.47\hsize,clip]{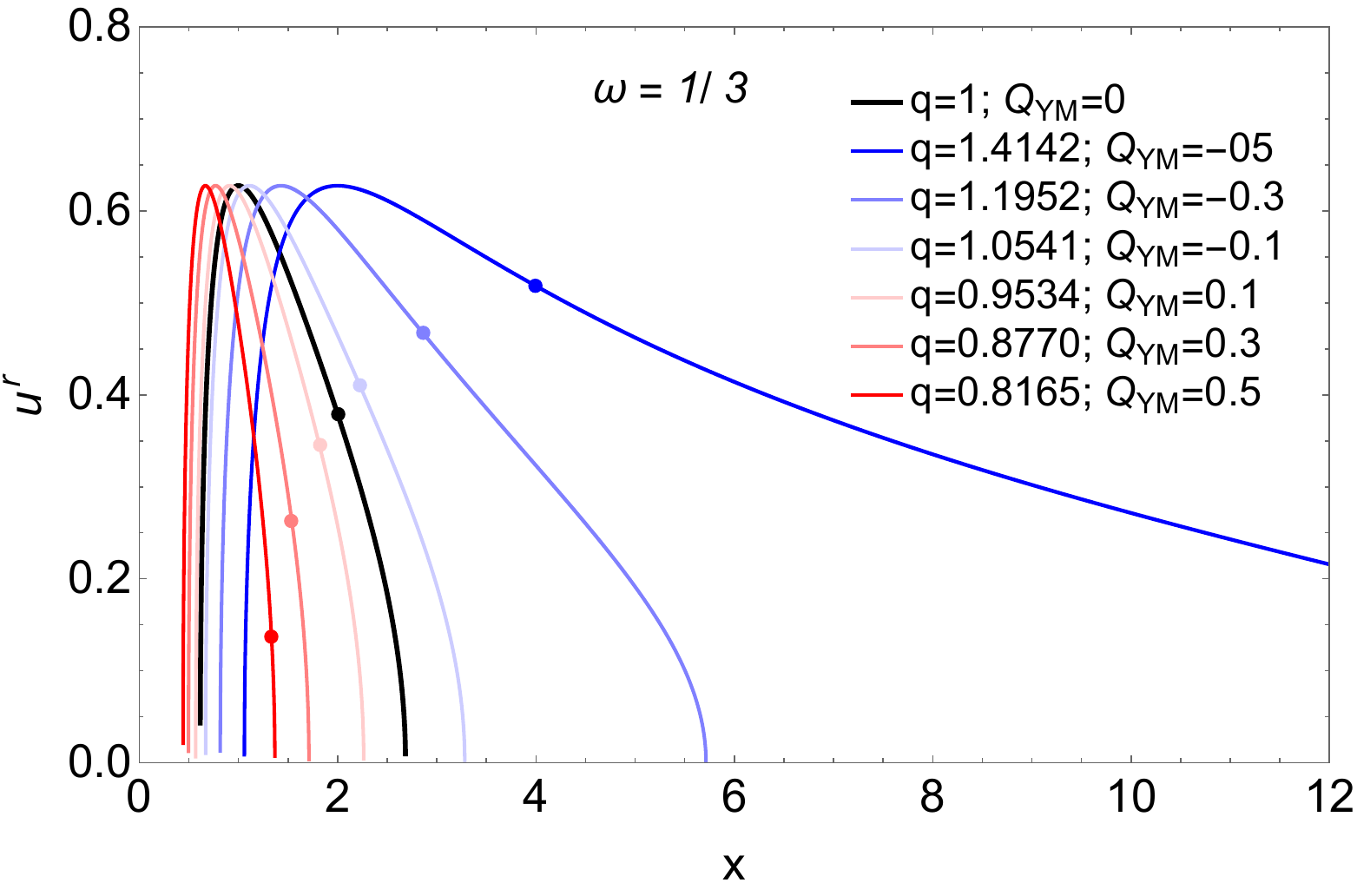}
\includegraphics[width=0.47\hsize,clip]{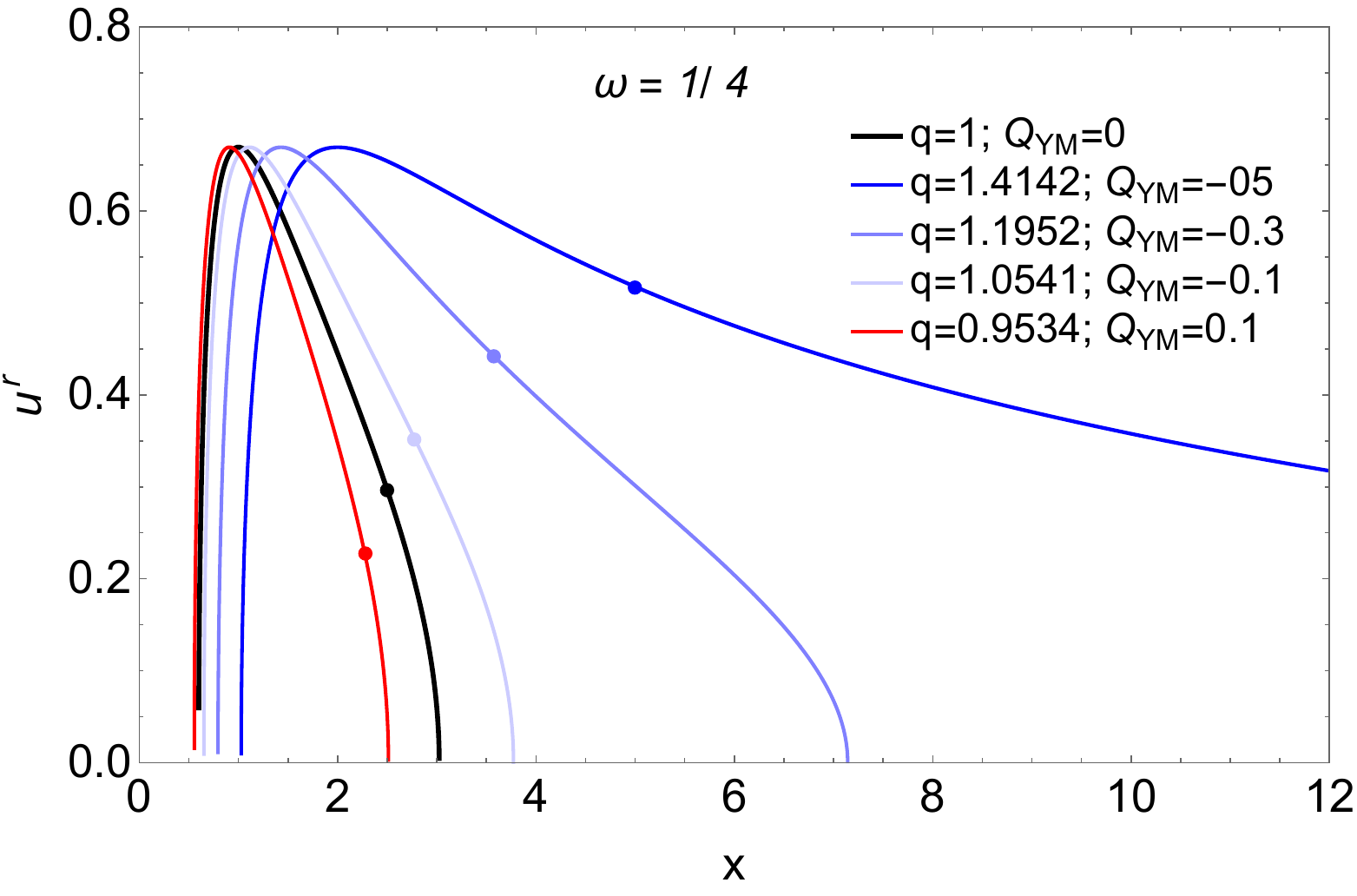}
\caption{Infall radial velocity for a couple of values $q$ and $Q_{\rm YM}$ that lead to the extremal BH solution of the theory. As the Yang-Mills charge is turned on, new radial velocities are allowed to transit around the critical points. See also Table \ref{tab:criticalpoints} for further information.} \label{fig:critical_velocity}
\end{figure*}

Finally, we focus on both charges values that lead to the extremal case for this new BH solution and pay attention on how the sequence of velocity curves transit for the same $\omega$-values as before. As a reference, the infall radial velocity for the standard extremal Reissner Nordstr\"{o}m case ($Q=1,Q_{\rm YM}=0$) is described by the black curve along with some cases characterized by positives and negatives Yang-Mills charges whose associated velocities are located respectively around it. This is shown in Fig.~\ref{fig:critical_velocity}. All points over the curves mark the position of the critical radius. Detailed information about the precise values of charges, critical points, and critical velocities can be found in Table.~\ref{tab:criticalpoints}. All the main features discussed above keep unalterably. Notice, however, that for the case $\omega=1$, the value of critical velocity does not change, unlike other cases.


\begin{table*}[htp]
\centering  
\begin{ruledtabular}
\caption{Critical radii and velocities for a couple of values $q$ and $Q_{YM}$, that correspond to the new extremal BH solution according to Eq.~(\ref{horizons}), for a given constant $\omega$. For $\omega=1$, the position of the critical point matches the corresponding event horizon as in the Schwarzschild and RN cases.}
\begin{tabular}{| c | c| c | c| c | c|}
 \multirow{2}{*}{$q$} & \multirow{2}{*}{$Q_{YM}$} & $\omega=1$ & $\omega=1/2$ & $\omega=1/3$ & $\omega=1/4$ \\ 
  &  & ($x_{c}, v_{c}$) &  ($x_{c}, v_{c}$) & ($x_{c}, v_{c}$) & ($x_{c}, v_{c}$) \\ \hline
1.41421 & -0.5 & (2, 0.41833) & (3, 0.505524) & (4, 0.51841) & (5., 0.517687)\\ 
1.19523 & -0.3 & (1.42857, 0.41833) & (2.14286, 0.483044) & (2.85714, 0.467706) & (3.57143, 0.442718)\\
1.05409 & -0.1 & (1.11111, 0.41833) & (1.66667, 0.459465) & (2.22222, 0.410789) & (2.77778, 0.352134)\\
1 & 0 & (1, 0.41833) & (3/2, 0.44721) & (2, 0.37914) & (5/2, 0.296645)\\ 
0.95346 & 0.1 & (0.90909, 0.41833) & (1.36365, 0.434606) & (1.81819, 0.344595) & (2.27274, 0.228028)\\
0.87705 & 0.3 & (0.76923, 0.41833) & (1.15389, 0.408229) & (1.53849, 0.262183) & --\\
0.81649 & 0.5 & (0.66666, 0.41833) & (1, 0.380048) & (1.33334, 0.136908) & --
\end{tabular}
\end{ruledtabular}\label{tab:criticalpoints}
\end{table*}
%

%%%%%%%%%%%%%%%%%%%%%%%%%%%%%%%%%%%%%%%%%%%%%%%%%%%%%
\subsection{Accretion rate for a polytropic fluid}
%%%%%%%%%%%%%%%%%%%%%%%%%%%%%%%%%%%%%%%%%%%%%%%%%%%%%

At this point, the properties of the steady transonic flow at the sonic point are known which allow us to compute the maximum accretion rate for the Einstein-Maxwell power-Yang-Mills theory. Before proceeding, it is very useful to rewrite radial velocity and sound speed at the critical point in terms of the known boundary conditions. It also demands knowledge of the equation of state. So we consider a non-relativistic baryonic gas with polytropic equation
\begin{equation}
    P = K n^{\gamma},\label{eqnofsate1}
\end{equation}
where $\gamma$ is the adiabatic index and $K$ is a constant. With this and from the first law of thermodynamics one can get
\begin{equation}
    \rho = m n + \frac{K}{\gamma-1} n^{\gamma}.\label{eqnofsate2}
\end{equation}
%

%%%%%%%%%%%%%%%%%%%%%%%%%%%%%%%%%%%%%
\subsubsection{Relativistic regime}
%%%%%%%%%%%%%%%%%%%%%%%%%%%%%%%%%%%%%

We provide here an exact expression for the accretion ratio in the relativistic regime for the theory in consideration. When studying accretion however most of the approaches focus on the asymptotic limit only, leaving thus incomplete the full picture of the problem. 
For our surprise, no work has investigated analytically the steady-state spherical accretion rate in alternative theories of gravity beyond GR in the fully relativistic regime. We follow closely Ref.~\cite{Richards:2021zbr}, where Bondi accretion of steady spherical gas flow onto a Schwarzschild black hole has been studied. 

From the polytropic equation, the sound speed can be written in terms of the mass density 
\begin{equation}
    c_{s}^{2}=\frac{\gamma K \rho_{0}^{\gamma-1}}{1+\gamma K\rho_{0}^{\gamma-1}/(\gamma-1)}.\label{soundspeed2}
\end{equation}
Evaluating this expression at the critical point and in the asymptotic region, it is easy to find the relation
\begin{equation}
    \rho_{0,s} = \rho_{0,\infty} \left( \frac{c_{s,c}^{2}}{c_{s,\infty}^{2}}\right)^{\frac{1}{\gamma-1}} \left( \frac{\gamma -1 - c_{s,\infty}^{2}}{\gamma -1 - c_{s,c}^{2}}\right)^{\frac{1}{\gamma-1}}.\label{criticaldensity2} 
\end{equation}
Finally, the relativistic Bernoulli equation, evaluated at the critical point, provides the relation
\begin{equation}
    (1+3c_{s,c}^{2})\left(1-\frac{c_{s,c}^{2}}{\gamma-1}\right)^{2} = \left(1-\frac{c_{s,\infty}^{2}}{\gamma-1}\right)^{2}.\label{Bernoullicriticaleqn}
\end{equation}
So, once the boundary condition at infinity is given, all the quantities at the critical point are uniquely determined.  Notice that the above equation is a cubic equation for $c_{s,c}^{2}$ but only one solution is (real) physical for a given polytropic equation of state. This is solved by using a root-finding procedure.

Considering all the above, the critical accretion rate
\begin{equation}
    \dot{M}=4\pi \rho_{0,s} u_{s} r_{s}^{2},\label{accretion2}
\end{equation}
can be computed easily
\begin{equation}
    \dot{M}_{\rm MRN}=4\pi \left(\frac{M}{c_{s,\infty}^{2}}\right)^{2} c_{s,\infty}^{2}\;\rho_{0,\infty}\;\lambda_{\rm MRN},\label{accretionDRN}
    \end{equation}
with a modified eigenvalue
\begin{align}
\begin{split}
   \lambda_{\rm MRN} \equiv  &\left( \frac{c_{s,c}^{2}}{c_{s,\infty}^{2}}  \right)^{\frac{5-3\gamma}{\gamma-1}} \left( \frac{\gamma-1-c_{s,\infty}^{2}}{\gamma-1-c_{s,c}^{2}} \right)^{\frac{1}{\gamma-1}} \times
   \\
   & \ 
   \frac{1}{4}\beta (1+3c_{s,c}^{2})^{3/2}.\label{eigenDRN}
\end{split}
\end{align}
Here the $\beta$ factor contains information of both charges whereby it  accounts for deviation from the Schwarzschild solution and the RN case as well. This is defined as
\begin{align}
\begin{split}
    \beta = &\frac{1}{4(1+Q_{\rm YM})^{2}} \times
    \\
    &
    \left[ 1+ \sqrt{1-\frac{8 c_{s,c}^{2}(1+c_{s,c}^{2})q^{2}(1+Q_{\rm YM})}{(1+3c_{s,c}^{2})^{2}}}\right]^{2}.\label{betafactor}
\end{split}
\end{align}
Computing the ratio of Schwarzschild to the modified Reissner-Nordstr\"{o}m accretion ratios is particularly convenient for quantifying the effect of both charges
\begin{equation}
   \frac{\dot{M}_{\rm MRN}}{\dot{M}_{\rm Sch}} = \beta. \label{accretionrations}
\end{equation}
The purely power Yang-Mills case is independent of the adiabatic index and  the boundary condition. This provides $\beta_{\rm YM} = \frac{1}{(1+Q_{\rm YM})^{2}}$. Considering $Q_{\rm YM}>0$ leads to $\beta<1$ and \textit{vice versa}. It means hence that $Q_{\rm YM}<0$ produce an enhancement of the accretion rate which does not happen in the standard Reissner Nordstr\"{o}m (see e.g. \cite{FreitasPacheco:2011yme,Ficek:2015eya}) and Kerr \cite{Aguayo-Ortiz:2021jzv} solutions: the mass accretion rate decreases as the charge and spin values increase. On the other hand, the accretion rate for the Schwarzschild solution is recovered when both charges are turned off, leading to $\beta\to 1$, as can be plainly checked. Solving the Bernoulli equation Eq.~(\ref{Bernoullicriticaleqn}), that is, defining the sound speed at the boundary condition and solving for $c_{s,c}^{2}$, one can fully compute the $\beta$ factor (Eq.~(\ref{betafactor})) to determine Eq.~(\ref{accretionrations}). This is shown in Fig.~\ref{fig:acretion}. 

It is important to mention that the larger the boundary sound speed, the
shorter the accretion rate variations are, keeping both the  adiabatic index $\gamma$  and the electric charge fixed. Also, taking $q$ small makes things less distinguishable for a different $\gamma$. So we  take $c_{s}^{2}=0.01$ and $q>0.3$ to see changes between different $\gamma$'s values. Intersections of all curves with the (gray) dashed vertical line $Q_{\rm YM}=0$ represent the Reissner Nordstr\"{o}m accretion rates (strictly for $q\leq1$) which in all cases is below one as is known. 
%
%\Gabriel{
Notice that, for $\gamma= 4/3, 5/3$, the effect of the electric charge is quite perceptible: compare $q=0$ (solid black curve) and $q=1$ (magenta curve) cases; unlike the case $\gamma=1$ where both curves are practically indistinguishable. 
All curves, however, converge as $Q_{\rm YM}\to -1$ independent of the electric charge and adiabatic index. In general, the accretion rate efficiency in the relativistic regime ($\gamma=4/3$) is only larger than its non-relativistic counterpart ($\gamma=5/3$) at sub-percent level. However, in the isothermal limit $\gamma\to1$, the difference is a few percent levels compared to the previous cases.
%} 
Even though the electric charge decreases the accretion rate, it is possible to have $\dot{M}_{\rm MRN}\geq\dot{M}_{\rm Sch}$ by taking $Q_{\rm YM}<0$. 

As a main result, the Yang-Mill charge, with $Q_{\rm YM}<0$, can correct the accretion rate deficiency of the electric charge or practically cease the accretion process for $Q_{\rm YM}>0$ because the critical point is well inside the event horizon. The enhancement of the accretion rate can be up to order $10^2$ for $Q_{\rm YM}\to -1$ in comparison to the standard Schwarzschild case. These results hold only for the power Yang-Mills case $p=1/2$. We expect to find however similar qualitative results for $p\neq1/2$ whenever it leads to two branches ($\pm$) of solutions for the Yang-Mills charge when finding the roots of the polynomial $Q_{\rm YM}= \frac{2^{p - 1} q_{\rm YM}^{2p}} {4 p - 3}$.

%%%%%%%%%%%%%%%%%%%%%%%%%%%%%%%%%%%%%
\subsubsection{Asymptotic limit}
%%%%%%%%%%%%%%%%%%%%%%%%%%%%%%%%%%%%%

\begin{figure}
\centering
\includegraphics[width=1\hsize,clip]{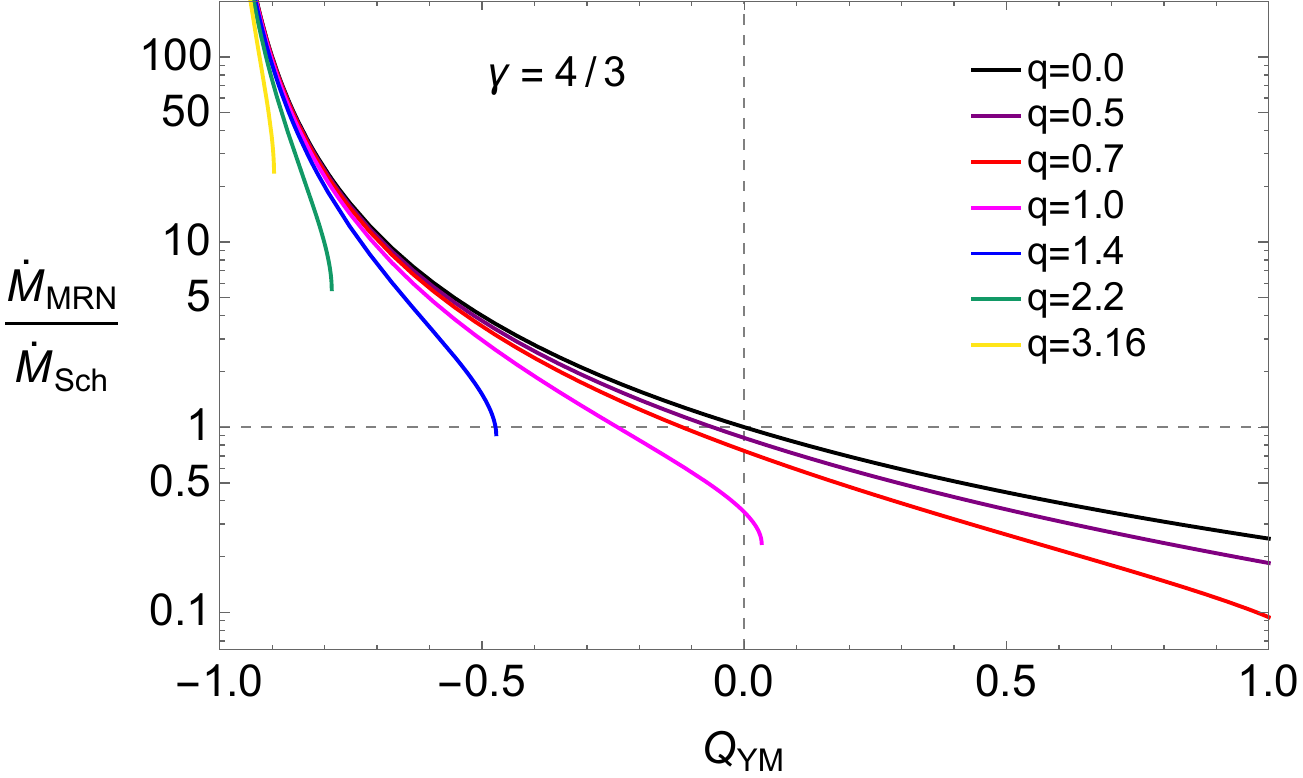} \\
\includegraphics[width=1\hsize,clip]{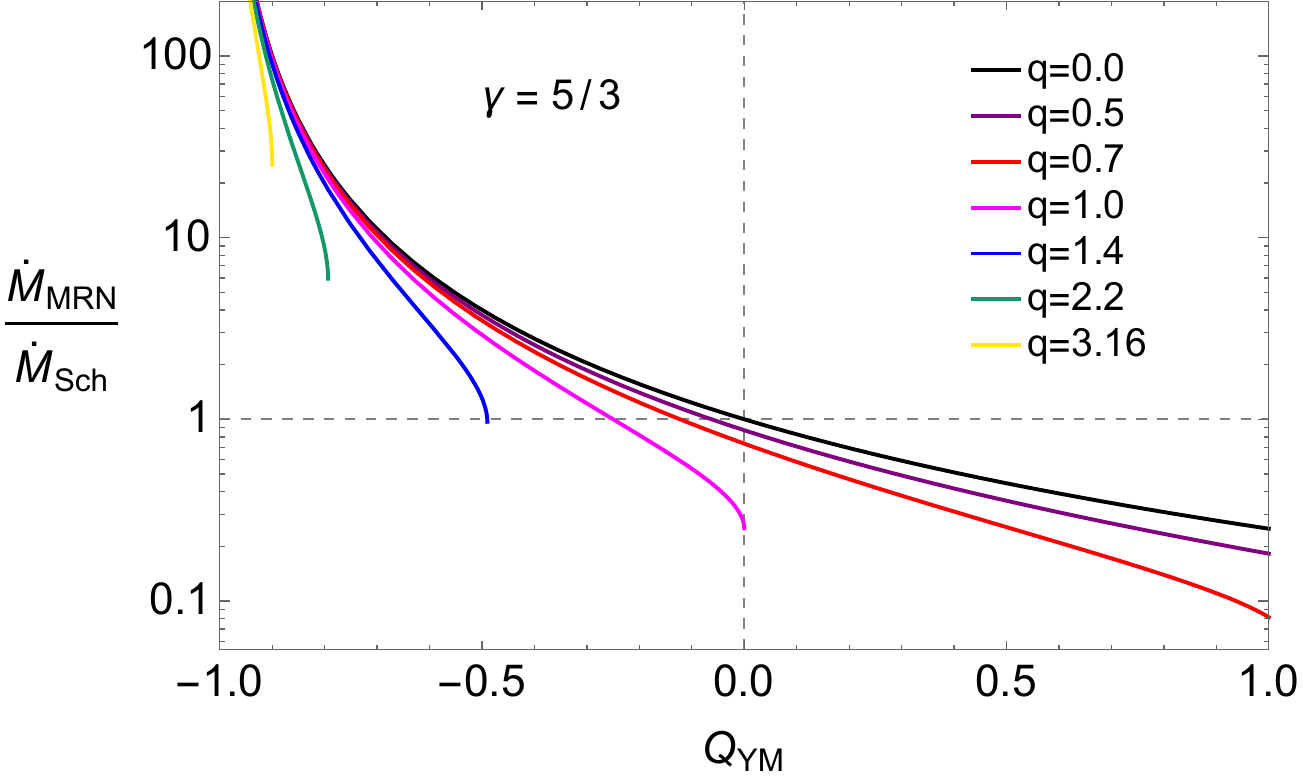} \\
\includegraphics[width=1\hsize,clip]{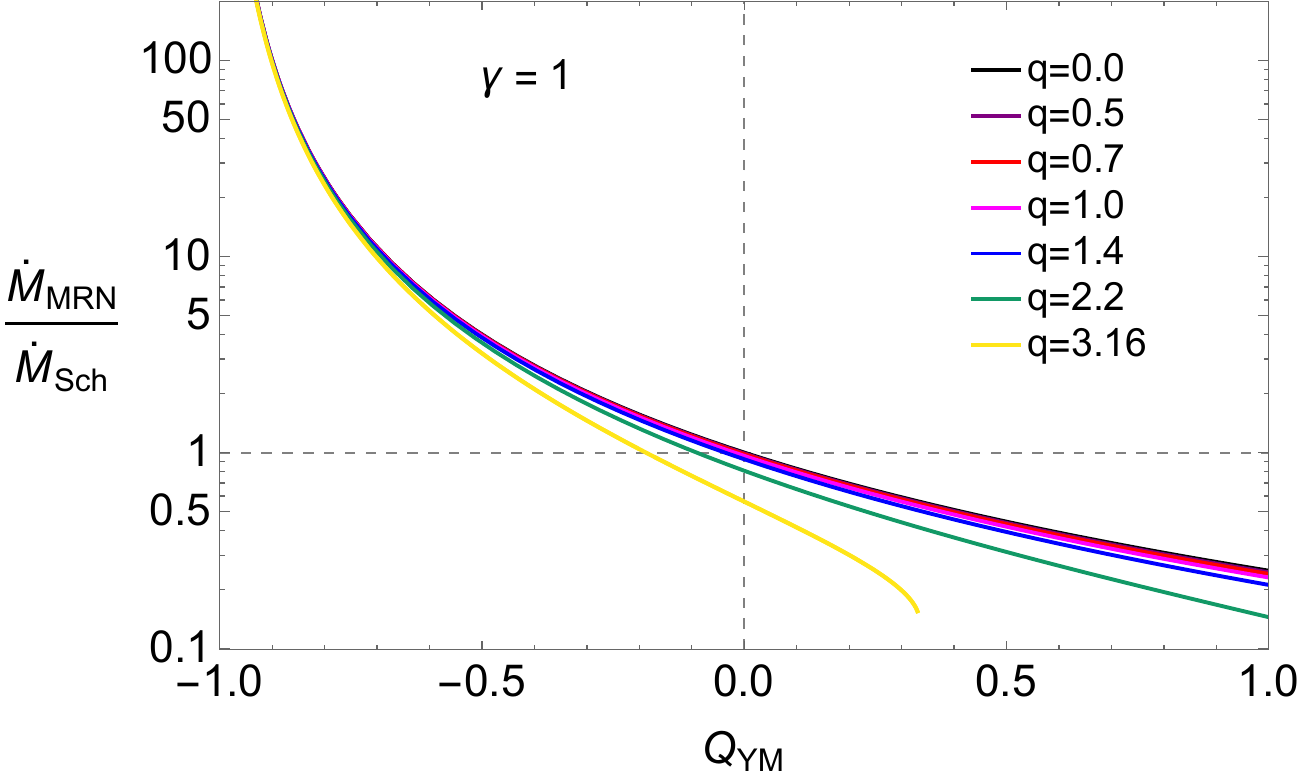}
\caption{Ratio of the mass accretion rates (Eq.~(\ref{accretionrations})) as a function of the Yang-Mills charge for different polytropic indexes and electric charges, as described by the legend. Notice that the Schwarzschild case is formally recovered when both charges vanish. However, as $\gamma\to1$, $Q_{\rm YM}\to0$ and $q<1$, $\dot{M}_{\rm MRN}$ approaches closely to $\dot{M}_{\rm Sch}$. We have chosen $c_{s,\infty}^{2}=10^{-3}$ as a reference value. The transonic condition for the critical velocity $u_{c}>c_{s,\infty}$ is fulfilled, where the maximum critical velocity is obtained for $\gamma=5/3$.} \label{fig:acretion}
\end{figure}

We do not describe in detail all derivations regarding the hydrodynamic equations for the non-relativistic case since most of them are independent of the metric background and can be found, for instance, in Ref.~\cite{Shapiro:1983du}. This description is strictly valid for a polytropic fluid Eq.~(\ref{eqnofsate1}) with adiabatic index $\gamma<5/3$ \cite{Shapiro:1983du,Aguayo-Ortiz:2021jzv}. The Bernoulli equation provides the relation
\begin{equation}
   c_{s,c}^2 \approx \frac{2 c_{s,\infty}^{2}}{(5-3\gamma)} ,\label{sub32:eqn3}
\end{equation}
for the non-relativistic condition $c_{s,c}\ll 1$ which holds for reasonable large radius, $r\gg r_{c}$, far away from the BH gravitational influence. The same condition leads to the simple relation 
\begin{equation}
   c_{s,c}^2 \approx K \gamma\; \rho_{0,c}^{\gamma-1} ,\label{sub32:eqn4}
\end{equation}
between the sound speed and the mass density at the critical point. This implies that the mass density can be expressed in terms of the sound speed at the infinity in view of Eq.~(\ref{sub32:eqn3}) to yield
\begin{equation}
   \rho_{0,c} \approx \rho_{0,\infty}\left(\frac{c_{s,c}^2}{c_{s,\infty}^2}\right)^{\frac{1}{\gamma-1}}.\label{sub32:eqn5}
\end{equation}
At this point all above is standard and the quantities does not receive contributions from the effective global charges $Q$ and $Q_{\rm YM}$ at the lowest order in $c_{s,c}$. This is not the case however for the critical radius Eq.~(\ref{eqn:criticalradius}) where the YM charge already appears explicitly at leading order, affecting the Schwarzschild critical radius as a non-linear correction. At the next leading order, both charges $Q$ and $Q_{\rm YM}$ appear independently of one another. Next correction, however, reveals the coupling between them as a non-linear manifestation of the Maxwell-Power Yang-Mills structure:
\begin{align*}
    r_{c} \approx & \frac{M}{2 (1+Q_{\rm YM}) c_{s,c}^{2}} + \frac{3 M^{2} - 2 Q^{2} (1+Q_{\rm YM})}{2 M (1+Q_{\rm YM})} + \\ \nonumber
    &\frac{2 Q^{2}(M^{2} - Q^{2} (1+Q_{\rm YM}))} {M^{3}} c_{s,c}^{2} + \mathcal{O}(c_{s,c}^{4}).\label{sub32:eqn6}
\end{align*}
Keeping only contributions up to next leading order and using Eq.~(\ref{sub32:eqn3}), the critical radius can be approximated to
\begin{equation}
   r_{c} \approx \frac{1}{\eta} \left(\frac{(5-3\gamma)}{4c_{s,\infty}^{2}} M + \frac{3 M^{2} - 2 Q^{2} \eta}{2 M} \right),\label{sub32:eqn7}
\end{equation}
where the dimensionless correction factor due to the YM charge has been defined simply as
\begin{equation}
   \eta = 1 + Q_{\rm YM},\label{sub32:eqn8}
\end{equation}
considering the restriction $Q_{\rm YM} \neq -1$. It is instructive to write the purely Power Yang-Mills case 
\begin{equation}
   r_{c}^{\rm YM} \approx \frac{M}{\eta} \left(\frac{(5-3\gamma)}{4c_{s,\infty}^{2}} + \frac{3}{2}\right),\label{sub32:eqn9}
\end{equation}
and the non-relativistic version of the Reissner-Nordstr\"{o}m case
\begin{equation}
   r_{c}^{\rm RN} \approx \frac{(5-3\gamma)}{4c_{s,\infty}^{2}} M + \frac{3 M^{2}-2Q^{2}}{2M}.\label{sub32:eqn10}
\end{equation}
Considering all the above, we can proceed now to compute the accretion rate for the Einstein-Maxwell-Power-Yang-Mills theory which corresponds to a non-relativistic flow with non-vanishing corrections to the Newtonian transonic flow. This results in
\begin{align}
\begin{split}
   \dot{M}_{\rm MRN}^{\rm NR} \approx 4\pi \rho_{0,\infty} &\left(\frac{M}{\eta}\right)^{2} c_{s,\infty}^{-3} \left(\frac{1}{2}\right)^{\frac{\gamma+1}{2(\gamma-1)}}
   \times
   \\
   &
   \left(\frac{5-3\gamma}{4}\right)^{\frac{3\gamma-5}{2(\gamma-1)}}\chi^{2},\label{sub32:eqn9}
\end{split}
\end{align}
where the function $\chi=1+\left(6-4\frac{Q^{2}}{M^{2}}\eta\right)\frac{c_{s,\infty}^{2}}{(5-3\gamma)}$ has been introduced for comparison purposes with the Schwarzschild accretion rate expression in the non-relativistic regime. This, besides, contains information of both electric and Yang-Mills charges even far outside the event horizon. The non-relativistic version of the Reissner-Nordstr\"{o}m solution is straightforwardly recovered in the limit $\eta\to 1$ and the one for the Newtonian accretion rate is achieved additionally when $Q\to 0$. Consequently, we can write the accretion rate in the compact form
\begin{equation}
   \dot{M}_{\rm MRN}^{\rm NR} = \left(\frac{\chi}{\eta}\right)^{2}\dot{M}_{\rm New}.\label{sub32:eqn9}
\end{equation}
In this way, the ratio $\left(\frac{\chi}{\eta}\right)^ {2}$ accounts for the deviation from the Newtonian accretion rate and it is the quantity we have to pay our attention to. First examination tells us that the effect of the electric charge is not so prominent ($\sim 0.02\%$) since it is attenuated by the squared sound speed which is taken to be very small\footnote{This is however somehow artificial because of the non-linear dependence of the accretion rate on the critical radius that leads to  write the accretion rate in the form Eq.~(\ref{sub32:eqn9}), i.e., as a factor of the Newtonian accretion rate.}, although it can be increased as $\gamma\to 5/3$. For values within the range $1<\gamma<5/3$, however, there do not exist appreciable differences when $q$ varies within the approximation $c_{s,\infty}^{2}\ll 1$. Interestingly, the factor $\eta$, which depends on the Yang-Mills charge, does produce, on the contrary, significant changes in the accretion rate, specially for very small values since accretion rate scales as $\eta^{-2}$. Hence, the largest differences will appear for charge values $Q_{\rm  YM}\sim \mathcal{O}(-0.1)$, or more precisely near the asymptotic value $Q_{\rm  YM}\to-1$, as happens for the fully relativistic case (see Fig.~\ref{fig:acretion}). 

%

%%%%%%%%%%%%%%%%%%%%%%%%%%%%%%%%%%%%%%
\section{Discussion and conclusions}
%%%%%%%%%%%%%%%%%%%%%%%%%%%%%%%%%%%%%%

In the present paper, we have investigated the spacetime structure of the resulting BH that arises from the Einstein-Maxwell Power-Yang-Mills theory with power $p=1/2$. The presence of the Yang-Mills charge modifies non-trivially the structure of the standard Reissner-Nordstr\"{o}m BH, and in the absence of the electric counterpart (the purely power Yang-Mills case), the Schwarzschild solution. This can be noticed through Eqs.~(\ref{horizons}) and (\ref{eqn:MSch}) that describe, respectively, the position of the event horizons in the mentioned cases. In particular, negative Yang-Mills charges can increase considerably the event horizon for a given electric charge that can take, in this new modified version of the Reissner-Nordstr\"{o}m BH, values up to  $\sim3.2$, but still preserving the weak cosmic censorship conjecture.
Interestingly, the ISCO radius can be smaller than the extremal Reissner-Nordstr\"{o}m case and larger than the Schwarzschild solution depending on the combination of the charges ($Q,Q_{\rm YM}$) as reported in Table \ref{tab:isco}. The introduction of the Yang-Mills charge, along with a specific electric charge value, can lead to an unexpected degeneracy of the event horizon and ISCO radii with respect to the Schwarzschild and Reissner-Nordstr\"{o}m cases. This is not the case in the purely power Yang-Mils case. This degeneracy may be broken, for instance, with the aid of the BHs Shadow using the Event Horizon Telescope observations of Sgr A$^{\star}$.

As a first astrophysical implication of this theory, we have investigated the steady, transonic properties of isothermal test fluids through extensive numerical examples that cover from the Einstein-Power-Yang-Mills ($q=0$ and $Q_{\rm YM}\neq0$) to the Einstein-Maxwell-Power-Yang-Mills case ($q\neq0$ and $Q_{\rm YM}\neq0$). 
Thus, broad numerical  solutions of sequence of velocity and density curves have been depicted for discrete values of the involving charges and given equation of state $\omega$.

The derived results point out that all critical points shift towards the horizon for $Q_{\rm YM}>0$, while for $Q_{\rm YM}<0$ such points move away from the event horizon. Nevertheless, for $\omega=1/3$ and $\omega=1/4$ isothermal fluids are affected distinctively from the other cases considered: for some positive values of $Q_{\rm YM}$, the radial critical velocity can not transit through the critical point which is a necessary condition to have a well-behaved transonic flow and to avoid singularities in it. 
It restricts the sign of the Yang-Mills charge to mostly negative values for $q\leq1/2$ and open, on the other hand, the possibility of having positive values for larger electric charge  $q>1$. In contrast, for $\omega=1$,  $\omega=1/2$ and $q\leq1/2$, positive values of $Q_{\rm YM}$ are allowed. Although there is a little preference for having negative Yang-Mills charges for $q$ small, this is not a conclusive outcome in the sense that the transonic condition depends sensibly on both the electric charge and the nature of the fluid.  It suggests that implementation of observational data is imperative to set the suitable sign of $Q_{\rm YM}$ for ensuring the astrophysical applicability of the theory, and more generally, to constrain more robustly the available parameter space beyond the theoretical bounds derived in this paper from physical grounds.

At the lowest order in $c_{s}$, or equivalently in the weak gravity regime, we have found that both the electric and the Yang-Mills charges contribute independently to the critical radius. The latter enters, however, into the accretion rate expression in a non-linear way. Furthermore, we have quantified the effect of the Yang-Mill charge, in the full non-linear gravity regime, on the accretion rate for a polytropic fluid. The effect of the Yang-Mills charge   results, in the more optimistic scenario, in an enhancement of a factor of up to $10^2$ for $Q_{\rm YM}\to -1$. This conclusion holds for the range $1<\gamma<5/3$ and it is quite independent of the electric charge. 
As a main result, the mass accretion rate efficiency can be considerably improved, with respect to the standard Reissner-Nordstr\"{o}m and Schwarzschild solutions, for negative values of the Yang-Mills charge. Physically, it can be understood as follows: as the location of the critical points are far from the BH for $Q_{\rm YM}<0$, which means that particles velocities reach the transonic flow far before than those located near the BH $Q_{\rm YM}\geq0$, a major contribution to the accretion flow of infalling particles towards the BH is naturally expected as a result of the spherical symmetry. This collective effect translates into larger accretion rate.

As a striking similarity to other BH solutions, we should mention that the Einstein-Maxwell Power-Yang-Mills black hole background for $p=1/2$ mimics a charged black hole solution surrounded by cloud of strings for concrete values of the parameters of such solution (see, for instance \cite{Toledo:2019amt}). In particular, the asymptotic structure of the metric potential is also modified, i.e., instead of recovering asymptotically flat solution $f(r \rightarrow \infty) \sim 1$, we obtain a slightly shifted modification, namely  $f(r \rightarrow \infty) \sim 1  +  Q_{\rm YM}$, in agreement with the predicted by the asymptotic behavior in a cloud of strings producing, $f(r \rightarrow \infty) \sim 1  +  C$, being $C$ a dimensionless constant. 
This similarity between non-linear charged black holes and a solution surrounded by a cloud of strings will be addressed in a future work.

A promising phenomenological approach to put constraints on the electric and Yang-Mills charges involves studying the main properties of the EHT's images of Sgr A$^{\star}$ and M87$^{\star}$ BHs within this new class of BH solution, such as the shadows and photon rings, surrounded by an optically and geometrically thin accretion disk. To achieve this, the Newman-Janis algorithm must be implemented to find the corresponding rotating charged BH solutions. 
%This is an immediate future work we plan to carry out.
However, we have recently made progress in the study of quasi-normal modes and shadows within this setup, as presented in our recent work \cite{Rincon:2023hvd}.

%%%%%%%%%%%%%%%%%%%%%%%%%%%%%%%%
\section*{Acknowledgments}
%%%%%%%%%%%%%%%%%%%%%%%%%%%%%%%%

G. G. acknowledges financial support from Vicerrector\'ia de Investigaci\'on, Desarrollo e Innovaci\'on - Universidad de Santiago de Chile, Proyecto DICYT,
C\'odigo 042031CM$\_$POSTDOC. A.R. is funded by the Generalitat Valenciana (Prometeo excellence programme grant CIPROM/2022/13) and by the Maria Zambrano contract ZAMBRANO 21-25 (Spain).

\bibliography{sample}

\end{document}